\documentclass[preprint]{aastex}

\newcommand{\Alf}{Alfv$\acute{\rm e}$n}
\newcommand{\bo}{B_{\rm 0}}
\newcommand{\bp}{B_{\rm 0\phi}}
\newcommand{\br}{B_{\rm R}}
\newcommand{\bz}{B_{\rm 0z}}
\newcommand{\cs}{c_{\rm s}}
\newcommand{\FB}{F_{\rm B}}
\newcommand{\GR}{G_{\rm R}}

\newcommand{\kr}{k_{\rm R}}
\newcommand{\kz}{k_{\rm z}}
\newcommand{\Of}{\Omega_{\rm f}}
\newcommand{\oo}{\omega}
\newcommand{\ooA}{\omega_{\rm A}}

\newcommand{\qA}{q_{\rm A}}
\newcommand{\qm}{q_{\rm m}}
\newcommand{\qR}{q_{\rm R}}
\newcommand{\qz}{q_{\rm z}}
\newcommand{\RA}{R_{\rm A}}

\newcommand{\Ree}{R_{\rm e}}
\newcommand{\Ri}{R_{\rm i}}
\newcommand{\Ro}{R_{\rm o}}
\newcommand{\rhoo}{\rho_{\rm 0}}
\newcommand{\toA}{\tilde{\omega}_{\rm A}}
\newcommand{\tom}{\tilde{\omega}}
\newcommand{\toF}{\tilde{\omega}_{\rm F}}
\newcommand{\tlt}{\tilde{\tau}}
\newcommand{\Vr}{v_{\rm R}}
\newcommand{\Voner}{v_{\rm 1R}}
\newcommand{\vor}{v_{\rm 1R}}
\newcommand{\Va}{v_{\rm A}}
\newcommand{\Vap}{v_{\rm A\phi}}
\newcommand{\Vaz}{v_{\rm Az}}
\newcommand{\xiR}{\xi_{\rm R}}
\newcommand{\xz}{x_{\rm z}}
\newcommand{\xr}{x_{\rm R}}

\shorttitle{Instabilities in MHD Winds and Disks}
\shortauthors{Kim \& Ostriker}

\begin{document}

\title{Magnetohydrodynamic Instabilities in Shearing, Rotating, 
Stratified Winds and Disks}

\author{Woong-Tae Kim and Eve C. Ostriker}
\affil{Department of Astronomy, University of Maryland,
College Park, MD 20742}

\email{kimwt@astro.umd.edu, ostriker@astro.umd.edu}

\slugcomment{Accepted for publication in the Astrophysical Journal
\hspace{0.3cm}}

\begin{abstract}

We investigate shear and buoyancy instabilities in radially stratified,
magnetized, cylindrical flows, for application to magnetocentrifugally
driven winds - such as those from protostars - and to magnetized accretion 
disks. Our motivation is to 
characterize the susceptibility of cold MHD disk winds to growing
internal perturbations, and to understand the relation of wind
instabilities to known accretion disk instabilities. 
Using four different linear analysis techniques, we
identify and study nine principal types
of unstable or overstable disturbances, providing numerical
and analytic solutions for growth rates for a wide range of parameters.

When magnetic fields are predominantly {\it toroidal}, as in 
protostellar winds far from their source, 
we find the system is susceptible to growth of five
different kinds of perturbations: axisymmetric fundamental and
toroidal resonance modes, axisymmetric and non-axisymmetric 
toroidal buoyancy modes, and non-axisymmetric
magnetorotational modes.  Winds having a sufficiently steep field gradient
($d\ln{B}/d\ln{R} < -0.75$ for a purely toroidal-field case) 
are globally unstable to the long wavelength fundamental mode
concentrated at small radii; these
promote the establishment of narrow dense jets in
the centers of wider winds. 
Long-wavelength outer-wind modes are all stable for power-law wind
equilibria. The toroidal buoyancy instabilities promote small-scale
radial mixing provided the equilibrium has nonzero magnetic forces.  
For low-temperature toroidal-${\bf B}$ winds, both axisymmetric and
non-axisymmetric magnetorotational instabilities have very low growth
rates. The stabilization of buoyancy instabilities by shear and of
magnetorotational instabilities by compressibility may be important
in allowing cold MHD winds to propagate over vast distances in space.

When magnetic fields are predominantly {\it poloidal}, as may occur in
protostellar winds close to their source or in astrophysical disks, 
we find the
system is susceptible to four additional growing modes: axisymmetric
magnetorotational (Balbus-Hawley), axisymmetric 
poloidal buoyancy, non-axisymmetric geometric buoyancy, and 
poloidal resonance modes.  The well-known
axisymmetric Balbus-Hawley mode has the fastest growth rate.  When the
magnetic field is nonuniform, the axisymmetric poloidal buoyancy mode
promotes radial mixing on small scales.  The geometric poloidal
buoyancy mode requires high $m$, thus is readily stabilized by shear.

Previous work on magnetorotational instabilities has concentrated on
near-incompressible systems (accretion disks or stellar interiors).
We extend this analysis to allow for compressibility (important in
winds). We introduce a ``coherent wavelet'' technique (a WKB temporal
approximation), and derive closed-form analytic expressions for 
instantaneous instability criteria,
growth rates, and net amplification factors for generalized non-axisymmetric
magnetorotational instabilities in compressible flows with both poloidal
and toroidal fields. 
We confirm that these are in excellent agreement with the results of
shearing-sheet temporal integrations, and that ``locally-axisymmetric''
perturbations have the largest amplifications only provided
$(\bold{k}\cdot\bold{\Va})/\Omega \lesssim 1$.

\end{abstract}


\section{Introduction}

The ubiquity of energetic molecular outflows and atomic jets from young
stellar objects (YSOs)
ranging from deeply embedded infrared sources to classical T Tauri stars
suggests that they are an inescapable by-product of star formation
\cite[e.g.,][and references therein]{ric00}.
Winds from YSO disks play an important role in shedding angular momentum
carried by inflowing material, thereby
permitting further accretion in order for the central objects to
attain stellar dimensions \citep{har82,pud86,shu88}.
These winds may also have strong effects
on the dynamical evolution of the parent cloud by providing a source of
turbulent energy \citep{nor80}, and may help to determine
the final masses of stars by reversing the infall of
surrounding gas \citep{shu87}. Therefore, understanding the physics
of protostellar winds is of essential importance to the theory of
star formation.

Among the various scenarios regarding the origin and nature of the
protostellar winds, the most promising is magnetohydrodynamic (MHD) models
in which winds are driven by the interaction of the centrifugal force with
open magnetic fields threading rapidly rotating disks. These
magnetocentrifugally driven
wind models can account for observed high mass and momentum losses
\citep{lad85, bac96} and the kinematic and structural
characteristics of bipolar molecular outflows in general \citep{li96},
as well as in specific cases \cite[e.g., HH111, cf.][]{nag97}.
It is, however, still controversial
whether the wind originates only from a small magnetosphere-disk interaction
region near the central star \citep{shu94,shu00}, or whether it
emanates from an extended region of the disk
(following the seminal model of Blandford \& Payne 1982;
see, e.g., K$\ddot{\rm o}$nigl \& Pudritz 2000).

Although the role of magnetic fields in driving protostellar winds is
by now well established (at least theoretically),
their complementary role in governing the
stability properties of winds is less well explored.  
In addition, azimuthal and vertical shear within winds may also
affect their stability properties.
Questions of
wind stability are potentially important for both large scale and
small-scale phenomena.  These include understanding the role of
magnetic fields and velocity shear
in (a) helping winds to propagate over enormous
distances (up to a factor $\sim 10^6$ in dynamic range) through the
ISM in parsec-scale giant HH flows \citep{rei97}; (b) creating
bright HH ``knots'' spaced throughout optical jets \citep{har00};
(c) governing the angular extent of the emergent wind and establishing
the momentum distribution for driving molecular outflows; and (d)
converting large-scale ordered flow energy to jet heating through
small-scale instabilities.  In addition, it is important to study the
dynamical properties of large classes of theoretical wind solutions to
test whether they are stable equilibria which can represent real
astronomical systems, or whether they are unstable equilibria which
should rarely be observed in nature because they evolve rapidly into
other configurations.

Previous studies of time-dependent behavior in steadily-input MHD
winds have focussed primarily on the Kelvin-Helmholtz instabilities
driven by the interaction of the wind surface layer with the ambient
medium \cite[see, e.g., the studies of][and references therein]{app92,
har92,ros99}. Generally, heavy jets containing strong toroidal
fields are relatively resistant to these instabilities. In addition to
``driven'' instabilities resulting from boundary conditions, winds may
also be subject to ``free'' instabilities in their interiors.  Whether
a given wind solution is internally unstable must depend on the
velocity shear, magnetic geometry, and internal stratification.  One route
to studying internal wind instabilities is via time-dependent numerical
simulations of winds.  Although such studies \cite[e.g.,]
[and references therein]{ouy97} have yielded intriguing results on 
the development of
episodic knots in MHD winds, the computational demands in carrying out
simulations precludes extensive exploration of parameter space, large spatial
dynamic range, or very long-term integration.  In addition, some of the
time-dependent internal features found in simulations may be introduced
by particular choices of inflow boundary conditions that are inconsistent
with a steady-state flow, rather than occurring as a result of intrinsic
instability of the wind.

Due to the importance of gaseous accretion disks in a wide variety
of astrophysical systems, major attention has focussed on disk dynamics,
and, in particular, the role of instability-driven turbulence in
angular momentum transport \cite[see, e.g.,][for reviews]
{bal98,sto00}. Saturated magnetorotational instabilities 
\cite[hereafter MRIs;][]{bal91,bal92,haw91,haw95}
represent perhaps the most important local
dynamical process affecting disk evolution. Magnetized disk winds
share many generic properties with disks, so it is interesting 
to investigate the potential importance of MRIs in winds.

In this work, we investigate the internal stability of rotating,
magnetized protostellar winds to (primarily) local shear and 
buoyancy modes. We also extend previous studies of local MRIs
in accretion disks.  The fundamental difference between
``wind'' and ``disk'' systems in our idealized models
is in the absence or presence of gravity
as a confining force.  These systems may also be distinguished by the
geometry of the magnetic field, with toroidally-dominant fields
expected in the wind case, but either poloidal or toroidal fields
possible for the disk case.
Our most general analysis and results apply to cold flows, but we
also perform separate calculations (see \S 7)
including thermal effects which specialize to local analysis
of MRIs.

Whether a wind originates from a narrow boundary layer or an extended
radial region, the radial expansion of the flow will lead to shear
in the azimuthal velocity field of the asymptotic state.
The total specific angular momentum of the flow,
$J = R(v_\phi - B_{\rm pol}B_\phi/(4\pi\rho v_{\rm pol}))$, is 
conserved along streamlines (where $B_{\rm pol}$, $B_\phi$ and $v_{\rm pol}$,
$v_\phi$ are the poloidal, toroidal components of the magnetic field 
and the flow 
velocity). If the \Alf\ mach number 
$M_{\rm A}\equiv v_{\rm pol}/v_{\rm A,pol} \gg 1$,
the kinetic part dominates the specific angular momentum 
and $v_\phi \approx J/R$. Thus, the angular velocity $\Omega =
v_\phi/R \approx J/R^2$ in the asymptotic wind will have a gradient
$d\ln \Omega/d\ln R= d \ln J /d\ln R - 2$. If the wind comes
from a boundary layer, then $d \ln J /d\ln R$ may be small;
if the wind originates from a large region with self-similar
scalings, then $d \ln J /d\ln R = 1/2$. In either case,
$d\ln \Omega/d\ln R$ is expected to be a negative, order-unity
quantity for wind systems. For thin disk systems (i.e.,
negligible pressure support), a Keplerian radial profile
$d\ln \Omega/d\ln R = -3/2$ is expected if the central
mass is dominant. For the analysis of this paper (except where
noted otherwise), we adopt
$d\ln \Omega/d\ln R=-3/2$ for both ``wind'' and ``disk'' systems,
but our qualitative results are insensitive to this assumption.
As discussed below, order-unity radial logarithmic gradients
may also be expected in the magnetic field strengths; we allow
for a range of magnetic gradients.

Significant shear may also exist in the poloidal velocities of jets
if they originate from an extended radial region. The asymptotic 
outflow speed $v_{\rm 0z}$ generally scales linearly with the rotational
speed at the footpoint $R_{\rm foot}$ of the streamline, 
so that 
\begin{displaymath}
\frac{\partial \ln v_{\rm 0z}}{\partial \ln R}=
\frac{\partial \ln v_{\rm 0z}}{\partial \ln v_{\rm foot}}
\frac{\partial \ln v_{\rm foot}}{\partial \ln R_{\rm foot}}
\frac{\partial \ln R_{\rm foot}}{\partial \ln R}
\sim -\frac{1}{2} \frac{\partial\ln R_{\rm foot}}{\partial \ln R}.
\end{displaymath}
If the range of footpoint radii is small compared to the range of
asymptotic radii (as for a wind from a boundary layer), then 
$|\partial\ln v_{\rm 0z}/\partial\ln R| \ll 1$; 
if the radial ranges are comparable,
then $\partial\ln v_{\rm 0z}/\partial\ln R$ is negative and order unity. 
For disks, the vertical velocity shear is negligible.
We allow for a range of vertical shear rates in the present analysis.

Our analysis consists of developing and solving sets of linearized MHD
equations for several general classes of background flows.  Both
axisymmetric and non-axisymmetric disturbances are explored.  Even in
linearized form, MHD problems present considerable technical
challenges. Thus, instead of attacking sophisticated sets of
steady-state wind solutions, for many specific examples we will take
as an unperturbed configuration one of the power-law cylindrical
equilibrium solutions recently identified by \citet{ost97} as
asymptotic states of self-similar disk winds. 
These have density $\rho \propto R^{-q}$, $\bold{B} 
\propto R^{-(1+q)/2}$, $\bold{v}
\propto R^{-1/2}$, and sound speed $\cs=0$. Although these adopted
initial configurations are relatively simple, they retain general asymptotic
characteristics of MHD disk winds in the sense that they have both
azimuthal and vertical magnetic field and velocity components with
arbitrary ratios, and all the physical variables have radial
gradients. These gradients may also be thought of as representing
local scalings within a more complex overall stratification.  We also
include models without vertical motion but with significant 
equilibrium gravity to study stability in magnetized
astrophysical disks. In our analysis of MRIs, we use equilibria with 
uniform $\bold{B}, \rho$, and $\cs\neq 0$, and take $\Omega\propto R^{-a}$
with arbitrary $a$.

Because the systems we are studying contain significant azimuthal
shear, an arbitrary initial spatial planform is not maintained
indefinitely. When $\Omega'\neq 0$ and the azimuthal wavenumber
$m\neq 0$ (and/or when $v_{\rm 0z}'\neq0$ and the vertical wavenumber
$\kz\neq0$; the prime represents a differentiation with respect to
$R$), spatial wavefunctions may become increasingly radially
corrugated in time due to the kinematic shearing of the wavefronts
imposed in the initial conditions. If we describe the radial 
spatial wavefunction in terms of the amplitudes of Fourier coefficients
with radial wavenumbers $\kr$, this corresponds at late times
to a secular increase in 
the amplitudes of large-$\kr$ terms and decrease in the amplitudes of
small-$\kr$ terms. As we shall show, all the disturbances 
we identify are stabilized at sufficiently large $\kr$. Thus, 
if $m\Omega'$ and/or $\kz v_{\rm 0z}' \neq 0$, the
net amplification factor for any arbitrary initial disturbance is
limited by the rate of kinematic growth of radial corrugation
compared to the growth rate of any dynamically-driven instabilities.

In previous work, two complementary analytical methods have been used
to study small-amplitude disturbances in shearing astrophysical
systems. One approach adopts the ``shearing-sheet'' formalism,
and integrates the local, time-dependent, linearized wave equations
directly to obtain the evolutionary behavior of shearing wavelets
treated as an initial-value problem \citep{gol65, jul66, bal92}.
An alternative analytical approach uses WKB techniques to derive
dispersion relations for spatial Fourier modes, superpositions of
which represent local wavefunctions \cite[e.g.,][]{shu74,shu92, 
ryu92,ter96}.
This paper includes analyses using both approaches, and also introduces
a hybrid technique which we term a ``coherent wavelet'' analysis.
We adopt the ``modal'' strategy in order to identify characteristic 
instantaneous growth rates and physical mechanisms for a wide
variety of spatial disturbances. By considering the time
over which a spatial pattern is altered by shear, we can estimate the
net amplification factor of a given initial modal disturbance.
We use the shearing-sheet formalism for studying magnetorotational
instabilities, which are cut off at relatively small values of
$R\kr/m$ (whereas the modal analysis applies to large $R\kr/m$), and
also for studying buoyancy instabilities in the high-$m$ regime where
modes are short-lived. We show that the results obtained from the
shearing-sheet integrations in both cases are in excellent
agreement with the predictions of a coherent wavelet analysis,
in which time-dependent growth rates $\gamma(t)$ are computed by
time-localizing the shearing-sheet equations and solving an
analytic dispersion relation.

The organization of this paper is as follows:
We begin by studying instabilities in cold, magnetized winds.
In \S 2, the basic MHD equations and the specific adopted wind equilibria
are described.
In \S 3, we analyze the stability of winds to the simplest perturbation
with $\kz=m=0$, where $\kz$ and $m$ are respectively the vertical and
azimuthal wavenumbers of the perturbation. We term these
the ``fundamental modes''; we present solutions for stable and unstable
global modes under the assumption of free Lagrangian boundary conditions.  
The modal analysis and general local dispersion relation for 
cold flows with arbitrary $\kz$
and $m$ are presented in \S 4.
We present numerical solutions of the
dispersion relation for both axisymmetric and
non-axisymmetric perturbations in \S 5.
In \S 6, we classify the unstable or overstable modes
and provide the physical interpretation for each mode.
Next, we include (variable) thermal pressure terms to compare
the susceptibility of cold winds vs. warm disks to shear instabilities.
In \S7 we analyze the axisymmetric Balbus-Hawley instability of 
poloidal fields and the non-axisymmetric MRI of
toroidal fields, 
discuss the respective instability mechanisms, and
provide the corresponding instability criteria. 
Here, we use the coherent wavelet technique to compute growth 
rates, and compare with direct shearing-sheet integrations.
The generalized instability criteria and net amplification factors for the
magnetorotational disturbances with both toroidal and poloidal 
background fields are also derived.
Finally in \S 8, we summarize and discuss conclusions of the present
work.

\section{Basic Equations and Cylindrical Equilibrium for Cold Wind}

We begin with the ideal MHD equations
\begin{equation}
\frac{\partial\rho}{\partial t} + \nabla\cdot (\rho \bold{v} ) = 0,
\end{equation}
\begin{equation}
\frac{\partial\bold{v}}{\partial t} + \bold{v}\cdot\nabla\bold{v}
= \frac{1}{4\pi\rho} (\nabla\times\bold{B})\times\bold{B}
-\frac{\nabla P}{\rho} - \nabla\Phi_{\rm G},
\end{equation}
\begin{equation}
\frac{\partial\bold{B}}{\partial t} = \nabla\times(\bold{v}\times\bold{B}),
\end{equation}
and
\begin{equation}
\nabla\cdot\bold{B} = 0,
\end{equation}
where $\rho$ is the density, $\bold{v}$ is the fluid velocity, $\bold{B}$ is
the magnetic field, $P$ is the thermal pressure, and $-\nabla \Phi_{\rm G}
\equiv -\bold{g}$ is the gravitational force due to a central object.
We ignore self-gravity in the flow.

We now consider cold, magnetized cylindrical flows. Since the flow velocity
in disk winds is always supersonic except in the vicinity of the disk
where the material is lifted by the thermal pressure \citep{bla82},
the thermal pressure term in eq.\ (2) can generally be neglected compared
to magnetic stress. Except for investigations of generalized
MRIs (\S 7), we shall drop the
thermal pressure term. We adopt standard cylindrical coordinates
$(R,\phi,z)$.

By assuming that $\Vr =\br =0$ and all variables are independent of $z$,
we have a general equilibrium condition from eq.\ (2)
\begin{equation}
\Omega^2R \equiv \frac{v_\phi^2}{R} = \frac{1}{4\pi\rho}
\left(\frac{B_\phi^2}{R} + \bold{B} \cdot
\bold{B}'\right) + g_{\rm R},
\end{equation}
where a prime denotes differentiation with respect to $R$. At a large
distance from the origin, the gravitational force due to the central
source can also be ignored on the grounds that magnetic and centrifugal
forces far exceed it. In this case, the magnetic hoop stress acting inward
is the only force that balances the outward centrifugal force and outward
magnetic pressure gradient force (under the assumption that the magnetic
field strength decreases outward).

As an initial equilibrium configuration of the wind, for specific cases we
will adopt the asymptotic solutions for cylindrically symmetric axial flows
presented by \citet{ost97}. All variables have power-law dependences on $R$:
$\rho \propto R^{-q}, B_{\phi} \propto B_{\rm z} \propto R^{-(1+q)/2}$,
and $v_{\phi} \propto v_{\rm z} \propto R^{-1/2}$. We define the local
pitch angle $i$ of the magnetic fields as
$i \equiv \tan^{-1}{(B_{\rm z}/B_{\phi})}$.
Neglecting $g_{\rm R}$ in eq.\ (5), radial momentum balance requires
\begin{equation}
v_{\phi}^2 = \frac{v_{\rm A}^2}{2}\left(\cos{2i} - q \right),
\end{equation}
where $v_{\rm A}$ is the local Alfv$\acute{\rm e}$n speed defined by
\begin{displaymath}
v_{\rm A}^2 \equiv v_{\rm A\phi}^2 + v_{\rm Az}^2\;\;\;\;
{\rm with} \;\;\;\; v_{\rm A\phi}\equiv
\frac{B_\phi}{\sqrt{4\pi\rho}}\;\;\;\;
{\rm and}\;\;\;\;v_{\rm Az}\equiv\frac{B_{\rm z}}{\sqrt{4\pi\rho}}.
\end{displaymath}

It is obvious from eq.\ (6) that there would be no such power-law solutions
if $q>1$. This is because when $q>1$, the magnetic field has so steep a
gradient that the corresponding pressure force always exceeds the tension.
Therefore, to ensure force balance and cylindrical collimation in winds
with power-law profiles, the magnetic field strength must decline with
$R$ more slowly than $R^{-1}$.

We can define an angular velocity $\Of \equiv 
\Omega-v_{\rm pol}B_\phi/(RB_{\rm pol})$ as that of a rotating frame in 
which the flow of winds is parallel to the local field line. $\Of$ is the 
rotation rate of the magnetic field lines thought of as rigid wires. 
In such a frame, the family of solutions can be completely described 
in terms of scaled values of the specific angular momentum $j$, 
the Bernoulli constant $e$, and $q$, where
\begin{displaymath}
j \equiv \frac{\Omega}{\Of}\left( 1 -
         \frac{(\Vap/R\Omega)^2}{1- \Of/\Omega}\right)\;\;\;\;{\rm and}
\;\;\;\;
e\equiv \frac{1}{(R\Of)^2}\left(
\frac{1}{2}\bold{v}^2 + \Phi_{\rm G} - R^2\Of\Omega \right),
\end{displaymath}
\citep{ost97}. The condition for a super-Alfv$\acute{\rm e}$nic outflow
velocity requires $0 < j \leq 1$. Generally speaking, the pitch angle $i$
does not depend on $q$, although one can parameterize $i$ in terms of $q$,
$e$, and $j$. However, for flows originating from a Kepler-rotating disk,
angular momentum and energy conservation requirements limit the range
of $i$ available to an equilibrium (asymptotic) magnetic field configuration.
Utilizing eqs.\ (15) to (23) of \citet{ost97} one can show that the maximum,
over all permitted values of $e$ and $j$, pitch angle $i_{\rm max}$ is
given by
\begin{equation}
\tan^2 i_{\rm max} = \left(\frac{4}{3+q}\right)^2 - 1,
\end{equation}
which is attained when $e=0$ and $j=1$. If $j>1$, the streamline never
reaches the Alfv$\acute{\rm e}$n radius.

\section{Fundamental Mode}

\subsection{Dynamical Equations}

We first consider the response of the equilibrium state when small,
axisymmetric perturbations with an infinite wavelength along a vertical
direction are imposed. We term the waves with $\kz=0$ and $m=0$
the ``fundamental modes'' analogous to eigenfunctions of oscillations without
any node. Let the subscripts 0 and 1 denote the equilibrium and perturbed
states, respectively. Linearizing the set of the dynamical equations (1)
to (4), we may write
\begin{equation}
\frac{\partial \rho_1}{\partial t} =
-\frac{1}{R}\frac{\partial}{\partial R} \left(R \rhoo v_{\rm 1R}\right),
\end{equation}
\begin{equation}
\frac{\partial v_{\rm 1R}}{\partial t}
= 2\Omega v_{1\phi} - \frac{1}{4\pi\rhoo}
\left\{
\frac{2\bp B_{1\phi}}{R} + \frac{\partial}{\partial R}
  \left( \bold{B}_{\rm 0} \cdot \bold{B}_1 \right)
     -\frac{\rho_1}{\rhoo}
  \left( \frac{\bp^2}{R} + \bold{B}_{\rm 0} \cdot \bold{B}_{\rm 0}'
\right) \right\},
\end{equation}
\begin{equation}
\frac{\partial v_{1\phi}}{\partial t} =
-\frac{\kappa^2}{2\Omega}v_{\rm 1R},
\end{equation}
\begin{equation}
\frac{\partial v_{\rm 1z}}{\partial t} =
-v_{\rm 0z}' v_{\rm 1R},
\end{equation}
\begin{equation}
\frac{\partial B_{1\phi}}{\partial t} =
-\frac{\partial}{\partial R}\left(B_{\rm 0 \phi} v_{\rm 1R}\right),
\end{equation}
\begin{equation}
\frac{\partial B_{\rm 1z}}{\partial t} =
- \frac{1}{R}\frac{\partial}{\partial R}\left(
RB_{\rm 0z}v_{\rm 1R}\right),
\end{equation}
and $B_{\rm 1R} = 0$. In eq.\ (10), $\kappa$ stands for the epicycle
frequency
$$
\kappa^2 \equiv \frac{1}{R^3}\frac{d}{dR}(R^4\Omega^2).
$$

Combining the perturbed equations (8)$-$(13) and eliminating all other
variables in favor of the perturbed radial velocity $\Voner$, one obtains
the wave equation
\begin{mathletters}
\begin{eqnarray}
\frac{1}{v_{\rm A}^2}\frac{\partial^2\Voner}{\partial t^2} &=&
\frac{\partial^2\Voner}{\partial R^2}
+ \frac{d\ln{(R\bo^2)}}{dR} \frac{\partial\Voner}{\partial R}
\nonumber \\
&-& \left[
\frac{\kappa^2}{v_{\rm A}^2} + \frac{1}{R^2}\left(
1 + \cos^2{i}\frac{d\ln{\rhoo}}{d\ln{R}}\right)
- \frac{\rhoo}{R\bo^2}\frac{d}{dR}
\left(\frac{R}{\rhoo}\bold{B}_0\cdot\bold{B}_0'\right)
\right]
\Voner.
\end{eqnarray}
\eqnum{14a}
\end{mathletters}
For power-law profiles, this becomes
\begin{equation}
\frac{1}{v_{\rm A}^2}
\frac{\partial^2\Voner}{\partial t^2} =
\frac{\partial^2\Voner}{\partial R^2} -
\frac{q}{R}
\frac{\partial\Voner}{\partial R} -
\left[\frac{\kappa^2}{v_{\rm A}^2} + \frac{1}{R^2}
     \left(\frac{1-q}{2} - q\cos^2i
     \right)
\right]
\Voner.
\eqnum{14b}
\end{equation}

To transform eqs.\ (14) to the ${\rm Schr\ddot{o}dinger}$ form, 
we define a new independent
variable $\Psi$ through
\begin{equation}
\Voner = \frac{\Psi(R)}{\sqrt{R\bo^2}}e^{i\omega t};\;\;\;{\rm or}\;\;\;
\Voner = R^{q/2}\Psi(R)e^{i\omega t}
\end{equation}
for the power-law case. Then, we have
\begin{equation}
\frac{d^2\Psi}{dR^2} + K^2(R)\Psi = 0,
\end{equation}
with $K(R)$ defined by
\begin{equation}
K^2(R) \equiv \frac{\omega^2-\kappa^2}{\Va^2}
-\frac{3}{4R^2}
-\frac{d\ln{\rhoo}}{dR}\left(\frac{\cos^2{i}}{R} +
\frac{\bold{B}_0\cdot\bold{B}_0'}{B_0^2}\right)
+ \left(\frac{\bold{B}_0\cdot\bold{B}_0'}{B_0^2}\right)^2,
\eqnum{17a}
\end{equation}
or
\begin{equation}
K^2(R) \equiv \frac{\omega^2-\kappa^2}{\Va^2} - \frac{1}{R^2}
\left(
\frac{1}{2} + \frac{q^2}{4} - q\cos^2i
\right)
\eqnum{17b}
\end{equation}
\setcounter{equation}{17}
for the power-law case.
Local (WKB) solutions to eq.\ (16) have $\Psi \sim e^{i\int \kr dR}$
with $R\kr \gg 1$ and $d\kr/dR \ll \kr^2$. 
In this case, $\Psi'' \rightarrow - \kr^2\Psi$,
and we can use $K^2(R) = \kr^2 $ to write a local dispersion relation 
\begin{equation}
\omega^2 = \Va^2 \kr^2, 
\end{equation}
which corresponds to MHD fast modes propagating along the radial direction.
When $|\omega|^2$ is comparable to or smaller than $\Va^2/R^2$, however,
modes are not localized, and solutions must be
sought as a global problem subject to boundary conditions.

\subsection{Global Analysis for the Fundamental Modes}

In the previous section we showed that there is no short wavelength 
(local) unstable 
fundamental mode with $\kz=m=0$ in self-similar MHD disk winds.
Here, we present the results of a global normal-mode analysis
performed with carefully chosen boundary conditions, and 
adopting the power-law equilibrium.  
Define the dimensionless radial variable $r \equiv R/\Ree$,
and dimensionless parameters
$\alpha \equiv  (q-1)\cos^2{i} + q(2-q)/4$
and $\sigma^2 \equiv \omega^2 \Ree^2/v_{\rm A}^2(\Ree)$,
with $\Ree$ being the position of the unperturbed outer edge of the wind.
Then eq.\ (16) can be cast into the form
\begin{equation}
\frac{d^2\Psi}{dr^2} + \left(\frac{\alpha}{r^2} + \sigma^2r\right)\Psi = 0.
\end{equation}
It is not difficult to show that the general solutions of eq.\ (19) are
\begin{displaymath}
\hskip 2.8cm
\Psi = \left\{ \begin{array}{ll}
A\sqrt{r}J_{\nu}(2\sigma r^{3/2}/3) + B\sqrt{r}Y_{\nu}(2\sigma r^{3/2}/3),
&{\rm if} \quad \sigma^2 > 0,
\hskip 2.15cm
{\rm (20a)}
\\
& \\
C\sqrt{r}I_{\nu}(2|\sigma| r^{3/2}/3) + D\sqrt{r}K_{\nu}(2|\sigma| r^{3/2}/3),
&{\rm if} \quad \sigma^2 < 0,
\hskip 2.15cm
{\rm (20b)}
\end{array} \right.
\end{displaymath}
\setcounter{equation}{20}
with $3\nu \equiv \sqrt{1-4\alpha} = \sqrt{(1-q)^2 + 4(1-q)\cos^2{i}}$.
In eqs.\ (20), $J_{\nu}$, $Y_{\nu}$, and $I_{\nu}$, $K_{\nu}$ are the
ordinary and modified Bessel functions of the 1st and 2nd kinds,
respectively, and the coefficients $A, B, C$, and $D$ are constants to
be determined from imposed boundary conditions.

Let us consider the case of a free Lagrangian boundary at which the total
pressure due to initial and perturbed fields balances with a fixed external
pressure at both inner and outer edges, 
which is equal to the unperturbed magnetic pressure. If the total
pressure at an edge of an outflow is different from the external pressure,
the boundary itself will move until a new balance exists. To first order,
this condition of constant pressure at the boundary is written
\begin{equation}
\frac{1}{2}\frac{dB_{\rm 0}^2}{dR}\frac{\partial R_{\rm b}}
{\partial t} + \bold{B}_{\rm 0}\cdot \frac{\partial \bold{B}_{\rm 1}}
{\partial t} = 0,
\end{equation}
where $R_{\rm b}$ is the location of the perturbed boundary. The first term
of eq.\ (21) represents the change in the total pressure due to the boundary
movement, while the second term arises from perturbed magnetic pressure itself.
All quantities are evaluated at the unperturbed boundary position. Using
eqs.\ (12), (13), and (15), and using $\partial R_{\rm b}/\partial
t= v_{\rm 1R}$ at boundaries, we find the desired boundary conditions are
\begin{equation}
\frac{d\Psi}{dr} + \frac{q/2+\sin^2i}{r}\Psi = 0, \;\;\;{\rm at}\;\;\;
r = r_{\rm i}\;\;{\rm and}\;\;1,
\end{equation}
where $r_{\rm i} \equiv \Ri/\Ree$ is the normalized distance of an 
inner boundary
from the axis. Together with the boundary conditions (22), eq.\ (19) forms
a Sturm-Liouville system. By employing the variational principle one can
show that $\sigma^2$ is real and that $\sigma^2(\Psi)$ is stationary subject
to an arbitrary variation of $\Psi$.

When $\sigma^2 > 0$ (stable modes), the oscillatory properties of $J_{\nu}$
and $Y_{\nu}$ guarantee the existence of discrete eigenvalues $\sigma_{\rm n}$
with $n$ denoting the number of nodes in the corresponding eigenfunction
$\Psi_{\rm n}$. The resulting eigenvalues for $r_{\rm i} = 10^{-1}$ and
$10^{-4}$, and $0< i < i_{\rm max}(q)$ are plotted in Fig.\ 1.
Only a few cases with small $n$ are shown. Eigenvalues associated with
different $i$'s fill each shaded area completely. When $r_{\rm i}= 10^{-4}$,
eigenfunctions which link inner and outer boundaries have to extend across
enormous changes in density and magnetic field strengths. In this case,
$B/A \ll 1$ and  eigenvalues become rather insensitive to the local
properties such as $q$ and $i$. When $r_{\rm i}=10^{-1}$, however, the
wind mimics a slender hollow cylinder. The variation in density and field
strengths over radius is slight, causing eigenvalues to be sensitive
to $i$ and $q$. In addition, the narrow width of the wind changes
the number of nodes in eigenfunctions. When $q>0.5$, for example,
the eigenfunctions with $r_{\rm i}=10^{-1}$ have almost the
same eigenvalues as, but one more node than, the $r_{\rm i}=10^{-4}$ case,
as seen in Fig.\ 1.

When $r_{\rm i} \ll 1$ and $\sigma^2 \gg 1$, the asymptotic solutions
to eqs.\ (20a) and (22) gives $ \sigma_{\rm n} = 3\pi n/2 + 3\pi(2\nu+1)/8$.
These are plotted with dotted lines in Fig.\ 1b, and show good agreement
with the values calculated without any assumption (even for $n=0$).
The case with $q=0.5$ and $i=0$ is special, because the slope of the
eigenfunction at the inner boundary is $-1/4$, which automatically satisfies
the boundary condition (22). In this case the asymptotic eigenvalues are
$\sigma_{\rm n}=3\pi n/2$, drawn as filled circles in Fig.\ 1b.
Eigenvalues have no upper bound as $n\rightarrow \infty$, which is a general
property of solutions to a Sturm-Liouville equation \citep{mor53}.

Now consider the unstable global solutions with $\sigma^2<0$.
Let $\psi_1$ and $\psi_2$ be the two linearly independent solutions
of $\Psi$: $\psi_1 \equiv \sqrt{r}I_{\nu}(2|\sigma| r^{3/2}/3)$
and $\psi_2 \equiv \sqrt{r}K_{\nu}(2|\sigma| r^{3/2}/3)$
such that $\Psi = C\psi_1 + D\psi_2$.
Because $\psi_1$ is a monotonically increasing function of $r$ (i.e.,
$\psi_1, \psi_1' > 0$ always) and $\Psi$ must have
a negative logarithmic slope at the inner and outer boundaries (cf.\
eq.\ [22]),
$\psi_1$ alone can not constitute eigenfunctions for global modes.
In addition,
since $\psi_1$ increases exponentially for a large value
of $|\sigma|r^{3/2}$, while $\psi_2, \psi_2' \rightarrow 0$,
the outer free Lagrangian boundary condition requires $C/D \rightarrow 0$.
Although $C$ is not strictly zero when $|\sigma|$ has a relatively
small value, the contribution of $\psi_1$ to global solutions near the
inner boundary is negligibly small. Thus, unstable eigenvalues, 
if they exist,
are essentially determined by the inner boundary condition imposed
on $\psi_2$.

As we move inward from the outer boundary, $\psi_2$ rapidly
increases asymptoting to
\begin{equation}
\psi_2 \sim r^{(1-3\nu)/2}
\left[ 1 - \frac{\pi\nu |\sigma|^{2\nu}}{3^{2\nu}\sin(\pi\nu)\Gamma^2(\nu+1)}
r^{3\nu} + \cdot\cdot\cdot \right],
\end{equation}
for $r \ll  1$ \cite[cf.][]{abr65}, where
$\Gamma(\nu+1)$ is a Gamma function.
In fact, $(1-3\nu)/2$ is the maximum logarithmic slope $\psi_2(r)$
can ever attain.
From the inner boundary constraint (22), the existence
of unstable global solutions is guaranteed if
$(1-3\nu)/2 > -(q/2+\sin^2{i})$, or, equivalently
\begin{equation}
q > 1 - \frac{(1+\sin^2{i})^2}{2},
\end{equation}
is satisfied. Eq.\ (24) is the global instability criterion for the
fundamental modes of self-similar, cold, magnetized winds, subject to
the free boundary conditions expressed by eq.\ (22). 
By putting $C=0$ and neglecting higher order
terms in $\psi_2$, we derive from eqs.\ (22) and (23) the approximate,
analytic expression for the eigenvalues of the global instability
\begin{equation}
|\sigma|r_{\rm i}^{3/2} =
\frac{|\omega| \Ri}{\Va(\Ri)} = 
3\left[\frac{\sin(\pi\nu)\Gamma^2(\nu+1)}{\pi\nu}
\frac{(1-3\nu+q+2\sin^2i)}{(1+3\nu+q+2\sin^2i)}\right]^{1/2\nu},
\end{equation}
which again shows that $|\sigma|$ has a positive real value if
eq.\ (24) holds.

In Fig.\ 2  we plot the approximate growth rates for
unstable modes from eq.\ (25) as dotted lines, 
as well as the exact growth rates numerically computed
for $r_{\rm i}=10^{-4}$ (thin solid lines) and for
$r_{\rm i}=0.1$ (dashed lines).
The curves shown are for $i=0^{\rm o}, 5^{\rm o},\cdot\cdot\cdot,35^{\rm o},
40^{\rm o}$ from right to left, and the uppermost thick lines are for
$i_{\rm max}$ calculated from eq.\ (7). Note that varying the width of 
outflow via $r_{\rm i}$ yields very little change in the plotted
solutions: $r_{\rm i}$-dependence of the growth rates appears mainly through
the product $|\sigma|r_{\rm i}^{3/2}$. Eq.\ (25) gives accurate
growth rates for relatively small values of $|\sigma| r_{\rm i}^{3/2}$,
while its estimates deviate up to $\sim 16\%$ from the exact values 
as $|\sigma| r_{\rm i}^{3/2}$ becomes larger.
In this case we need to include next order terms in $\psi_2$
(cf.\ eq.\ [23]) to obtain more accurate results.
For a given set of equilibrium parameters, we note that whereas
there exist an infinite set of stable eigenmodes, there is (at most)
a unique unstable eigenmode.
Noting
$|\omega| \equiv |\sigma|v_{\rm A}(\Ree)/\Ree = |\sigma|r_{\rm i}^{3/2}
v_{\rm A}(R_{\rm i})/R_{\rm i}$, we expect from Fig.\ 2 that the system
is typically globally unstable within $\sim$5 crossing times of \Alf\ waves
at the inner boundary.

Once the ratios of the coefficients and the eigenfrequencies are found
for the fundamental modes,
one can easily construct radial solutions for the
perturbed variables: $\rho_1$, $\Voner$,
$v_{1\phi}$, and $B_{1\phi}$. These are plotted in Fig.\ 3
for $i=0^{\rm o}$ and $r_{\rm i}=10^{-3}$. 
Fig.\ 3a corresponds to a stable case with
$q=0.4$ and $\sigma_0=2.42$, while Fig.\ 3b depicts an unstable case with
$q=0.6$ and $|\sigma|r_{\rm i}^{3/2}=0.11$.
Although normalization is arbitrary, we note
that for the unstable modes, 
the negative radial velocity case drives the entire system into
a more stable configuration (with lower magnetic energy)
when the equilibrium magnetic field is predominantly toroidal.
This can be shown as follows:
Let $\delta M$, $\delta E_{\rm B}$, and $\delta \Phi_{\rm B}$ denote 
the mass, magnetic energy, and toroidal magnetic flux 
per unit height in a local flux tube. Then we have
$\delta E_{\rm B} = (\pi/2)(\delta \Phi_{\rm B}/\delta M)^2 \delta M
\rho R^2$. For a given flux tube, 
$\delta M$ and $\delta \Phi_{\rm B}/\delta M$
are constant in time and $\delta M>0$. Thus,
${\rm sgn}\; d(\delta E_{\rm B}) /dt = {\rm sgn} \; d(\rho R^2)/dt =
{\rm sgn}\; [\rho_0 R^2 (\vor/R - \partial \vor/\partial R)]$ from
the equation of continuity. 
If $\vor/R$ dominates $\partial \vor/\partial R$ and $\vor<0$, 
then ${\rm sgn}\; d(\delta E_{\rm B}) /dt <0$; magnetic 
energy decreases with time, meaning that the system evolves into a
more stable state.
Thus we scale
$v_{\rm 1R}/v_{0\phi} = 1$ at $r=1$ for Fig.\ 3a and
$v_{\rm 1R}/v_{0\phi} = -1$ at $r=r_{\rm i}$
for Fig.\ 3b, respectively.
Note that stable eigenfunctions have their largest amplitude 
near the outer boundary, while the inner, high density region is nearly 
static during oscillation. Unstable eigenfunctions, on the other hand,
are almost zero except in the region close to the inner boundary.
The respective inner-region vs.\ outer-region predominance of unstable 
vs.\ stable eigenfunctions reflects the respective characteristic 
frequencies as well: the inner-wind unstable modes grow at large rates
comparable to \Alf\ frequencies in the interior, whereas outer-wind
stable modes oscillate at low frequencies comparable to the 
\Alf\ frequencies in the exterior.

We remark that there is no globally unstable fundamental mode when one
adopts rigid boundaries with $\Psi(r)=0$ at both $r=r_{\rm i}$ and
$r=1$, instead of the free Lagrangian boundaries,
since both $\psi_1$ and $\psi_2$ are monotonic functions of $R$.
If $\Psi'(r)=0$ is imposed at both boundaries \cite[cf.][]{dub93},
however, we still have unstable fundamental modes
with a different instability criterion\footnote{
In this case, eq.\ (24) would become
$(1-3\nu)/4=\alpha > 0$, corresponding to
$R^2(K^2(R) - \omega^2/\Va^2) > 0$.} and different growth rates.

We discuss the significance of fundamental modes to protostellar 
outflows in \S 8.3.

\section{Local Analysis for Cold Winds}

We now consider general non-axisymmetric Eulerian perturbations with small
amplitudes. Neglecting the effects of thermal pressure and external
gravity due to a central object, we linearize eqs.\ (1)$\sim$(4)
\begin{equation}
\frac{d\rho_1}{dt} = - \frac{1}{R}\frac{\partial}{\partial R}(R\rho_0 \vor)
         - \frac{1}{R}\frac{\partial}{\partial\phi}(\rho_0 v_{1\phi})
         - \frac{\partial}{\partial z}(\rho_0 v_{1z}),
\end{equation}
\begin{equation}
\frac{d\vor}{dt} = 2\Omega v_{1\phi}
+\frac{1}{4\pi\rho_0}\left[
-\frac{2\bp B_{1\phi}}{R}
+ \frac{\bp}{R}\frac{\partial B_{\rm 1R}}{\partial \phi}
+ \bz\frac{\partial B_{\rm 1R}}{\partial z}
-\frac{\partial}{\partial R} (\bold{B}_0\cdot\bold{B}_1)
+\frac{\rho_1}{\rho_0}
\left(\frac{\bp^2}{R} + \bold{B}_0 \cdot \bold{B}_0' \right)
\right],
\end{equation}
\begin{equation}
\frac{dv_{1\phi}}{dt} = - \frac{\kappa^2}{2\Omega}\vor
+\frac{1}{4\pi\rho_0}\left[
\left(\bp'+\frac{\bp}{R}\right)B_{\rm 1R}
+\bz\left(\frac{\partial B_{1\phi}}{\partial z}
- \frac{1}{R}\frac{\partial B_{\rm 1z}}{\partial \phi}\right)
\right],
\end{equation}
\begin{equation}
\frac{dv_{\rm 1z}}{dt} = - v_{\rm 0z}' \vor
+\frac{1}{4\pi\rho_0}\left[
\bz'B_{\rm 1R} + \bp\left(
\frac{1}{R}\frac{\partial B_{\rm 1z}}{\partial \phi}
-\frac{\partial B_{1\phi}}{\partial z}\right)
\right],
\end{equation}
\begin{equation}
\frac{dB_{\rm 1R}}{dt} = \frac{\bp}{R}\frac{\partial \vor}{\partial \phi}
+\bz\frac{\partial \vor}{\partial z},
\end{equation}
\begin{equation}
\frac{dB_{1\phi}}{dt} = R\Omega'B_{\rm 1R}
-\frac{\partial}{\partial R}(\bp\vor)
+\bz\frac{\partial v_{1\phi}}{\partial z}
-\bp\frac{\partial v_{\rm 1z}}{\partial z},
\end{equation}
\begin{equation}
\frac{dB_{\rm 1z}}{dt} = v_{\rm 0z}' B_{\rm 1R}
-\frac{1}{R}\frac{\partial}{\partial R}(R\bz\vor)
+\frac{\bp}{R}\frac{\partial v_{\rm 1z}}{\partial \phi}
-\frac{\bz}{R}\frac{\partial v_{1\phi}}{\partial \phi},
\end{equation}
where the Lagrangian time derivative is denoted by 
\begin{equation}
\frac{d}{dt} \equiv \frac{\partial}{\partial t}
+ \Omega(R)\frac{\partial}{\partial \phi}
+ v_{\rm 0z}\frac{\partial}{\partial z},
\end{equation}
and again the subscripts
0 and 1 indicate the equilibrium and perturbation
variables, respectively.

Since all the coefficients of the perturbed variables in eqs. (26)$\sim$(32)
do not depend on the coordinates $\phi$ and $z$, we may look for
solutions having sinusoidal dependence on $\phi$ and $z$.
Furthermore, if there exist any normal modes, 
we can write eigenfunctions in the form
\begin{equation}
\chi_1(R,\phi,z,t) = \chi_1(R)e^{i(m\phi + \kz z - \omega t)},
\end{equation}
where $\chi_1$ refers to any physical variable of perturbations.
Substituting eq.\ (34) into the set of
eqs.\ (26)$\sim$(32) and eliminating all other variables in terms of
the radial Lagrangian displacement $\xiR \equiv - \Voner/i\tom$ with
a Doppler shifted frequency 
\begin{displaymath}
\tom \equiv \omega - m\Omega - \kz v_{\rm 0z},
\end{displaymath}
we obtain the second order differential equation
\begin{equation}
\frac{d^2\xiR}{dR^2} + \frac{d}{dR}\ln\left(RB_{\rm 0}^2\frac{\toA^2}
{\toF^2}\right)\frac{d\xiR}{dR} + \frac{H(R)}{\toA^2}\xiR = 0,
\end{equation}
where
\begin{mathletters}
\begin{eqnarray}
H(R) &\equiv &
 \toF^2 \left\{ \frac{\toA^2-\kappa^2}{\Va^2} -
\FB\frac{d\ln{\rho_{\rm 0}}}{dR}
- \frac{1}{R^2} + \frac{1}{R\bo^2} \frac{d}{dR}
\left(R \bold{B}_0 \cdot \bold{B}_0' \right)
\right\} \nonumber
\\
&-& 4\Omega\left\{
\tom\left( \frac{m}{R}\FB + \frac{\bz}{R\bo}G_+ \right)
+\Omega k^2 \frac{\bz^2}{\bo^2}
- \Va^2 k^2 \FB \frac{\bp(\bold{k}\cdot\bold{B}_0)}{\tom\bo^2}
\right\}
\\
& - & \left(\frac{d}{dR}\ln{\frac{\toF^2}{R\bo^2}}
       - \frac{d}{dR} \right)
\left\{
\Va^2 \FB G_-^2 + 2\Omega G_-\tom\frac{\bz}{\bo} + \frac{\Va^2}{R}G_+G_-
\right\}
\nonumber 
\\
& - &\Va^2 \left\{
k^2\FB^2 \frac{\toA^2}{\tom^2} + \frac{2}{R} \FB G_+ G_-
+ \frac{G_+^2}{R^2}
\right\}, 
\nonumber
\eqnum{36a}
\end{eqnarray}
\end{mathletters}
with
\begin{equation}
\FB \equiv
\frac{1}{\bo^2}\left(\frac{\bp^2}{R} + \bold{\bo} \cdot \bold{\bo}'\right),
\quad
G_{\pm} \equiv \frac{1}{\bo}\left(\frac{m}{R}\bz \pm \kz\bp \right),
\eqnum{36b}
\end{equation}
\begin{displaymath}
\toA^2 \equiv \tom^2 - v_{\rm A}^2(\bold{k}\cdot\bold{b}_0)^2,
\quad
\toF^2 \equiv \tom^2 - v_{\rm A}^2k^2,
\quad {\rm and} \quad
\bold{k} \equiv (0, \frac{m}{R}, \kz).
\end{displaymath}
\setcounter{equation}{36}
Here, $\FB$ represents the equilibrium magnetic force, and $\toA$ and
$\toF$ are frequencies connected to the Alfv${\acute{\rm e}}$nic and fast
magnetosonic modes in cold MHD fluids, respectively. $\bold{b}_0\equiv
\bold{\bo}/|\bold{\bo}|$ is the
unit vector along an equilibrium  field direction, 
and finally $\bold{k}$ is a vector
wavenumber. When $m=\kz=0$, only the terms in the first bracket in the
definition of $H(R)$ do not vanish, recovering the radial wave equation
(eq.\ [16], [17a]) for the fundamental mode.

The second order differential equation (35) has a singularity at $\toA^2=0$,
but if we treat a fully general problem including the thermal effects of
compressible gas, we will find another singularity (a so called 
cusp singularity) at the positions
where  Doppler shifted frequencies of traveling waves match with
slow MHD wave frequencies of the medium \cite[cf.][]{rob85}. 
For an incompressible medium,
$\toA^2=0$ singularities are often referred to as shear \Alf\ singularities
where because of resonances the characteristics of waves propagating
radially would be modified to be either absorbed into or amplified by
background medium, if considered as a boundary value problem
\citep{ros82, cur96}. 
As pointed out by \citet{app74}, the locations with $\toF^2=0$ in eq.\ (35)
are not singularities; these cut-off points
in our local analyses appear as resonance waves with
frequencies having relatively small imaginary parts, suggesting potential
attenuation or amplification of amplitudes.

To remove the second term in eq.\ (35) we further define
\begin{equation}
\xiR \equiv \Psi
\left(RB_{\rm 0}^2\frac{\toA^2}{\toF^2}\right)^{-1/2}.
\end{equation}
Then eq.\ (35) is reduced to the standard ${\rm Schr\ddot{o}dinger}$ form of 
eq.\ (16), with generalized $K^2(R)$ defined by
\begin{equation}
K^2(R) \equiv  \frac{H(R)}{\toA^2} - \frac{1}{2}\frac{d^2}{dR^2}
\ln{\left(RB_{\rm 0}^2\frac{\toA^2}{\toF^2}\right)}
-\frac{1}{4}\left[
\frac{d}{dR}\ln{\left(RB_{\rm 0}^2\frac{\toA^2}{\toF^2}\right)}\right]^2.
\end{equation}

Generally speaking, $K(R)$ is a function of $R$ for
fixed values of $m$, $\kz$, and $\omega$. 
However, we can still consider the behavior
in a local sense near some fixed $\Ro$, such that $K$ is close to
$K(\Ro)$. This is mathematically formalized as described in \citet{lin93}
and \citet{ter96}. Let us consider in the nonuniform background the spatially
localized wave packet of the form
\begin{displaymath}
\Psi = \psi(R-R_{\rm o})e^{i\kr(R-R_{\rm o})} + O(\frac{1}{\kr}),
\end{displaymath}
where $\psi(r)$ is a function which is non-zero only in a small neighborhood
of $r \equiv R-\Ro =0$. The scale over which $\psi(r)$ varies significantly
must tend to zero as $\kr\rightarrow \infty$, but no faster than
$\kr^{-1}$. Then, to leading order, $d^2\Psi/dR^2 \approx -\kr^2\Psi$,
and the solution $\kr^2 = K^2(R,\tom,m,\kz)$ of the
${\rm Schr\ddot{o}dinger}$ equation
yields a local dispersion relation
with the right hand side evaluated at a reference point $\Ro$,
provided $\kr$ is limited to a sufficiently large value 
(i.e., $\Ro\kr \gg 1$). We may invert this dispersion relation to
find $\tom = W(m,\kz,\kr,R)$, so that $\kr$ now plays the role of
an independent parameter and the dispersion relation yields the 
Doppler-shifted frequency $\tom$ of a wave near a position $\Ro$
having local wavevector $\bold{k}$.
This is equivalent to a standard WKB approximation in the radial direction.

Solution of the local dispersion relation near $\Ro$ yields
\begin{equation}
\omega = W(m,\kz,\kr,\Ro) + m\Omega(\Ro) + \kz v_{\rm 0z}(\Ro) 
+ {\cal O}(r),
\end{equation} 
where the ${\cal O}(r)$ term is
$(m\Omega'(\Ro) + \kz v_{\rm 0z}'(\Ro))r$.
Defining 
${d\kr}/{dR} \equiv - ({\partial W}/{\partial R}) /
({\partial W}/{\partial \kr}) \sim \kr/R$, 
the WKB condition $|d\kr/dR| \ll \kr^2$ will be satisfied for $\kr R\gg 1$. 
For normal mode solutions, the ${\cal O}(r)$ term in $\omega$ must 
provide a negligible contribution to the phase;
this requires that we must have
$|m\Omega'(\Ro) + \kz v_{\rm 0z}'(\Ro)|t \ll \kr$. 
For axisymmetric modes with negligible vertical shear, this is always
satisfied. However, for $m\neq 0$ disturbances, or flows with
non-negligible $\kz v_{\rm 0z}'$, 
spatially localized wavepackets maintain a characteristic radial
wavenumber for only a limited time, altering their spatial pattern because
of the background shear. 
For a wavepacket with initial wavenumber $\kr(0)$, the radial wavenumber 
at time $t$ becomes, upon inclusion of $-t$ times the ${\cal O}(r)$ term
in $\omega$ in the phase,
\begin{equation}
\kr(t) = \kr(0) - (m\Omega'+\kz v_{\rm 0z}')t.
\end{equation}
Thus, for example, with $\kz=0$, the pitch
$\tan p \equiv m/R\kr$ of a spiral pattern changes by a fraction
$\epsilon = |d\kr|/\kr$ over time
$t = \epsilon \kr /|m\Omega'|$. If $R\kr \gg m$, the pattern
changes slowly compared to the orbit time. 
Among nonaxisymmetric disturbances, the wavepackets with low $m/R\kr$ 
have the largest temporal range for which they remain close to normal
modes of the system.
In the following two sections, we present solutions for the growth
rates of unstable disturbances determined from a local modal 
analysis (i.e., producing solutions $\tom = W({\bold k},\Ro))$, 
with the understanding that when 
$|m\Omega' + \kz v_{\rm 0z}'| \neq 0$, the modal growth
(i.e., $\sim e^{|\tom|t}$) with fixed pattern holds only for a limited time.
In assessing the potential of the non-axisymmetric 
instabilities we shall identify to affect flow dynamics, we will
consider their total amplification over times $< \kr R/m\Omega$, for which
the spatial pattern changes little.

For simplicity let us define dimensionless variables
\begin{displaymath}
\sigma \equiv \tom\Ro / {\Va(\Ro)},\; \xz \equiv \kz \Ro,\;
\xr \equiv \kr \Ro,\;
\kappa \equiv {\Ro\kappa_{\rm o}}/{\Va(\Ro)},
\end{displaymath}
\begin{displaymath}
\Omega \equiv {\Ro\Omega_{\rm o}}/{\Va(\Ro)}, \;{\rm and}\;
\zeta \equiv  {\Ro v_{\rm 0z}'(\Ro)}/{\Va(\Ro)}.
\end{displaymath}
Here, $\zeta$ measures the amount of shear in the vertical velocity
of the winds, and the ``o'' subscripts in the equilibrium epicyclic and
rotation frequencies denote evaluation at the reference point $\Ro$.
We adopt the power-law equilibria of \S 2. We now organize
the terms in eq.\ (38) finally to get a 12th-degree polynomial
\begin{equation}
0 = \sigma^{12} +
\sum_{j=0}^{10} f_j(q,i,\xr,\xz,m,\Omega,\zeta) \sigma^j.
\end{equation}
The functional dependences of the coefficient $f_j$'s on the parameters are
so complicated that it is not illuminating to write down the whole
expression here. We may obtain more simplified forms for $f_j$'s by
sorting out terms and taking the limit of $\xr\gg 1$. Because there are
various interesting modes which demand different regimes of parameters,
however, we keep all the terms as the general local dispersion relation 
(with $\xr$ large) and 
compute numerical growth rates by solving eq.\ (41) for
$\sigma$ as a function of the other variables.
We present these numerical results in \S 5.
In \S 6, we will classify individual
(either unstable or overstable) modes and provide their limiting 
dispersion relations.

\citet{ter96} considered only incompressible modes, for disk applications
where thermal pressure is considerable, by taking a divergence-free
displacement vector as a perturbation eigenvector, thus obtaining a 4th-order
polynomial. Our dispersion relation, for applications to supersonic MHD winds
in which thermal pressure is negligible hence motions are compressible,
contains information about all possible, either oscillating or unstable,
modes of cold MHD winds.

\section{Numerical Solutions of Modal Dispersion Relation}

The derived dispersion relation (41) is a 12th-order polynomial with real
coefficients, indicating that solutions appear as complex conjugate pairs.
Solutions having non-zero real and imaginary parts are overstable modes, 
and solutions with vanishing real parts are unstable modes.
In this section, we present both types of solutions
which are consistent with the local analysis by fixing $\xr=10$.
As we shall show, both higher values of $\xr$ (more spatially localized)
and lower values of $\xr$ (less spatially localized) give
qualitatively the same family of solutions as with $\xr=10$.

\subsection{Axisymmetric Modes of Instabilities}

First, we consider the axisymmetric case with $m=0$.
For 4 selected sets of parameters, we plot the real and imaginary parts of
the unstable and overstable modes in Fig.\ 4. A Keplerian rotation gradient
with $\kappa^2 = \Omega^2$ is assumed and vertical shear is neglected
except in $\tom$.
We take $\Omega$ as arbitrary rather than using the relation (6), by allowing
that the gravitational force from a central object also contributes to the
equilibrium rotation velocity.
Then, from eq.\ (5), the normalized angular velocity becomes
\begin{equation}
\Omega^2 = \kappa^2 = \cos^2{i} - \frac{1+q}{2} + G_{\rm R},
\end{equation}
where $\GR \equiv g_{\rm R}(\Ro)\Ro/\Va^2(\Ro) > 0$ is the normalized
gravitational acceleration. Note that for $1+q>0$ (i.e., magnetic fields
decreasing outward), equilibrium solutions with $i$ approaching 90$^{\rm o}$
require non-zero gravity (because hoop stresses do not confine a
primarily-poloidal flow). Also note that as $\GR$ strengthens, the initial
equilibrium is maintained by the balance between centrifugal and
gravitational forces, implying that the magnetic force is negligible.

The behavior of the solutions shown in Fig.\ 4 (and similar behavior
for other parameters) allows us to identify 4 different axisymmetric mode
families: a toroidal resonance mode (TR), an axisymmetric toroidal 
buoyancy mode (ATB), a poloidal buoyancy mode (PB), and a Balbus-Hawley (BH) 
mode. One (TR) of these is an overstable mode and the 
others (ATB, PB, and BH) are purely growing modes.

Fig.\ 4a and 4b correspond to a disk wind at large distance from the source,
where magnetic fields are dominantly toroidal (small $i$) and centrifugal
force balances magnetic force (small $\GR$), while Fig.\ 4c and 4d correspond
to an accretion disk or inner part of a wind where magnetic fields are
poloidal (large $i$) and centrifugal force is balanced by the gravity from
a central object (relatively large $\GR$). In each frame, solid and dotted
lines represent the imaginary and real parts of the frequencies, respectively.
Fig.\ 4a shows the TR mode which splits into two branches in the presence
of (arbitrarily small) poloidal magnetic fields (Fig.\ 4b and 4c). This TR mode
is not a generic instability mode because it has a far
larger real part (associated with ordinary MHD oscillations), 
indicating an overstability.
With the presence of poloidal field components, there exist two different
types of buoyancy modes, namely ATB and PB modes. When the pitch angle
of the magnetic field is relatively small, the buoyancy
instabilities are driven by the interplay of the centrifugal force with the 
hoop stress of toroidal fields, so we call these ATB modes. 
Since, as explained in \S 6.1.2, the ATB modes need non-zero poloidal 
fields as well to be unstable, they disappear when $i=0$. On the 
other hand, the PB instability modes arise when the fields are 
predominantly poloidal so that the pressure gradient forces of poloidal
fields and the gravity from a central object are main driving forces,
similar in character to the Parker instability.
ATB and PB are pure instability modes with Re($\sigma$)=0, as shown 
in Fig.\ 4. These instability modes operate even in the 
arbitrarily high-$\kz$ limit because of our cold MHD assumption; 
otherwise sound waves would stabilize
short wavelength perturbations, as they do in the Parker instability.
Fig.\ 4d shows that the BH instability mode appears when $\GR \gg 1$, 
corresponding to dynamically weak magnetic fields in the equilibrium; 
with reduced $\GR$ (also shown in Fig.\ 4d), BH is stabilized by 
radial MHD wave motions when $\xr$ is large. 
As discussed in \S 6.1.3 and \S 7.1, one interesting finding in our work is
that the {\it compressible} axisymmetric BH mode is strongly suppressed 
even for $\GR$ large when the
toroidal field is sufficiently strong; in a cold, Kepler-rotating  
MHD flow, it is fully
stabilized when the pitch angle $i<30^{\rm o}$ (see \S 6.1.3).

Fig.\ 5 shows how the characteristics of unstable/overstable modes change
as $i$ and $\GR$ vary for the fixed values of $x_{\rm z}=4$, 
$x_{\rm R}=10$, and $q=\zeta =0$.
For a pure toroidal field configuration with $i=0^{\rm o}$, we observe only 
overstable TR modes that are almost independent of $\GR$.
As $i$ increases, ATB emerges but is stabilized by rotation with $\GR$ large.
When $i=45^{\rm o}$ and $q=0$, the buoyancy mode disappears because with these
parameters the net force from the background magnetic fields
vanishes (cf. eqs.\ [5], [6], and [42]). When $i>30^{\rm o}$
the BH mode strengthens as $\GR$ increases. This is because in our 
normalization higher values of $\GR$ correspond to weaker equilibrium 
magnetic fields, with which the BH instability operates efficiently. 
At a pure poloidal configuration of magnetic fields, BH and PB modes remain 
unstable (Fig.\ 5d). Dotted lines at very small $\GR$ in Fig.\ 5c and 5d
mark the minimum value of $\GR$, available for given values of $q$ and $i$,
below which no initial equilibrium exists (cf.\ eq.\ [42]).

\subsection{Non-Axisymmetric Modes of Instabilities}

When non-axisymmetric perturbations are applied, the cold MHD system responds
with 3 more modes which are either unstable or overstable, in addition
to the axisymmetric modes. We shall refer to these as 
non-axisymmetric toroidal buoyancy (NTB),
geometric poloidal buoyancy (GPB), and poloidal resonance (PR)
modes. The PR modes are MHD waves which
have non-zero azimuthal wavenumbers and become overstable when there
is a radial gradient of the axial field, analogous to the TR modes.
In addition to the above modes, systems with toroidal magnetic field 
configurations and non-zero sound speed are also
subject to significant non-axisymmetric magnetorotational instability (NMRI)
modes. Unlike the other three non-axisymmetric modes, the NMRI mode arises 
due to a differential rotation with $d\Omega/dR <0$, where $\Omega$ is an
angular velocity of the rotation. 
More than anything else,
the fact that the NMRI in $B_\phi$-dominated systems needs a finite 
sound speed to be unstable
distinguishes it from the axisymmetric Balbus-Hawley instability,
which can be unstable regardless of temperature for purely axial fields.
As we shall discuss later, the basic mechanism for the onset of
NMRI is quite different from that of axisymmetric BH instability.
We reserve the discussion of NMRI modes for \S 7, concentrating 
here on numerical results for our basic cold MHD system.

Fig.\ 6 shows the unstable and overstable solutions of the dispersion
relations for two combinations of selected parameters: Fig.\ 6a
corresponds to a disk wind with small $i$ and small $\GR$,
while Fig.\ 6b is for the near-disk case with large $i$ and
large $\GR$. We assume a Keplerian rotation and take an arbitrary $\Omega$
once again using eq.\ (42). For all cases, we chose $\xr=10$, $q=\zeta=0$,
and $m=1$, and confirmed that changes to these parameters do not
appreciably affect the qualitative results. Solid and dotted lines in
Fig.\ 6 represent imaginary and real parts of the normalized wave
frequencies, respectively. When $i=\GR=0$, Fig.\ 6a shows the presence of
unstable NTB and overstable TR modes which are split by the non-axisymmetry.
NTB modes are nearly like PB modes in their physical basis and have an
almost constant growth rate over a wide range of $\xz$. But they depend
sensitively on the logarithmic gradients of the density and magnetic 
structures (i.e., $q$; see Fig.\ 7).
For an intermediate value of $i$, both TR and PR modes coexist.
At some wavenumber $\xz$, they combine to simply vanish,
but overall they give rise to complicated behavior of Im($\sigma$).
When $i=90^{\rm o}$, we observe
three unstable GPB, BH, and PB modes, and overstable PR modes (Fig.\ 6b).
GPB modes are driven by a buoyancy force together with the geometrical
effect. Note that the real parts of PR modes are linearly proportional
to the vertical wavenumber $\xz$, as TR modes are, indicating
that they are really overstable modes.
Since $\xr \gg m$, however, there exists only a small contribution
from non-axisymmetric effects to the axisymmetric BH and PB modes
(cf.\ Fig.\ 4d).
Remember that when $m\gg\xr$, the normal mode assumption rapidly breaks down
because such high-$m$ modes lose their spatial pattern very quickly;
we investigate $m\gg \xr$ cases using different methods in \S\S7 and 8.

Fig.\ 7 shows how the characteristics of the buoyancy modes change 
as $\xz$, $i$, and $q$ vary. When $\GR$=0, an initial equilibrium exists 
only for a limited range of $i < i_{\rm crit}\equiv \cos^{-1} \sqrt{(1+q)/2}$ 
from eq.\ (42), with toroidal field components dominating over poloidal
field components. In Fig.\ 7, therefore, the unstable modes with 
$i<i_{\rm crit}$ correspond to toroidal buoyancy modes, while 
poloidal buoyancy modes have $i>i_{\rm crit}$.
Generally speaking, with the assumption of extremely cold medium, 
smaller-scale buoyancy modes with high $\xz$ have larger growth rates.
When $i$ is very small, as seen in Fig.\ 7, ATB modes are stable
because they need the aid of poloidal fields to be unstable, 
while NTB modes become unstable for {\it all} $i<i_{\rm crit}$. This 
reflects the physically different driving mechanisms between ATB and 
NTB instabilities. PB modes become more unstable
with higher $q$ (steeper background gradients), while ATB/NTB modes are more 
efficient with smaller $q$. Greater instability is simply associated with
higher background magnetic force in the respective cases (cf.\ the
initial equilibrium condition [42]). 

The $\kr$-dependence of the unstable/overstable modes are summarized in
Fig.\ 8. Here we fix $\xz=2$ for all cases and choose $q=0.8$, $\GR=0$,
and small $i$ for Fig.\ 8a and 8c, corresponding to disk wind-like systems, 
and $q=0$, $\GR=5$, and large $i$ for Fig.\ 8b and 8d, 
corresponding to accretion disks or disk winds near their sources. 
The BH instability modes are completely suppressed by MHD waves when
$\xr \gtrsim 3$; we will show that this is consistent with the 
prediction of the asymptotic dispersion relation. 
All the other modes extend with smaller growth rates
to larger $\xr$, with Im($\sigma)\sim \xr^{-1}$, which we will show
agrees well
with the asymptotic dispersion relations (43), (45), (50), and (52) for the 
PB, ATB,
TR, and NTB modes, respectively. For the PR modes the asymptotic dispersion
relation (56), showing Im($\sigma)\sim \xr^{-2/3}$, is valid only when
$R\kz \gg m$, which is not  consistent with the parameters adopted in
Fig.\ 8d. When $\kr \gg \kz \sim m/R$, one can confirm analytically that
the PR modes also behave as Im($\sigma)\sim \xr^{-1}$. 
In the shearing-wavelet point of view with eq.\ (40),
Fig.\ 8 shows that kinematic shear arising from the background
flows ultimately stabilizes both unstable and overstable modes,
as $\kr$ grows secularly increases in time.
Although the local approximation breaks down if $\xr$ is 
not large, Fig.\ 8 indicates that the BH mode exists and may show
interesting behavior for small $\xr$. 
In addition, Fig.\ 8 also suggests larger growth rates when $\xr$ is small
for other modes, although the assumptions of this section
of a radially-local, slowly-changing pattern are not
self-consistent when $\xr$ is small.
To study dynamical growth of disturbances which occurs when $\xr \ll m$,
we use direct integrations of the shearing-sheet equations. 
We present these results in \S 8.2 (for the NTB modes) and \S\S 7.2 
and 7.3 (for the NMRI modes and generalized MRIs).

\section{Mode Classification}

The cold MHD system we are investigating has 8 distinct local modes with
Im$(\sigma)>0$. Some of them (TR and PR) have larger Re$(\sigma)$ 
corresponding to overstability, while the others (PB, ATB, BH, NTB, GPB, 
and NMRI) have negligible Re$(\sigma)$, indicating pure instability. 
The NMRI modes do not appear in the numerical solutions because of the 
cold MHD assumption we made. Detailed discussion of the NMRI modes will
be separately given in \S 7.2. In this section we describe the physical
nature of the individual cold-fluid modes and present the respective 
dispersion relations under some limiting approximations.

\subsection{Axisymmetric Modes}

\subsubsection{Poloidal Buoyancy Mode }

Consider a system with pure axial fields. 
If gravitational forces are large, then they may balance the combined 
outward radial centrifugal force and pressure gradient force
of outward-decreasing $B_{\rm 0z}(R)$; otherwise, if $g_{\rm R}=0$, 
then the strength in the magnetic fields
must increase outward for an equilibrium to exist.
In an initial state, at any point in the system the magnetic pressure
force acting outward is balanced by the difference between gravity 
and the centrifugal force acting inward.
If perturbed, a denser fluid element experiences reduced magnetic forces
but unchanged centrifugal and gravitational forces per
unit mass, and thus it would tend to sink radially
inward dragging the field line with it; a lighter fluid element would
correspondingly tend to float outward. Then, in a  frozen-in-field condition,
the neighboring gas finds itself on sloping lines of force and thus slides
inward to add its weight and to cause field lines to bend more, expediting
the instability. This poloidal buoyancy  mode is analogous to
the Parker instability \citep{par66}, with the driving force role
of external gravity in Parker's instability replaced by 
combination of gravity and the centrifugal force in the PB.
The PB mode can occur for both axisymmetric and non-axisymmetric 
disturbances.

Putting $B_{0\phi}=m=0$ and considering short wavelength perturbations
with $\Vaz^2\kz^2 \gg \tom^2$ in eq.\ (38), one can find
the dispersion relation for the poloidal buoyancy mode is
\begin{equation}
\tom^2 = -\left(\frac{1+q}{2}\right)^2
\frac{\Vaz^2\kz^2}{R^2(\kz^2+\kr^2)}
= - \left(\frac{|g_{\rm R} -R\Omega^2|^2}
{\Vaz^2}\right)\frac{\kz^2}{\kz^2+\kr^2}.
\end{equation}

Eq.\ (43) states that there is no preferred
length scale as long as $\kz$ is large. However, the inclusion of thermal
effects would stabilize the PB mode with shorter wavelengths, as in the
Parker instability\footnote{Also by taking a local approximation and by
neglecting density stratification and the effects of thermal and cosmic ray
pressures, one can simplify eq.\ (III.12) of \citet{par66} to get the
asymptotic (${\bold k} \rightarrow \infty$) dispersion relation
\begin{displaymath}
\omega^2 = - \left(\frac{g^2}{v_{\rm A\parallel}^2}\right)
\frac{k_{\parallel}^2}{k_{\perp}^2+k_{\parallel}^2},
\end{displaymath}
where $g$ is the gravity perpendicular to the galactic plane, 
$v_{\rm A\parallel}$
is the Alfv$\acute{\rm e}$n speed of initial fields parallel to
the galactic plane, and $k_{\parallel}$ and
$k_{\perp}$ are perturbation wavenumbers in the respective directions 
parallel and perpendicular to the galactic disk and magnetic field.
Comparing the above with eq.\ (43), we may write $g_{\rm eff} 
\equiv g_{\rm R} - R\Omega^2$ for the PB modes,
with wavenumber correspondence $\kz \leftrightarrow k_\parallel$
and $\kr \leftrightarrow k_\perp$.}
Also, sufficiently large $\kr \gg \kz$ stabilizes the mode. 

\subsubsection{Axisymmetric Toroidal Buoyancy Mode}

Now consider a system with weak poloidal but strong toroidal field 
components and negligible gravity. When magnetic fields are predominantly 
toroidal (i.e., when $i< \cos^{-1} \sqrt{(1+q)/2}$ from eq.\ [42]), 
an initial equilibrium state is maintained by the balance mainly between
the centrifugal force acting outward, and magnetic hoop stresses
which act inward. With sinusoidal density
perturbations with $\kz$ imposed on the equilibrium, a heavier blob of
material would tend to float radially outward under the action of 
unchanged centrifugal forces per unit mass but reduced specific 
magnetic forces; a lighter element would correspondingly tend to sink.
The radial motions of the heavier and lighter blobs are in opposite
directions and thus cause the poloidal field lines to bend, creating
radial perturbed fields. 

The azimuthal fluid motion is slightly accelerated by the tension force 
exerted by the initial toroidal and the perturbed radial fields 
(cf.\ $\bp B_{\rm 1R}/R$ term in eq.\ [28], associated with spiral magnetic
field line projections in the $z$=constant plane). This causes the initial 
poloidal component of field lines to bend now in the azimuthal direction,
creating bands of perturbed azimuthal fields with signs alternating in
the $\hat{\bold{z}}$ direction. The resulting total azimuthal fields are 
distributed in such a way that the heavier (lighter) blob in the initial
perturbation has a lower (higher) toroidal field strength. Induced
motions due to the vertical magnetic pressure gradient force
carry the matter from under dense to over dense regions, closing
the loop and amplifying the initial perturbation. 

By setting $m=v_{\rm 0z}=0$ and taking the $\Va^2\kz^2 \gg \tom^2$ limit,
we obtain from eq.\ (38) the following dispersion relation 
\begin{equation}
0 = \tom^4 - \left[ \Vaz^2\left(\kz^2+\kr^2\right) + \kappa^2
-4\Omega^2\frac{\bz^2}{\bo^2}
\right]\tom^2
-4\Omega
\Vap\Vaz\FB\kz\tom - \Va^2\Vaz^2\FB^2\kz^2,
\end{equation}
with $\FB=(\cos^2 i - (1+q)/2)/R$. When $\Vap=0$ (i.e., with pure poloidal
fields), eq. (44) immediately recovers eq. (43), the limiting dispersion 
relation for PB modes. On the other hand, if $\Vaz=0$, there is no
unstable ATB mode, clearly demonstrating that ATB modes operate by
bending poloidal field lines.
From Fig.\ 5b, we note that ATB modes are
stabilized by rotation (larger $\GR$ corresponds to stronger rotation).
It can be shown from eq. (44) that when $R\kz, R\kr \gg \tom$, 
the critical wavenumbers are 
$(\kz^2 + \kr^2)_{\rm crit}  = -(d\Omega^2/d\ln{R})/(\Va^2\sin^2{i})$,
below which the system is stable against the ATB modes. 
For $\kz \gg k_{\rm z,crit}$, eq. (44) is further reduced to
\begin{equation}
\tom^2 = - \left(\cos^2 i - \frac{1+q}{2}\right)^2 
\frac{\Va^2\kz^2}{R^2(\kz^2+\kr^2)}
= - \left(\frac{|g_{\rm R} -R\Omega^2|^2}{\Va^2}\right)
\frac{\kz^2}{\kz^2+\kr^2}.
\end{equation}
Since ATB instabilities are axisymmetric modes, they can persist 
without being disturbed by the kinematic growth of $\kr$ due to 
shear, if $v_{\rm 0z}'=0$.

Among well-known plasma modes, the pinch or sausage mode of a plasma
column is most similar to the ATB in overall geometry and effect.
Both are axisymmetric and require the radial tension force from
predominantly toroidal magnetic fields to drive the instability.
For both the plasma pinch mode and the ATB of cold cylindrical
winds, the net effect is that matter tends to be ejected radially
in bands alternating with contracting magnetic field loops.
However, in pinch modes, the plasma is generally unmagnetized
and surrounded by {\it external} toroidal fields, and axial
fields tend to suppress the instability. In the ATB, on the other
hand, {\it internal} toroidal magnetic fields permeate the fluid,
and non-zero axial fields are required for instability.

\subsubsection{Compressible Balbus-Hawley Mode}

In the presence of axial magnetic fields, a differentially rotating disk
is unstable to an axisymmetric incompressible perturbation
(Balbus \& Hawley 1991; see also Velikhov 1959 and Chandrasekhar 1960).
Because this Balbus-Hawley instability\footnote{Often referred to
magnetorotational instability, or briefly, MRI.}
has a rapid growth time (comparable
to the local rate of rotation) and exists for arbitrarily weak magnetic 
field strength,
it is believed to provide a powerful mechanism for the generation
of the effective viscosity in astrophysical accretion disks. Through numerical
simulations, \citet{haw91} argued that the roles of compressibility
and toroidal fields are not significant as long as the total field
strength is subthermal.  Also, \citet{bla94} studied the effect of
toroidal fields on the compressible axisymmetric BH instability and showed 
that toroidal fields do not modify the instability criterion, 
while reducing growth rates slightly if $\Vap<\cs$.
We find the striking
result that under extremely cold conditions (i.e., $\Va\gg\cs$),
compressibility prohibits the axisymmetric BH instability from 
occurring if the toroidal fields are as strong as the poloidal fields.

By taking the weak magnetic field limit ($\Omega \gg \Va\kr, \Va\kz$)
and $m=0$, we
obtain from eq.\ (38) the following dispersion relation for the compressible
axisymmetric BH instability in a cold MHD flow
\begin{equation}
\omega^4 - \omega^2[\Va^2(\kz^2+\kr^2) + \Vaz^2\kz^2 + \kappa^2]
+ \Va^2\kz^2[\Vaz^2(\kz^2+\kr^2) + \kappa^2 - 4\Omega^2\sin^2 i] = 0,
\end{equation}
and thus from the last term in eq.\ (46) we obtain the instability criterion
\begin{equation}
\Vaz^2(\kz^2+\kr^2) + \kappa^2 - 4\Omega^2\sin^2 i < 0.
\end{equation}
With an $\Omega \propto R^{-a}$ rotation profile, eq.\ (47) implies 
that if $\sin^2 i < 1-a/2$, we anticipate no BH instability in a cold flow. 
For a Keplerian rotation law with $a=3/2$, for
instance, no axisymmetric BH instability occurs if $i < 30^{\rm o}$!; 
{\it when the magnetic field strength is superthermal, 
the inclusion of toroidal fields tends to suppress the growth of the BH 
instability}. With a steeper rotation profile (as would occur, for example, 
in winds from boundary layers), there is an increase in the range of $i$ 
for which a system is BH-unstable.

We defer the full discussion on the BH instability until \S 7.1,
where we explicitly include pressure terms in the dynamical equations.

\subsubsection{Toroidal Resonance Mode}

Consider a system having pure toroidal fields without rotation.
If the initial fields are homogeneous in space, magnetosonic waves,
driven solely by magnetic pressure (with the assumption of the cold
medium), would propagate without any interruption in the plane
whose normal is perpendicular to the magnetic field direction.
In an inhomogeneous medium, however, MHD waves
no longer maintain a sinusoidal planform, and the characteristics
of the waves change through the interaction with the background medium.
The amplitudes of the waves may sometimes increase as they propagate,
or sometimes they may become evanescent and decay at a resonance,
or even may be trapped between two resonance points \cite[cf.][]{rae82}.
In such a strongly structured medium, the classification of MHD waves
is not in general possible.

Our local treatment of MHD waves can provide some insight on the
amplification or evanescence of propagating MHD waves in a structured
medium.
For $\toF^2 \rightarrow 0$ and $B_{0\rm z}=\Omega=0$,
the local wave equation (35) can be simplified as
\begin{equation}
\frac{d^2\xiR}{dR^2} - \frac{d\ln{\toF^2}}{dR}
\frac{d\xiR}{dR} + \frac{\toF^2}{\Vap^2} \xiR = 0.
\end{equation}
Again we define $\xiR \equiv (\toF^2)^{1/2}\Psi$, then eq.\ (48) takes the
form of eq.\ (16) with $\kr\equiv K(R)$ defined by
\begin{equation}
\kr^2 \equiv \frac{\toF^2}{\Vap^2}
- \frac{3}{4}\left(\frac{\Vap^2\kz^2}{R\toF^2}\right)^2,
\end{equation}
where $\Vap \propto R^{-1/2}$ was assumed.
Thus, considering limiting cases of $\kz$ and $\kr$, one can find the
dispersion relation for this mode near the resonance frequencies
(i.e., $\tom \approx \Vap\kz$)
\begin{equation}
\tom^2 = \left\{ \begin{array}{ll}
\Vap^2\kz^2\left[1 + (3/4)^{1/2}e^{\pm\pi i/2}(R\kr)^{-1}\right],
&{\rm for}\;\; R\kr \gg R\kz \gg 1, \\
& \\
\Vap^2\kz^2\left[1 + (3/4)^{1/3}e^{\pm2\pi i/3}(R\kz)^{-2/3}\right],
&{\rm for}\;\; R\kz \gg R\kr \gg 1, \\
\end{array} \right.
\end{equation}
showing that the imaginary part of toroidal resonance mode vanishes quickly
as $\kr \rightarrow \infty$, while its real part gets bigger as $\kz$
increases. Therefore, it is not adequate to regard TR modes as a true local
instability mode. Though the TR mode is not a local instability, it
suggests potential for waves to have global instabilities in which the
magnetosonic resonance ($\tom = \Vap\kz$) plays a similar role to the
Lindblad resonance in rotating disks. Thus, waves of fixed frequency
propagating with a radial component of $\bold{k}$ into their magnetosonic
resonances may be amplified or reflected.

The modification of traveling waves due to the inhomogeneity of the medium
is mediated through the magnetic pressure. A similar effect would occur when
hydrodynamic waves propagate into an inhomogeneous medium.

\subsection{Non-Axisymmetric Modes}

\subsubsection{Non-Axisymmetric Toroidal Buoyancy Mode}

The non-axisymmetric toroidal buoyancy mode is very similar to the 
PB mode in its physical
mechanism, in spite of the different field geometry. For toroidal-field
dominated cases, an equilibrium can exist with the net inward magnetic
stresses balancing the outward centrifugal force.
When the system is perturbed
non-axisymmetrically, the instability would develop similarly to
PB modes, as described in \S6.1.1.
For the $B_\phi$-dominated case, however, over-dense regions float outward
and under-dense regions sink, because the inward magnetic forces are 
enhanced when the density drops.

Setting $\bz=v_{\rm 0z}=0$ and assuming $R\kz \ll m, R\kr$, one can show that
the general dispersion relation (38) is reduced to the following quartic
equation in terms of $\tom$
\begin{equation}
0 = \tom^4 - \left[ \Vap^2\left(\frac{m^2}{R^2}+\kr^2\right) + \kappa^2
\right]\tom^2
-4\Omega\frac{m}{R}
\FB\Vap^2\tom - \Vap^4\FB^2\left(\frac{m}{R}\right)^2,
\end{equation}
with $\FB=(1-q)/2R$, The negative last term in
eq.\ (51) guarantees the existence of unstable NTB modes. The third term
(caused by the coupling of the rotation with the background fields) 
tends to stabilize NTB modes. Thus, if
\begin{displaymath}
\frac{m^2}{R^2} + \kr^2 \; < \; \frac{(4\Omega^2-\kappa^2)}{\Vap^2},
\end{displaymath}
there is no unstable NTB mode. 
Note that for wind equilibria with $\bz =0$, from eq.\ (6)
the RHS of the above equals $3(1-q)/2R^2$; thus the NTB instability
will be present at all $m$ when $1/3<q<1$.

In the limit of large $m$, we obtain the asymptotic dispersion relation
for the NTB mode
\begin{equation}
\tom^2 = - \left(\frac{1-q}{2}\right)^2 \frac{\Vap^2m^2}
{R^2(m^2+R^2\kr^2)} =
- \left(\frac{|g_{\rm R}-R\Omega^2|^2}{\Vap^2}\right)\frac{m^2}{m^2+R^2\kr^2}.
\end{equation}
Here for the second equality, eq.\ (6) with $i=0^{\rm o}$ is used.
Eq.\ (52) is akin to eq. (43), the dispersion relation for the PB mode,
and to eq.\ (45), the dispersion relation for the ATB mode, reflecting
the common origin in buoyancy forces of all three. In fact, Fig.\ 7 
clearly shows how the various buoyancy modes extend and smoothly join
at intermediate pitch angles.

\subsubsection{Geometric Poloidal Buoyancy Mode}

Now suppose a system with pure vertical fields. When perturbed azimuthally,
a fluid element becomes over dense and tends to move inward due to the
decreased background magnetic pressure force per unit mass if $0<q<1$.
This geometrically converging motion of fluid increases density and field
strength by factors of $(1-q)$ and $(1-q)/2$, respectively. On the other
hand, the magnetic field enhancement induces diverging motions of the
fluid in the azimuthal direction by building up a pressure gradient, 
tending to lower the density. The net effect of these two processes is a
density increase by a factor of $(1-q)/2$, accelerating the inward motion of
the heavier element. When $m \gg R\kz$, the dispersion relation for this
GPB mode is found to be
\begin{equation}
\tom^2 = - \left(\frac{1-q^2}{4}\right) \frac{\Vaz^2m^2}{R^2(m^2+R^2\kr^2)}.
\end{equation}
When $q=1$, there is no instability. This is because the initial
configuration of the density and the field is such that the mass and
magnetic flux contained in a thin ring with the thickness $dR$ and the
radius $R$ are constant over $R$, and no gain from the geometrical effect
is possible.

\subsubsection{Poloidal Resonance Mode}

The physical basis for the poloidal resonance mode is quite similar to that
of the
toroidal resonance mode. The only difference between them is the background
field geometry. In the presence of pure axial fields, MHD waves with non-zero
$m$ are easily influenced by radial magnetic pressure gradients.

To derive the dispersion relation near the resonance frequencies
(i.e., $\toF \approx 0$), let us suppose a system with pure axial fields and
neglect the vertical velocity shear. The system is also assumed to rotate
slowly enough that the effects of rotation may not be important in the wave
dynamics (i.e., $m\Vaz \gg R\Omega$).
For $\tom^2 \rightarrow \Vaz^2(\kz^2+m^2/R^2)$, we are left from eq.\ (35)
with
\begin{equation}
\frac{d^2\xiR}{dR^2} + \frac{d}{dR}
\ln{\left(\frac{\toA^2}{\toF^2}\right)}
\frac{d\xiR}{dR} + \frac{\toF^2}{\Vaz^2} \xiR = 0.
\end{equation}
We now define $\xiR \equiv (\toF^2/\toA^2)^{1/2}\Psi$ to simplify eq.\ (54)
into eq.\ (16) with $\kr\equiv K(R)$ defined by
\begin{equation}
\kr^2 \equiv \frac{\toF^2}{\Vaz^2}
- \frac{m^2\Vaz^2}{R^2\toF^2}\left(\frac{d\ln{\toF^2}}{dR}\right)^2,
\end{equation}
where we took the limit of $R\kz \gg m$ and
assumed $\Vaz \propto R^{-1/2}$. Solving eq.\ (55) for two limits
of $\kr$, we obtain the dispersion relation near the resonance frequencies
\begin{equation}
\tom^2 = \left\{ \begin{array}{ll}
\Vaz^2\kz^2\left[1 + e^{\pi i/3}
(ma_{\rm sh})^{2/3}(R^2\kr\kz)^{-2/3}\right],
&{\rm for}\;\; R\kr \gg R\kz \gg m, \\
& \\
\Vaz^2\kz^2\left[1 \pm e^{\pi i/2} (ma_{\rm sh})^{1/2}
(R\kz)^{-1}\right],
&{\rm for}\;\; R\kz \gg R\kr \gg m, \\
\end{array} \right.
\end{equation}
where $a_{\rm sh} \equiv 1\pm 3m\Omega/\Vaz\kz$, showing again a rapidly
declining imaginary part as $\kr$ increases, at which the local
approximation is valid. Thus, just like TR modes, PR modes are not
strictly local instability modes.

\section{Magnetorotational Instability}

\subsection{ Axisymmetric BH Instability}

In the preceding section, we briefly discussed the axisymmetric
BH instability in a cold, differentially rotating medium and
found that the BH instability can be suppressed by the azimuthal component 
of magnetic fields, if the medium is cold enough.
Incompressibility has generally been adopted in the study
of the BH instability in an accretion disk on the grounds that in such
a system the magnetic fields are subthermal and thus acoustic waves
can maintain the incompressible condition over many rotation periods.
For magnetocentrifugally driven winds, however, sound waves play
a minor role in controlling the dynamics and thus the incompressible
approximation is inapplicable.
In addition, since an initial equilibrium is attained through the 
balance between the centrifugal and magnetic forces (cf.\ eq.\ [5]),
the \Alf\ crossing time scale is comparable to the rotation time
scale ($\Va \sim R\Omega$); in this case, the fields are not weak
and the unstable range of wavenumbers becomes narrow.

We generalize the previous discussion of the axisymmetric 
compressible BH instability
by explicitly including the thermal pressure terms in the momentum 
equation and exploring the role of compressibility to the development 
of the Balbus-Hawley instability. 
We consider a cylindrical flow threaded by both vertical and azimuthal
magnetic fields, ignoring the radial variations in the initial
configuration except $\Omega=\Omega(R)$ and neglecting the vertical velocity.
We assume the medium is isothermal and take the WKB ($R \kz \gg 1$)
approximation. Through the standard approach to linear analyses,
we arrive at the dispersion relation for the compressible version of
the BH instability
\begin{equation}
(\ooA^2 - \kappa^2)f(\omega^2) = \kr^2\ooA^2((\cs^2 + \Va^2)\omega^2
-\cs^2\Vaz^2\kz^2) + 4\Omega^2\Vaz^2\kz^2(\omega^2-\cs^2\kz^2),
\end{equation}
where $\cs$ is the isothermal sound speed of the medium,
$\ooA^2 \equiv \oo^2 - \Vaz^2\kz^2$, and $f(\omega^2)$ is defined by
\begin{displaymath}
f(\omega^2)   \equiv
\omega^4 - \omega^2 (\cs^2+\Va^2)\kz^2 + \cs^2\Vaz^2\kz^4.
\end{displaymath}
Eq.\ (57) is a sixth-order equation for $\omega$ with only even terms.
When $\kr=0$, eq.\ (57) is identical to 
eq.\ (64) of \cite{bla94} or eq.\ (99) of \citet{bal98}.
Now let us take the two opposite limits of $\cs$ to obtain the following
dispersion relations
\begin{mathletters}
\begin{eqnarray}
\ooA^4 - (\kappa^2 + \Va^2\kr^2 + \Vap^2\kz^2)\ooA^2
+(\kappa^2\Vap^2 - 4\Omega^2\Vaz^2)\kz^2 &=& 0,
\quad{\rm for}\quad \cs\rightarrow 0, \\
(1 + \kr^2/\kz^2)\ooA^4 - \kappa^2\ooA^2 - 4\Omega^2\Vaz^2\kz^2 &=& 0,
\quad{\rm for}\quad \cs\rightarrow \infty,
\end{eqnarray}
\end{mathletters}
and the corresponding instability criteria\footnote{In fact,
from eq.\ (57) the formal instability criterion (59b)
is generic for any value of $\cs\neq0$; it may be
written as $\Vaz^2(\kz^2+\kr^2) + \kappa^2 - 4 \Omega^2 <0$.
However, when $\cs/\Va \ll 1$, growth rates for small $i$ are very low.}
\begin{mathletters}
\begin{eqnarray}
\Vaz^2(\kz^2 + \kr^2) + \kappa^2 - 4\Omega^2\sin^2i &<& 0,
\quad{\rm for} \quad \cs\rightarrow 0, \\
\Vaz^2(\kz^2 + \kr^2) + d\Omega^2/d\ln R &<& 0,
\quad{\rm for} \quad \cs\rightarrow \infty.
\end{eqnarray}
\end{mathletters}
Note that eq.\ (58b) is the same as the original dispersion relation of
the incompressible BH instability (eq.\ [2.9] of Balbus \& Hawley 1991
without the Brunt-${\rm V\ddot{a}is\ddot{a}l\ddot{a}}$ frequency).
The instability criterion (59a) in the extremely compressible limit
depends explicitly on the
local pitch angle, showing that as $i$ departs from $90^{\rm o}$ the
instability becomes gradually confined to smaller values of $\kz$.
For a cold Keplerian flow, no instability occurs when $i$ is smaller
than $30^{\rm o}$.

To examine what role thermal pressure plays to the growth of the
BH instability and why the instability criterion depends on $i$,
we plot the unstable solutions of eq.\ (57) as functions of
$\qA \equiv (\bold{k}\cdot\bold{v}_{\rm A})/\Omega$ $(= \kz \Vaz /\Omega$
for the axisymmetric case) and $\beta \equiv \cs^2/\Va^2$ in Fig.\ 9.
For the time being, we confine our discussion to the $\kr=0$ case.
When $i=90^{\rm o}$, the instability criterion from eqs.\ (59a,b)
is $\Va^2\kz^2 < - d\Omega^2/d\ln{R}$ and the growth rate is independent of
$\beta$, implying that the compressibility does not alter the instability
(Fig.\ 9a). This can be understood as follows: when magnetic fields are
mainly axial, sound waves propagating along a vertical direction decouple
completely from the magnetic fields and are undisturbed by rotation.
But transverse MHD waves which are intrinsically incompressible are influenced
by rotation to become ultimately unstable for a range of $\kz$ when
$d\Omega /dR <0$. Therefore, $i=90^{\rm o}$ is a very special case.

On the other hand, when both vertical and azimuthal fields are present,
toroidal perturbed fields generated by an initial azimuthal displacement or by
sheared motion following a radial perturbation of the initial axial fields
tend to cause vertical oscillations, but in a cold assumption, mainly due
to the magnetic field gradient terms, $-\bp(\partial B_{1\phi}/\partial z)$.
This oscillatory vertical motion tries to distribute the perturbed fields as
uniformly as possible, thereby tending to suppress 
the growth of the disturbances.
However, the vertical magnetic pressure gradients are not strong enough to
create significant vertical motions if thermal pressure is large:
a compressed region tends to expand vertically but with little change in
the strength of the toroidal fields, thus providing a favorable condition
for the development of the BH instability. This explains why higher $\beta$ 
cases have higher growth rates at fixed $i$, and why the growth rate
decreases as $i$ decreases at fixed $\beta$ (Fig.\ 9b$\sim$9d).

Although the instability criterion (59b) is completely independent
of the strength of the azimuthal fields provided that $\beta\neq 0$,
indicating as noted by \citet{bla94} that to all orders, azimuthal 
fields do not modify the stability criterion,
the corresponding 
growth rates drop progressively as $\beta$ decreases if $i\neq 90^{\rm o}$. 
When $\beta \gtrsim 1$, any change of an inclination angle $i$ from
$90^{\rm o}$ does not bring significant reductions in growth rates, 
implying that the characteristics of the instability are essentially 
the same as the pure poloidal case.
If $\beta \ll 1$, however, we observe dramatic stabilizing effects from
toroidal fields, as illustrated in Fig.\ 9.

A few comments should be devoted to the effect of $\kr$. Radial wave motions
do nothing but add another restoring force to perturbations. This in turn
means that thermal pressure has a stabilizing influence on the growth of
the BH instability. Thus there are two competing processes of thermal
pressure: thermal pressure associated with vertical wave motion promotes
the BH instability, while thermal pressure controlling radial motion opposes
it. It turns out that for $i \neq 90^{\rm o}$ the former process always
dominates. For $i=90^{\rm o}$, only the latter effect exists,
giving higher growth rates for smaller $\beta$, when $\kr\neq0$.

Notice the stabilizing effect of $\kr$ in eqs.\ (59). 
If the background vertical flow has significant shear, the local radial
wavenumber would increase with time (cf.\ eq.\ [40]),
suppressing the instability. Thus when
$v_{\rm 0z}' \neq 0$, the compressible BH instability will exhibit a
transient growth, as must happen to all modes if $\kz v_{\rm 0z}' \neq 0$
and/or $m\Omega' \neq 0$.

In conclusion, we have found that compressibility has a
stabilizing effect on the axisymmetric BH instability. 
Even though its effect is small
if the sound speed is super-Alfv$\acute{\rm e}$nic, compressibility must
be considered whenever the Alfv$\acute{\rm e}$n speed is comparable to or
even exceeds the thermal sound speed, as is expected in winds and
also in disk coronae \cite[cf.][]{mil99}.

The above discussion applies only for axisymmetric perturbations.
It was also found that an accretion disk with purely toroidal fields is
subject to non-axisymmetric instability \citep{bal92, ter96},
but we will show in the following section that 
the physical role of compressibility in
that case is completely different, in spite of the same quantitative
instability criteria.

\subsection{Non-Axisymmetric MRI: Coherent Wavelet Analysis}

\citet{bal92} found that a differentially rotating disk of incompressible
fluid with embedded toroidal magnetic field is unstable to non-axisymmetric
perturbations. Adopting shearing sheet coordinates (see below), 
they integrated a set
of the perturbed equations and showed that perturbations with an
intermediate azimuthal wavenumber $m$ can exhibit transient, but enormous
growth over a time scale of several percent of $\Omega^{-1}$.

An alternative approach was taken by \citet{ter96} to study a similar
instability to that identified by \citet{bal92}. They solved the problem
using the local WKB approximation.
They started from a general compressible equation of state, but
subsequently they supposed divergence-free poloidal Lagrangian
displacements, which made their treatment essentially incompressible.
They derived a sufficient condition for the instability
which is exactly the same form as that of axisymmetric BH instability
(i.e., $d\Omega^2/d\ln R < 0$). 
Noting that azimuthal shear is the main driving
mechanism and bending of the field lines provides a stabilizing 
restoring force,
they suggested the non-axisymmetric instability of toroidal
magnetic fields might resemble the original BH instability.

We argue in this work that the underlying physical mechanisms for
non-axisymmetric toroidal-${\bf B}$ MRI (which we refer to as ``NMRI'') and 
axisymmetric poloidal-${\bf B}$ MRI (which we refer to as ``BH'') are in
fact quite different 
from each other.
In this section, we analyze the NMRI by looking at ``coherent wavelet''
solutions in which every physical variable, localized in both
space and time, oscillates or grows 
with the same space-time dependence, 
and provide quantitative results in detail.

\subsubsection{Localization in Space and Time}

We begin by considering a shearing, rotating disk with uniform density, 
and magnetic fields with only an azimuthal component. 
We ignore any unperturbed vertical motion in the medium.
We include thermal pressure effects
with an isothermal equation of state to obtain the explicit dependence
of the NMRI on the temperature, but neglect effects
of cylindrical geometry. This configuration is the same as
Balbus \& Hawley's (1992), except that they considered only the 
incompressible case with the Boussinesq approximation, and allowed for 
vertical equilibrium gradients yielding buoyant oscillations. 
Adopting the shearing sheet coordinates
$(\tilde{R},\tilde{\phi},\tilde{z})$ such that $\tilde{R} = R$,
$\tilde{\phi} = \phi - \Omega(R) (t-t_{\rm o})$, and $\tilde{z}=z$ 
\citep{gol65, jul66, bal92}, 
we consider the time development of an initial plane-wave 
disturbance which preserves
sinusoidal variation in the local rest frame of the equilibrium shearing, 
rotating flow 
\begin{equation}
\chi_1(R,\phi,z,t)
= \chi_1(t)e^{im\tilde{\phi} + i\kz \tilde{z} +i\kr(t_{\rm o})\tilde{R}},
\end{equation}
where $\kr(t_{\rm o})$ is a radial wavenumber at a fiducial time $t=t_{\rm o}$.
The linearized form of the MHD eqs.\ (1)$\sim$(4), can be written in
dimensionless form as
\begin{equation}
\frac{d \alpha}{d\tau}
=  -\qR u_{\rm 1R} - \qm u_{1\phi} -\qz u_{\rm 1z},
\end{equation}
\begin{equation}
\frac{d u_{\rm 1R}}{d\tau}  =  2 u_{1\phi}
                       - \qm b_{\rm R} + \qR  (\beta\alpha + b_{\phi}),
\end{equation}
\begin{equation}
\frac{d u_{1\phi}}{d\tau}
= -\frac{\kappa^2}{2\Omega^2}u_{\rm 1R} + \beta \qm \alpha,
\end{equation}
\begin{equation}
\frac{d u_{\rm 1z}}{d\tau}=
-\qm b_{\rm z} + \qz (\beta\alpha + b_{\phi}),
\end{equation}
\begin{equation}
\frac{d b_{\rm R}}{d\tau}=\qm u_{\rm 1R},
\end{equation}
\begin{equation}
\frac{d b_{\phi}}{d\tau}
= \frac{d\ln{\Omega}}{d\ln{R}} b_{\rm R} - \qz u_{\rm 1z} - \qR  u_{\rm 1R},
\end{equation}
\begin{equation}
\frac{d b_{\rm z}}{d\tau} =\qm u_{\rm 1z},
\end{equation}
where the dimensionless Lagrangian derivative is denoted by
\begin{equation}
\frac{d}{d\tau} = \frac{1}{\Omega}
\frac{\partial}{\partial t} + \frac{\partial}{\partial \phi}.
\end{equation}
In eqs.\ (61)$\sim$(68), all perturbed variables are dimensionless and
defined by $\alpha \equiv \rho_1/\rho_0$,
$\bold{u_1} \equiv i \bold{v_1}/\Vap$,
$\bold{b}\equiv \bold{B_1}/\bp$, and 
$\tau \equiv t\Omega $, and dimensionless parameters are 
$\beta \equiv \cs^2/\Vap^2$,
$\qm \equiv \Vap m/R\Omega$,
$\qz \equiv \Vap\kz/\Omega$,
and
\begin{equation}
\qR(\tau) \equiv \frac{\Vap\kr(t)}{\Omega} = 
\frac{\Vap}{\Omega}
\left[\kr(t_{\rm o}) - m(t-t_{\rm o})\frac{d\Omega}{dR}\right]
= -mt\frac{\Vap}{\Omega}\frac{d\Omega}{dR}
= -\tau\qm\frac{d\ln\Omega}{d\ln R},
\end{equation}
where the third equality holds when 
$t_{\rm o}\equiv -\kr(t_{\rm o})/m\Omega'$.
Eqs.\ (65)$\sim$(67) yield the
divergence free condition for the perturbed magnetic fields.

Since $\qR$ has a $\tau$-dependence,
the linear system of eqs.\ (61)$\sim$(67) does not form an eigenvalue problem; 
kinematics of shear
wrap a given disturbance by increasing its radial wavenumber linearly with
time. In the original shearing sheet formalism, 
the fate of a system exposed to perturbations is analyzed through
direct integrations of linearized equations over time.
In doing so, one may
observe transient amplification or decay of applied disturbances depending
on their stability. One can say that a system is unstable if some physical
variables grow sufficiently over certain time scales.
The efficiency of instability for a system is identified by computing the
response of the system to variation of
parameters input to temporal integrations.
This approach was adopted by \citet{bal92} in their identification
of the NMRI.

Here, we instead analyze the NMRI by
proceeding one more step from the original shearing sheet formalism
to find solutions which are localized in time as well as in space.
First, we note that there exist two distinct time scales: the growth
time of instabilities determined by the inverse of the dimensionless 
instantaneous growth rate, 
\begin{equation}
\gamma(\tau) \equiv \frac{d}{d\tau}\ln{\chi_1(\tau)} =
\frac{1}{\Omega}\frac{d}{dt}\ln{\chi_1(t)},
\end{equation}
and the dimensionless shearing time, 
$(d\ln\qR /d\tau)^{-1}=\qR |\qm d\ln \Omega/d\ln R|^{-1}$, 
as a typical time scale of the linear growth of the radial wavenumber.
If the shearing time is much longer than the growth time, that is,
if $\qm |d\ln \Omega/d\ln R|/(\gamma\qR) \ll 1$
(the ``weak shear limit''),
the time dependence of $\qR$ in eq.\ (69)
can be neglected, 
and thus normal mode solutions having
an exponential or oscillatory behavior
can be sought.
\citet{shu74} applied this technique to investigate the effects of
a differential rotation on the Parker instability.
Also, \citet{ryu92} obtained an algebraic dispersion relation for
the convective instability in differentially rotating disks, by
assuming that $\qR$ is time-independent.

Because the convective and the Parker instabilities arise from hydrodynamic
and magnetic buoyancy effects, respectively, independent of the rotation
of a disk, one can always find
a regime in which the weak shear limit is applicable.
In some cases, however, as for example in the axisymmetric 
poloidal or the non-axisymmetric toroidal 
MRIs with weak magnetic fields,
the instabilities result directly from
a differential rotation with $\Omega' <0$.
In such cases, 
peak growth rates are of the same order as rotational 
frequencies \citep{bal98}, 
and thus the weak shear is not a good approximation for 
these non-axisymmetric instabilities.

However, we can still look for coherent solutions in which
{\it all} perturbed variables vary as
$e^{\gamma \tau}$ with time,
provided the variation of the instantaneous growth rate $\gamma(\tau)$
over the growth time $\gamma^{-1}$
is relatively small, i.e.,
\begin{equation}
\left|\frac{d\ln \gamma(\tau)}{d\tau}\right| \ll \gamma(\tau).
\end{equation}
We refer to the solutions under this approximation
as ``coherent wavelet solutions''
because all physical quantities localized in both space and
time grow at the same instantaneous rate.
If the condition (71) holds, the changes in $\gamma(\tau)$
can be neglected over a short time interval, and the set of dynamical
equations (61)$\sim$(67) constitutes an eigensystem instantaneously.
This is equivalent to the WKB method in the time dimension.
Since $\gamma^{-1}d\ln \gamma(\tau)/d\tau =
-\qm(d\ln\Omega/d\ln R) (d\ln \gamma/d\ln\qR) /(\gamma\qR)$, eq.\ (71) 
is satisfied if either
$\qm|d\ln\Omega/d\ln R|/(\gamma\qR) \ll 1$ (the weak shear limit), or
$|d\ln\gamma/d\ln\qR| \ll 1$ (instantaneous growth rates are relatively
insensitive to the radial wavenumber);
the condition (71)
is less restrictive and in fact is the generalization of the weak shear limit.
Of course, we need to check the self-consistency of this coherent
wavelet approximation by examining
{\it a posteriori} whether resulting solutions satisfy the condition (71).
For incompressible media, \citet{bal92} mapped the regime of instability
in [$({\bf k\cdot\Va})^2$, $|{\bf k}|/\kz$] space 
using WKB methods similar to those we adopt.

\subsubsection{Coherent Wavelet Dispersion Relation}

Upon substituting eq.\ (70) into eqs.\ (61)$\sim$(67) and applying the 
approximation (71) so that $d\chi_1/d\tau \rightarrow \gamma\chi_1$, 
one can form a matrix equation $\gamma Q = {\cal M}Q$,
where  $Q=(\alpha,u_{\rm 1R},u_{1\phi},u_{\rm 1z},b_{\rm R},
b_\phi,b_{\rm z})^{\rm T}$ is a column vector and ${\cal M}$ is a $7\times$7 
matrix whose components are determined by the coefficients of $Q$
in the right hand sides of eqs.\ (61)$\sim$(67). By solving
the condition det(${\cal M} - \gamma{\cal I})=0$, where ${\cal I}$ is the
identity matrix,
we obtain a seventh order polynomial in $\gamma$.
As a further approximation, however, 
if at least one of the conditions, $\qz \gg \qR$, $\gamma \gg \tau$,
or $\gamma\tau \gg 1$, is satisfied,
all even order terms
that depend linearly on $\tau$ and $\qm$ but are independent of $\qz$, 
can be neglected compared to the remaining terms.
The first two conditions apply when
the radial wavenumber is not significant, either because disturbances
are highly localized in the vertical direction ($\qz \gg \qR$)
or simply because we are looking at modal behaviors at the time $\tau \sim 0$, 
while the third condition holds when net amplification of perturbations 
is large. 
This simplification yields a trivial solution $\gamma=0$ (this arises 
from the fact that perturbed magnetic fields, $b_{\rm R}$, $b_\phi$, 
and $b_{\rm z}$, are linearly dependent via the divergence-free 
condition) and a third-order polynomial in $\gamma^2$ which is 
the resulting instantaneous dispersion relation for NMRI
\begin{mathletters}
\begin{eqnarray}
0 = \gamma^6 &+& \gamma^4 
\left[(1+\beta)q^2 + \qm^2 + \frac{\kappa^2}{\Omega^2} \right]
\nonumber \\
&+& \gamma^2 \left[(1+2\beta)q^2\qm^2 +
\frac{\kappa^2}{\Omega^2}(\qm^2 + (1+\beta)\qz^2)\right]
+\beta \qm^2 \left[q^2\qm^2 + \frac{d\ln{\Omega^2}}{d\ln R}\qz^2\right],
\eqnum{72}
\end{eqnarray}
\end{mathletters}
where the amplitude of the total wavenumber defined by $q^2(\tau) 
\equiv \qR (\tau)^2 + \qm^2 + \qz^2$ is a function of $\tau$ through
eq. (69). Combining eqs.\ (69) and (72), one can evaluate a local,
instantaneous growth rate at a given time.

With vanishing magnetic fields and thermal pressure, 
we would obtain from eq.\ (72)
stable epicyclic oscillations. In the limit of strong 
magnetic fields and no rotation, eq.\ (72) is immediately reduced to
$(\gamma^2+\qm^2)(\gamma^4 + (1+\beta)q^2\gamma^2 + \beta\qm^2q^2)=0$,
the usual dispersion relations for the \Alf\ waves and the fast and slow MHD
waves in a medium embedded with toroidal magnetic fields. 
In the presence of rotation with non-vanishing but weak fields,
however, these \Alf\ and MHD modes are coupled to exhibit generally 
complex modal behaviors. They can be stable or unstable depending on
the parameters, but it is always a slow MHD wave that becomes 
unstable because it has the lowest frequency so that there is a plenty
of time during which destabilizing forces (centrifugal forces for NMRI) 
act on it.
We have instantaneously growing solutions
with real positive values of $\gamma$ provided that 
the last term in eq.\ (72) is negative. Thus, when $\beta\qm \neq 0$,
the local, instantaneous
instability criterion in terms of the dimensionless variables for the
NMRI can be written
\begin{equation}
\qm^2\left(1 + \frac{\qR (\tau)^2 + \qm^2}{\qz^2}\right)
+ \frac{d\ln\Omega^2}{d\ln R} < 0,
\end{equation}
demonstrating that $d\Omega^2/d\ln{R} = \kappa^2 - 4\Omega^2<0$ is indeed a
sufficient condition for the instability to arise when the magnetic field
strength is negligible \citep{bal92,fog95,ter96}. 
Eq.\ (73) recovers the results of \citet{bal92} for the instability regime
for toroidal-field NMRI.
The result of eq.\ (73) can be compared
to the poloidal-field BH instability criterion of eq.\ (59b). Notice
that the NMRI (with $B_{\rm 0z}=0$) 
vanishes completely as $\beta\rightarrow 0$,
while the axisymmetric BH instability still exists even when $\beta=0$
(see \S 7.1).

Although they share the same instability criterion, the operating 
mechanisms for the NMRI of toroidal fields are quite different
from the axisymmetric BH instability of poloidal fields.
Both arise via destabilization of the
slow mode. The NMRI mode, just as the
poloidal BH mode, depends on  shear to generate azimuthal fields from
radial perturbations of the background fields. But there is more
to the story. The key mechanism for the NMRI instability lies in the
vertical MHD wave motions driven by the gradient of the total (initial
plus perturbed) azimuthal fields, as schematically illustrated in Fig.\ 10.
Since we suppose perturbations which are
sinusoidal in both vertical and azimuthal directions, the perturbed
azimuthal fields are also periodic in both directions. Rapid vertical
motions with high $\kz$, generated by $-\bp(\partial B_{1\phi}/\partial z)$
stress, would produce over- and under-dense regions which regularly
alternate along the azimuthal direction (Fig.\ 10a). And then,
azimuthal fluid motions
are induced, according to the equation of continuity, from over-dense
regions to under-dense regions (Fig.\ 10b).
Depending on the direction ($\mp \hat{\phi}$) of these induced motions,
the coriolis and/or centrifugal force would alter the paths,
radially inward or outward (Fig.\ 10c).
Under the condition of field freezing, these radial motions would produce
radial magnetic fields with a small amplitude from the background
toroidal fields (Fig.\ 10d). These radial fields would in turn be sheared
out to generate (positive or negative) perturbed azimuthal fields,
due to the differential rotation of the background flows.
When $d\Omega/dR <0$, the resulting azimuthal fields from initial and
perturbed ones would be distributed (Fig.\ 10e) such that they reinforce
the applied initial vertical perturbations (Figs.\ 10a and 10f),
implying the MRI; the entire system
would just oscillate with rotation-modified MHD frequencies if
$d\Omega/dR >0$. This explains how the NMRI operates.
 
When $\kz$ is large, the stabilization of the NMRI occurs when a 
magnetic tension from
radially bent field lines exceeds the centrifugal or coriolis force
(Figs.\ 10b and 10c). Shear \Alf\ waves with radial polarization
can suppress the instability if the field lines are sufficiently strong  
or if the azimuthal wavenumber is large enough, as 
expressed by the dimensionless parameter $\qm^2$ outside the 
parentheses in eq.\ (73).
When $\qz \ll \qR$, on the other hand, MHD waves propagating along the radial 
direction stabilize the NMRI, as indicated by the terms inside
parentheses in eq.\ (73);
$\qR(\tau)$ clearly reflects the stabilizing effect of the 
background shear.

The maximum instantaneous growth rate is achieved when $\qR\approx 0$.
In this case, eq.\ (73) implies instability if
\begin{displaymath}
q^2_{\rm m} < q^2_{\rm m,crit} \equiv  \frac{1}{2}
\left[-\qz^2\! +\! \sqrt{\qz^4-
4\qz^2\frac{d\ln{\Omega^2}}{d\ln{R}}}\;\right].
\end{displaymath}
Note that $q^2_{\rm m,crit} \rightarrow -d\ln{\Omega^2}/d\ln{R}$, for
$\qz \gg 1$, while $q^2_{\rm m,crit} \rightarrow \qz \sqrt{-d\ln \Omega^2/
d\ln R}$, for $\qz \ll 1$.
Numerical solutions of eq.\ (72) with $\qR=0$ are presented
in Fig.\ 11. As both eq.\ (73)
and Fig.\ 11 show, the maximum growth rates for 
toroidal-field background states are attained when $\qz\rightarrow
\infty$, which is a sharp contrast to the axisymmetric 
poloidal-field BH instability that has fastest
growing mode at moderate $\qz$'s \cite[cf.\ Fig.\ 9 and see also
discussion in][]{bal98}. But both forms of the MRI 
have the same maximum growth rates at the same $\qA \equiv
(\bold{k}\!\cdot\!\bold{v}_{\rm A})/\Omega$. Fig.\ 11b shows how growth rates
depend on the sound speed. As the sound speed increases, the medium becomes
more unstable. This reflects the incompressible nature of the
NMRI.
Even though the marginally critical wavenumber is independent of temperature
(for $\beta \neq 0$), the virulence of the instability is greatly inhibited
as $\beta$ decreases. 
For $\qm \ll 1$, one can find from eq.\ (72) the temperature 
dependence of the limiting growth rate 
\begin{equation}
\gamma = \qm\sqrt{-\frac{\beta}{\kappa^2(1+\beta)}\frac{d\Omega^2}{d\ln R}},
\end{equation}
or $\gamma = \qm\sqrt{3\beta/(1+\beta)}$ for a Keplerian rotation.
Eq.\ (74) gives slopes of the growth rates for small $\qm$ (Fig.\ 11).
For magnetocentrifugally driven winds which are
as cold as $\beta < 0.01$, the NMRI is not expected to play
a significant role;
the growth rate in dimensional units is $\sqrt{3}\cs m/R$.

When the medium is incompressible ($\beta \rightarrow \infty$), eq. (72)
allows the analytic expression for the instantaneous growth rate
for pure toroidal-field background states,
\begin{equation}
\gamma^2 = \left\{ \begin{array}{ll}
\frac{\qz^2}{2q^2}\frac{\kappa^2}{\Omega^2}\left[\sqrt{
1 + 16\frac{q^2\qm^2}{\qz^2}\frac{\Omega^4}{\kappa^4}} -1 \right] - \qm^2,
&{\rm if\;\;\;} \frac{q^2\qm^2}{\qz^2} + \frac{d\ln\Omega^2}{d\ln R} < 0, \\
& \\
0.
&{\rm otherwise}, \\
\end{array} \right.
\end{equation}
When $\qz \gg 1$, one can derive the maximum growth rate
$\gamma_{\rm max} = |d\ln\Omega/d\ln R|/2$,
which is achieved when 
$q^2_{\rm m,max} =  -(d\ln\Omega^2/d\ln R) /2 - \gamma^2_{\rm max}$.
It can be shown from eq.\ (72) or (75) that
$d\gamma^2/d\tau \sim \qm^3\qR/q^2 \rightarrow 0$  as $\qz \rightarrow
\infty$. This proves that the coherent wavelet approximation is 
self-consistent for the NMRI with high $\qz$.

\subsubsection{Comparison With the Shearing Sheet Formalism}

In order to compare the coherent wavelet solutions for
toroidal-field NMRI with the results from
the shearing sheet approximation, we directly integrate eqs.\ 
(61)$\sim$(67) over time, with given sets of initial conditions.
In Fig.\ 12, we display the time evolution of all perturbed  
variables for $\qm=0.1$, $\qz=1$, and $\beta=100$, which are
the same parameters as chosen for Fig.\ 3 of \citet{bal92}.
We adopt a Keplerian rotation profile in what follows. 
The initial amplitudes are 0.1 for every variable except
$b_{\rm R}=0.01$ and $b_{\rm z}=0.4$, and the initial $\tlt$,
where the orbit number $\tlt \equiv \tau/2\pi = t\Omega/2\pi$, is 
allowed to be determined from the divergence free condition of 
the initial, perturbed magnetic fields. When $\tlt < -20 $, 
the system responds with MHD wave motions before they start to grow.
During this relaxation stage, fast MHD modes having large $|\qR|$
are nearly longitudinal acoustic waves,  affecting $u_{\rm 1R}$ 
and $\alpha$, 
while $u_{1\phi}$ and $b_\phi$ are mostly influenced by transverse 
slow modes.
As time increases, $|\qR|$ gradually decreases, permitting
rotational shear to affect the overall dynamics.
Once the condition (73) is satisfied, shear drives the slow modes to 
be unstable, following the process
illustrated in Fig.\ 10. Even though the growth of disturbances shows
a transient nature due to the kinematic growth of $\qR$, 
the net amplification is about 9 orders of magnitude,
a bit higher
than Balbus \& Hawley's result. This is because the integration interval
in Balbus \& Hawley covered a slightly smaller part of the unstable time range.
At later time when $\qR$ has a large value, the system
exhibits stable oscillations with the slow MHD wave frequency.
Fig.\ 12 also shows the predicted amplification magnitude
(thick solid line) from the coherent
wavelet approximation (see below).

In Fig.\ 13, we plot the numerical growth rates for each variable 
calculated from Fig.\ 12
based on the direct numerical integrations in the shearing sheet formalism,
together with the growth rate of the corresponding coherent wavelet solution. 
Here, a dimensionless instantaneous growth rate $\tilde{\gamma}$ 
as a function of $\tlt$ is
defined through $10^{\tilde{\gamma}\tlt}=e^{\gamma \tau}$,
(or $\tilde{\gamma}(\tlt) = 2\pi \gamma(\tau) \log e)$. 
Note that the heavy solid line for $\tilde{\gamma}$ drawn from
eq.\ (72) fits well with various curves computed from
the direct numerical integrations.
The instantaneous growth rates 
are almost symmetric with respect to their maxima near $\tlt=0$,
as expected. 
Growth of the modes occurs only when $|\tlt| < 18.3$,
which is in good agreement with the results of the direct integrations,
demonstrating the validity of the coherent wavelet approximation.

We define a dimensionless amplification magnitude as
\begin{equation}
\Gamma(\tlt) \equiv \int_{-\infty}^{\tlt}
\tilde{\gamma}(\tlt')d\tlt'
= \log e \int_{-\infty}^{\tau} \gamma(\tau')d\tau'.
\end{equation}
Then, $\Gamma(\tlt)$ is an order of magnitude measurement of
the amplification
of an unstable mode during the time interval ($-\infty, \tlt$).
The total amplification is given by $10^{\Gamma(\infty)}$.
When eq.\ (75) is substituted,  the analytic evaluation of the integral
in eq.\ (76) is not an easy
task. In view of a shape of $\tilde{\gamma}(\tlt)$ (Fig.\ 13),
we further approximate $\tilde{\gamma}$ with a simple form
\begin{equation}
\tilde{\gamma}= \left\{ \begin{array}{ll}
\tilde{\gamma}_{\rm o}(1- |\tlt|/\tlt_{\rm c})^{1-\qm/2},
&{\rm if\;\;\;} |\tlt|<\tlt_{\rm c}, \\
0,
&{\rm otherwise}, \\
\end{array} \right.
\end{equation}
where
\begin{equation}
\eqnum{78a}
\tilde{\gamma}_{\rm o} \equiv
\sqrt{2} \pi\log e \left\{\frac{\qz^2}{\qm^2+\qz^2}\frac{\kappa^2}{\Omega^2}
\left[\sqrt{
1 + 16\frac{(\qm^2+\qz^2)\qm^2}{\qz^2}
\frac{\Omega^4}{\kappa^4}} - 1 \right] - 2\qm^2\right\}^{1/2},
\end{equation}
and the termination epoch of growth $\tlt_{\rm c}$ is defined by
\begin{equation}
\eqnum{78b}
\tlt_{\rm c} \equiv
\frac{1}{2\pi\qm}\left|\frac{d\ln\Omega}{d\ln R}\right|^{-1}
\sqrt{-\left(\frac{d\ln \Omega^2}{d\ln R} \right)
\frac{\qz^2}{\qm^2}-\qm^2-\qz^2}.
\end{equation}
\setcounter{equation}{78}
Notice that eq.\ (77) is valid only if $\tlt_{\rm c}$ is real,
that is, only if the condition (73) is satisfied.
From eqs.\ (76) and (77), the total amplification magnitude is easily found
to be
\begin{equation}
\Gamma(\infty) = \frac{4\tilde{\gamma}_{\rm o}\tlt_{\rm c}}{4-\qm},
\end{equation}
which is illustrated with solid contours in Fig.\ 14.
Also shown with dotted contours are the direct results from numeral
integration of eq.\ (75), which are in excellent agreement with
$\Gamma(\infty)$.
The thick contour is the locus of $\tlt_{\rm c}=0$, demarcating
the stable and unstable regions: the system is stable at
the right hand side of the thick contour.
In the $\qm-\qz$ plane, the total amplification tends to be greater
as $\qz$ becomes larger and as $\qm$ becomes smaller. This
is because the NMRI with $B_{\rm 0z}=0$ 
acquires maximum instantaneous growth
rates at $\qz=\infty$ (Fig.\ 11a) and because the
shearing time is longer with smaller $\qm$ (cf.\ eq.\ [78b]).
For comparison,
we also include in Fig.\ 14 the results from the shearing sheet equations
for four parameter sets:
($\qm$, $\qz$) = (0.03, 0.1), (0.1, 1), (1, 10), and
($\sqrt{2}$, 100$\sqrt{2}$), and $\beta$=100 for all cases: these
are marked with dots on $\qm-\qz$ plane, labeled by the respective
exact and estimated (in parentheses) amplification magnitudes.
Note that all of the estimated amplification magnitudes are within 5\%
of the results of direct shearing sheet integrations.
This indicates that eq.\ (79) is an excellent analytic estimate for 
the amplifications of incompressible NMRI modes.

\subsection{Generalized MRI}

Motivated by the success of the coherent wavelet method in finding 
the solutions of the NMRI with purely toroidal background fields,
we now generalize both the axisymmetric BH and NMRI
instabilities by considering non-axisymmetric perturbations applied to
the rotating medium threaded by both vertical and azimuthal magnetic fields.
We include the effect of thermal pressure and allow the angular velocity
$\Omega$ to be a function of $R$, but ignore any other radial variations
in the initial state. We adopt the shearing sheet coordinates as before,
and linearize eqs.\ (1)$\sim$(4). After applying perturbations in the form 
of eq.\ (60), we assume that the perturbations evolve with time as
$e^{\gamma (t) t}$ with the coherent wavelet condition 
(i.e., $d\ln \gamma(t)/dt \ll \gamma(t)$).
Following the same procedure as \S 7.2, 
we obtain the general instantaneous dispersion relation for the MRI
(now written in dimensional form)
\begin{mathletters}
\begin{eqnarray}
0 = \gamma^6 &+& \gamma^4
\left[(\cs^2+\Va^2)k^2 + \kappa^2 + (\bold{k}\cdot\bold{\Va})^2 \right]
\nonumber \\
&+& \gamma^2 \left[(2\cs^2 + \Va^2)(\bold{k}\cdot\bold{\Va})^2k^2 + 
\kappa^2\left(\cs^2\kz^2 + \Vap^2(\frac{m^2}{R^2}+\kz^2) \right)
+(\bold{k}\cdot\bold{\Va})\kz\Vaz \frac{d\Omega^2}{d\ln R}
\right]
\nonumber \\
&+& \cs^2 (\bold{k}\cdot\bold{\Va})^2 
\left[(\bold{k}\cdot\bold{\Va})^2 k^2 + 
\frac{d{\Omega^2}}{d\ln R}\kz^2\right],
\eqnum{80} 
\end{eqnarray}
\end{mathletters}
where $k^2\equiv k_{\rm R}^2(t) + m^2/R^2 + \kz^2$ with the radial
wavenumber defined by $k_{\rm R}(t) = - mtd\Omega/dR$
when we choose $t_{\rm o} = -\kr(t_{\rm o})/m\Omega'$.
When either $m=0$ (axisymmetric case) or $\Vaz=0$ (pure toroidal field
case), eq.\ (80) becomes identical respectively with eq.\ (57) for the BH
modes or eq.\ (72) for the NMRI modes.

From eq.\ (80), we obtain the instantaneous
instability criteria for the generalized MRI modes
\begin{mathletters}
\begin{eqnarray}
\Va^2k^2(t)(\bold{k}\cdot\bold{v}_{\rm A})^2 +
\kappa^2\Vap^2\left(\frac{m^2}{R^2}+\kz^2 \right)
+ (\bold{k}\cdot\bold{v}_{\rm A})
\Vaz\kz\frac{d\Omega^2}{d\ln R}\; & < \;0,
\quad{\rm for} \quad \cs = 0,
\\
(\bold{k}\cdot\bold{v}_{\rm A})^2
\left(1 + \frac{\kr^2(t) + m^2/R^2}{\kz^2} \right) +
\frac{d\Omega^2}{d\ln R}\; & <\; 0,
\quad{\rm for} \quad \cs\neq 0,
\end{eqnarray}
\end{mathletters}
which are obviously the generalizations of eqs.\ (59) and (73).
It can also be shown that when $i=90^{\rm o}$, both
equations (81a) and (81b) become identically
$\Vaz^2(\kr^2(t)+m^2/R^2 + \kz^2) + d\Omega^2/d\ln R <0$.
With an $\Omega \propto R^{-a}$ rotation profile, 
eq.\ (81a) gives the sufficient condition for the instability in
an extremely cold medium: $\sin i > (4-2a)/(4-a)$. Keplerian flows 
for example become unstable only if $i\gtrsim24^{\rm o}$, indicating that 
cold, $B_\phi$-dominated media are not subject to the generalized 
non-axisymmetric MRI disturbances, just as we found earlier that 
axisymmetric BH modes are also stable in cold flows for small $i$.
We remark that the case with $a\rightarrow 2$, as potentially possible
in MHD winds from boundary layers, can just barely satisfy the cold 
medium instability criterion for $i\rightarrow 0$.
Note that unlike the NMRI mode with $\Vaz=0$, maximum growth rates 
in the $\cs\neq 0$ case are not achieved
as $\kz\rightarrow \infty$. In fact, high-$\kz$ or high-$m$ disturbances
are efficiently stabilized by \Alf\ and/or MHD waves whenever {\it both} 
poloidal and toroidal fields are present.
However small they may be, therefore, inclusion of poloidal fields 
would yield a different result from the case with pure toroidal fields
(this point was previously noted by Balbus \& Hawley 1998). 
Again the stabilizing effect of kinematic
shear appears through the time dependence of $\kr^2(t)$, when $m\neq0$.

Fig.\ 15a shows how the compressible BH modes are stabilized by azimuthal
magnetosonic waves. Here, we confine consideration to the radial wavenumber 
$\kr=0$. 
As $\qm (= \Va m/R\Omega)$ increases, both the growth
rates and the ranges in $\qz (=\Va \kz/\Omega)$ of unstable modes decrease.
This is because if $\qm\neq0$, azimuthally displaced material feels 
relatively stronger restoring forces due to both thermal and magnetic pressures
of the medium as well as stronger tension forces from bent field lines.
Non-axisymmetric poloidal-field BH instability modes become
stabilized with increasing values of $m$.
When $\qm > \sqrt{-d\ln \Omega^2/d\ln R}$ ($=\sqrt{3}$ for a Keplerian
rotation), the instability is strictly cut off,
even when the effect of kinematic shear is not taken into account.

We remark that the role of temperature of the medium
to the BH instability is different between axisymmetric
(with $\qm=0$ and $i\neq 90^{\rm o}$; see Fig.\ 9) and
non-axisymmetric (with $\qm\neq 0$ and $i=90^{\rm o}$; see Fig.\ 15) cases.
When $i=90^{\rm o}$, as already explained in \S 7.1, the axisymmetric BH
instability with $\qm=0$ is independent of $\beta$,
because only Alfv$\acute{\rm e}$n and sound waves exist
and they do not interact with each other.
When $\qm=0$ and $i\neq 90^{\rm o}$, magnetic pressure induces vertical
MHD wave motions which tend to stabilize the system when $\beta$ is small.
If $\beta \gg 1$, however, the vertical wave
motions become nearly acoustic, leaving the toroidal component of
perturbed fields unaffected and permitting higher growth rates.
When $\qm\neq 0$ and $i=90^{\rm o}$, on the other hand,
the coupling of thermal pressure with magnetic pressure occurs
through azimuthal MHD motions,
and the growth rate depends only weakly on $\beta$.

Fig.\ 15b shows loci of equi-growth rate on the $\qz\!-\!\qm$ plane for
$i=10^{\rm o}$ and $\beta=0.01$ (dotted contours) and
$\beta=100$ (thin solid contours). For $0<\qm<1.17$, there exist
upper and lower critical vertical wavenumbers, $q_{\rm z,u}$ and
$q_{\rm z,l}$ such that the system is unstable with
$q_{\rm z,l}<\qz<q_{\rm z,u}$. When $\qz> q_{\rm z,u}$,
disturbances are stabilized by MHD waves propagating mainly
along vertical direction, while perturbations with
$\qz<q_{\rm z,l}$ approach stable Alfv$\acute{\rm e}$n waves.
Each contour has
a slope of $\sim -\tan i$ ($= -0.18$ for $i=10^{\rm o}$) at both ends.
Note that dotted contours with lower $\beta$ are labeled with much smaller
growth rates than solid ones  with higher $\beta$, even though they are
similar in shape. Compared to Fig.\ 9 or Fig.\ 15a, this implies that the
NMRI instability with a toroidal field configuration is more sensitive
to temperature than the axisymmetric/non-axisymmetric BH instability
with poloidal fields.

When $\beta \rightarrow \infty$, from eq.\ (80) we have the 
instantaneous growth rates for the generalized MRI modes
\begin{equation}
\gamma^2 = 
\frac{\kappa^2\kz^2}{2k^2}\left[\sqrt{
1 + 16({\bf k\cdot\Va})^2\frac{k^2\Omega^2}{\kz^2\kappa^4}} -1 \right] - 
({\bf k\cdot\Va})^2,
\end{equation}
for $({\bf k\cdot\Va})^2k^2/\kz^2 + d\Omega^2/d\ln R < 0$, which is also
a generalization of eq.\ (75). With weak magnetic field strength,
one can show from eq.\ (82) that
$\gamma^{-3}d\gamma^2/dt \sim -m\kr({\bf k\cdot\Va})/(\Omega\kz^2)
\rightarrow 0$ as $({\bf k\cdot\Va})/\Omega \rightarrow 0$.
Thus, we see that for the generalized MRIs,
the coherent wavelet approach is self-consistent 
in the weak field limit.
When the field strength is moderate,
on the other hand, we obtain 
$d\gamma^2/dt \sim m\Omega'\kr({\bf k\cdot\Va})^2/k^2$.
Since $\gamma\sim\Omega\sim ({\bf k\cdot\Va})$ in this case,
the coherent wavelet condition is met only when
$m|\Omega'| \ll \kr\gamma$ (the weak shear limit).
Of course, 
the predominantly toroidal-field case that becomes 
unstable with $\kz\gg 1$ also satisfies 
the coherent wavelet condition, since $k^2$ becomes arbitrarily large
without increasing the $({\bf k\cdot\Va})$-term,
as discussed in \S 7.2.2.

Comparing eq.\ (75) with eq.\ (82), we note that the incompressible
MRI can be generalized simply by replacing the dimensionless azimuthal 
wavenumber $\qm$ with $({\bf k\cdot\Va})/\Omega$. Therefore, we can
write the net amplification magnitude for the generalized 
incompressible MRI as
\begin{equation}
\Gamma(\infty) = \frac{ 4\log e}
{4-\qA}\;\gamma_{\rm o}t_{\rm c}
\end{equation}
where the dimensional peak growth rate $\gamma_{\rm o}$ and 
the cut-off time of the instability $t_{\rm c}$ are defined by
\begin{equation}
\frac{\gamma_{\rm o}^2}{\Omega^2} \equiv
\frac{\kappa^2\cos^2\theta}{2\Omega^2}\left[
\sqrt{1 + 16\frac{\qA^2}{\cos^2\theta}\frac{\Omega^4}{\kappa^4}}
-1 \right] - \qA^2
\end{equation}
and
\begin{equation}
t_{\rm c}\equiv
\frac{1}{\sin\theta}\left|\frac{d\Omega}{d\ln R}\right|^{-1}
\sqrt{-\frac{d\ln\Omega^2}{d\ln R}\frac{\cos^2\theta}{\qA^2}
-1},
\end{equation}
respectively. The net amplification can thus be 
completely determined by the two parameters:
the dimensionless wavenumber $\qA \equiv ({\bf k \cdot \Va})/\Omega$ 
projected in the direction of initial equilibrium magnetic fields, and 
the angle $\theta \equiv \tan^{-1}(m/R\kz)$ of the wavenumber vector
with respect to the vertical axis. Note that $\gamma_{\rm o}\rightarrow 0$
with vanishing $\qA$, while $t_{\rm c}\rightarrow \infty$ as 
$\theta\rightarrow 0$,
indicating that low-$m$ instabilities show higher net
amplifications than high-$m$ disturbances as long as $\qA \neq 0$.
The total amplification magnitudes, eq.\ (83), are plotted in 
Fig.\ 16 with thin solid contours. For comparison, we also plot 
the numerical results from eqs.\ (76) and (82) with dotted contours. 
We assume a Keplerian rotation profile.
The heavy curve with $\qA^2=3\cos^2\theta$ draws the locus of the marginal
stability. 
In the limit of a weak magnetic field strength (i.e., $\qA\rightarrow 0$),
it can be shown from eqs.\ (83)$\sim$(85) that
$\Gamma(\infty) = (2\log e)\Omega/(\kappa\tan\theta)$, inversely proportional
to $\theta$ (for $\theta \ll 1$) 
but independent of $\qA$, as illustrated in Fig.\ 16.
Also shown in Fig.\ 16 are the results from the direct
temporal integrations of shearing-sheet equations with $\beta=100$ as
filled circles (for $i=90^{\rm o}$), filled triangles (for $i=30^{\rm o}$),
and open circles (for $i=0^{\rm o}$), labeled by the
respective exact and estimated (in parentheses) amplification
magnitudes. These two results agree very well, implying that
the coherent wavelet approach indeed provides excellent 
approximations to the solutions for amplification of generalized MRIs.

From Fig.\ 16, it is apparent that it is the locally
near-axisymmetric (in the sense $m/R\kz = \tan\theta \ll 1$) disturbances 
that experience maximum amplification, with the amplification magnitude 
only weakly dependent on $\qA = ({\bf k \cdot \Va})/\Omega$
within the unstable regime ($\qA \lesssim 1$). The increase in 
amplification factor with $R\kz/m$ predicted from linear theory may
in part explain the larger amplitudes of power spectra for modes with
larger ${\bf \hat{k}\cdot\hat{z}}$ measured from nonlinear 
simulations of the saturated MRI \cite[cf.][]{haw95}.
In addition, for the case of pure toroidal fields, Fig.\ 16 suggests
only low amplification factors unless $\kz$ is very large, which may
help explain why \citet{haw95} found lower magnetic field
saturation amplitudes in cases with initial $B_{\rm z} =0$.

\section{Summary and Discussion}

\subsection{General Conclusions Based on Linearized Analysis}

Through linear analyses of the ideal MHD equations, we have explored the
stability of shearing, rotating flows to a wide range of (primarily local)
disturbances. The chief motivation for this study is to characterize
the internal instabilities that could develop in
disk winds that emanate from an extended region of a
differentially rotating protostellar disk around a young star.
The dynamics of such winds has inspired intensive theoretical
effort because they may be responsible for observed YSO jets and
outflows. In our analysis,
we include both results based on generic density, magnetic field,
and flow profiles, and results which adopt as initial equilibrium
configurations the power-law asymptotic solutions of self-confined
cylindrically symmetric winds presented by Ostriker (1997):
$\rho \propto R^{-q}, B_{\phi} \propto B_{\rm z} \propto R^{-(1+q)/2}$,
and $v_{\phi} \propto v_{\rm z} \propto R^{-1/2}$.
For most of our analysis (\S\S 2-6), the flows were assumed to be cold 
enough that thermal effects can be ignored compared with magnetic forces.
To make contact with other studies of shear-induced MHD instabilities
in rotating disks, we also consider stability of specific models which 
include non-zero thermal pressure (\S7). For the lowest-order 
``fundamental'' modes, we employ a normal-mode analysis with free 
Lagrangian boundary conditions to find
eigenvalues and eigenfunctions of both stable and unstable modes (\S 3).
For higher-order modes, we employ three different local techniques to
study growth of unstable disturbances: In \S\S 4-6, we present
numerical and analytic solutions of dispersion relations obtained
from normal mode analyses in the $R\kr \gg 1$ WKB limit. These are
exact for $m\Omega'+\kz v_{\rm 0z}'=0$ disturbances and
are valid for a limited time for weakly-shearing circumstances
where $\kr\gg m|\Omega'/\Omega|, \kz|Rv_{\rm 0z}'/v_{\rm 0z}|$
(see also \S 8.2 below). In \S 7, we employ temporal integrations
of the shearing-sheet equations to study MRI modes (which are
cut off for $R\kr \gg 1$). We also introduce, in \S 7.2,
a ``coherent wavelet'' formalism which adapts modal analyses for 
situations where shear is considerable (i.e., small $R\kr/m$); 
the coherent wavelet analysis
is equivalent to a WKB approach in the temporal domain.
We include a
comparison of results from the shearing-sheet and coherent-wavelet
techniques applied to MRIs, in \S 7.2.3.

Applying these techniques we have identified a total of nine different
unstable or overstable families of disturbances that 
occur for a wide range of flow
parameters: five (FM, BH, ATB, PB, and TR) of them
are axisymmetric and the other four (NTB, GPB, PR, and NMRI) are
non-axisymmetric. Table 1 summarizes the properties of these modes.
The main general conclusions drawn from the analysis in
this work can be summarized as follows:

(1) Systems having a primarily azimuthal magnetic field, for example,
disk winds far from their source, are susceptible to the
fundamental (FM), axisymmetric (ATB) and non-axisymmetric (NTB) toroidal 
buoyancy, non-axisymmetric magnetorotational instability (NMRI) 
and toroidal resonance (TR) modes. 
Unstable fundamental modes (see \S 3.2) are concentrated in the central parts
of jets, and occur in $B_\phi$-dominated flows when the logarithmic
gradient of the magnetic field is steeper than $\approx -0.75$ 
(cf.\ eqs.\ [24] and [25]). Growth rates of unstable FM are comparable
to inner-wind \Alf\ frequencies. Long wavelength modes with large amplitudes
at large radii are all stable, for power-law wind profiles. 
The TR mode (see \S 6.1.4) is an overstability, 
with growth suppressed when $\kr$ increases through shear
of the vertical velocity, simply becoming oscillatory MHD waves.
The axisymmetric toroidal buoyancy mode (ATB; \S 6.1.2) is activated
initially by the buoyancy force and subsequently by bending poloidal
magnetic fields. In geometrical form, it is locally similar to the sausage
mode of a plasma column confined by toroidal fields, and leads to radial
mixing. Because growth rates are larger on smaller scales, ATB can 
contribute to the generation of local turbulence in disk winds.
The non-axisymmetric toroidal buoyancy mode
(NTB; see \S 6.2.1) is much like the Parker instability, but with the
centrifugal force replacing the role of external gravity.
Although the normal-mode analysis for NTB has the largest temporal validity 
at small $m/R\kr$, the instantaneous growth rate increases with increasing
$m/R\kr$ (cf.\ eq.\ [52] and Fig.\ 8c). We thus return to the NTB in \S 8.2, 
below, applying time-dependent techniques to study the $R\kr/m \ll 1$ limit. 
Because the NTB is present whenever radial magnetic forces are non-zero, it 
may be important in promoting radial mixing. Both of the toroidal 
buoyancy instabilities require non-zero magnetic forces in the equilibrium
state. The rarefied and cold conditions
of disk winds do not favor the development of the NMRI 
(see \S 7.2.2 and \S 7.3). Like the original (poloidal field) 
magnetorotational (BH) instability,
NMRI requires $d\ln{\Omega^2}/d\ln{R} <0$, but also requires
a relatively incompressible medium, as is provided by the relatively
dense and warm ($\cs \gtrsim \Va$) conditions in an accretion disk.
We show the NMRI vanishes in the limit of $\cs/\Va \rightarrow 0$.
We further discuss perturbations in cold, $B_\phi$-dominated flows
in \S\S8.2 and 8.3, below.

(2) Systems having primarily axial magnetic fields, for example,
accretion disks or winds very near their origin, are susceptible to
the Balbus-Hawley (BH), poloidal buoyancy (PB and GPB), and
poloidal resonance (PR) modes.
The well-known axisymmetric Balbus-Hawley instability (BH; see \S 6.1.3 and
\S 7.1) is the most efficient member of the family of magnetorotational
instabilities (MRIs; see \S 7.3).
It will work to produce channel flows, eventually
generating fully-developed MHD turbulence through coupling to
non-axisymmetric disturbances in the non-linear stage.
Driven by background magnetic pressure and the centrifugal force,
the axisymmetric poloidal buoyancy mode (PB; see \S 6.1.1)
requires a gradient in the magnetic field strength to be
unstable. If the field distribution is steep enough, the poloidal buoyancy
modes would also work effectively to generate radial mixing and turbulence
over much smaller scales than the BH instability. Because of their
overstable characteristics, the impact on the system of poloidal
resonance modes (PR; see \S 6.2.3) would be best evaluated with a global 
rather than local formalism. Configurations with shallower background 
magnetic gradients $(q<1)$
are also subject to a non-axisymmetric poloidal buoyancy instability
(GPB; see \S 6.2.2) which arises in part from geometric effects.

(3) In distinction to the original, incompressible, axisymmetric BH 
instability, we found
that the compressible axisymmetric BH mode is strongly stabilized by
the presence of an azimuthal magnetic field if the medium has
substantially sub-Alfv$\acute{\rm e}$nic sound speeds. For example, 
in a cold rotating flow with $\Omega \propto R^{-3/2}$, 
the axisymmetric BH instability would be completely suppressed
if the local pitch angle $i \equiv \tan^{-1}(B_{\rm z}/B_\phi)$ is 
less than 30$^{\rm o}$ (cf.\ eq.\ [47]). In an incompressible medium
(as provided by a disk with $\cs \gtrsim \Va$), faster sound waves preserve
perturbed toroidal fields from being dispersed by MHD wave motions, thereby
providing a favorable condition for the BH instability.
When the field configuration is purely poloidal,
the compressible BH instability is identical with its incompressible
counterpart, independent of temperature (cf.\ eqs.\ [58] and [59]).

(4) Even though they share the same instability criterion (cf.\ eqs.\ [59b]
and [73]), the operating mechanisms for the NMRI
of purely toroidal $\bold{B}$-fields is entirely different from the 
axisymmetric BH instability of primarily poloidal $\bold{B}$-fields. 
In the NMRI (see \S 7.2), vertical MHD wave motions driven by magnetic 
pressure play an essential role in the feedback loop for induced radial 
disturbances, while the axisymmetric BH instability tends to be stabilized 
by vertical wave motions. Faster sound speeds produce higher growth rates 
in both instabilities, but for different reasons: in the NMRI by activating
azimuthal fluid motions preceded by the vertical MHD wave motions;
in the BH instability by maintaining the perturbed azimuthal fields generated
by shear (when $\bp\neq 0$).
Because of their non-axisymmetric nature, the NMRI
has a transient growth, stabilized by the growth of $\kr$ from kinematic
azimuthal shear.
For the NMRI mode, we show explicitly by comparison to direct temporal
integrations of the shearing sheet equations that the growth rate
at $\kr=0$ can be used to provide a good estimate of the net amplification
magnitude (see \S 7.2.3).

(5) The coherent wavelet formalism we develop (\S7.2) may be used to compute 
instability criteria and net amplification factors for generalized MRI
disturbances with arbitrary magnetic field and wavevector orientations 
(\S7.3). Eq.\ (80) gives the instantaneous dispersion relation for 
generalized MRIs. For strongly compressible flows ($\cs/\Va\rightarrow 0$),
instability does not occur in $B_\phi$-dominated configurations
(cf.\ eq.\ [81a]); in this case, flows with an $\Omega\propto R^{-3/2}$
rotation law can be unstable only when the magnetic pitch angle 
$i>24^{\rm o}$. Because MHD disk winds generally have very small pitch 
angles, this result has the important implication that such {\it winds 
will not be subject to the development of strong internal turbulence 
that occurs as a consequence of nonlinear MRIs in disks}. 
The absence of MRIs in cold, $B_\phi$-dominated winds may be crucial in
enabling them to propagate over large distances from their sources.
High-$\kz$ and/or high-$m$ modes of the generalized MRI are stabilized
by MHD waves, which is a sharp contrast with the NMRI of 
{\it purely toroidal} fields in which maximum growth rates are 
attained at $\kz \rightarrow \infty$.  
For incompressible flows, the amplification 
factor for all MRIs can be written analytically in terms of 
$\qA = ({\bf k \cdot \Va})/\Omega$ and $\theta = \tan^{-1}(m/R\kz)$
(eqs.\ [83]$\sim$[85]); within the unstable regime ($\qA <
|d\ln\Omega/d\ln R|^{1/2}$, from eq.\ [81b]),
the amplification is $\sim \exp[2\Omega/(\kappa\tan\theta)]$,
favoring ``locally-axisymmetric'' disturbances.

\subsection{Effect of Shear on Dynamical Growth of 
Buoyancy Instabilities}

Apart from the results of \S 7 where we adopted the shearing sheet
formulation of the dynamical equations to study MRIs, the results in
this work have been elicited on the basis of the local normal mode analyses.
As described in \S 4, these modes may have a limited range of temporal 
validity, due to the effects of background shear. For axisymmetric 
disturbances with negligible vertical shear 
(i.e., $m\Omega'+\kz v_{\rm 0z}'\rightarrow 0$), the results presented
in \S 5 and \S 6 are acceptable for all time; the modes with pure
imaginary $\omega$ will show an exponential growth without interruption
over arbitrarily long time until nonlinearity sets in.
However, for non-axisymmetric disturbances, or for flows with non-negligible
vertical shear, unstable modes identified in \S5 and \S6 are
not purely growing. As time evolves, the differential velocities build up
the radial wavenumber through the kinematic shear (cf.\
eq.\ [40]), which in turn tends to stop the further growth of disturbances.
This can be seen directly through the suppression of instabilities in the 
local analysis when $\kr$ is large (cf.\ Fig.\ 8).  
The characteristic time for the wave pattern to change by a fraction
$\epsilon$ is $t=\epsilon\kr/|m\Omega'+\kz v_{\rm 0z}'|$; over
this interval, the disturbance will be amplified by a factor
$\exp(\epsilon\kr{\rm Im}(\tom)/|m\Omega'+\kz v_{\rm 0z}'|)$.
When $\kr \gg m/R, \kz$, Fig.\ 8 shows that Im($\tom) \propto \kr^{-1}$,
so that the net amplification factor is nearly independent of $\kr$.
Since, however, Im($\tom$) is not larger than 
$\sim|m\Omega'+\kz v_{\rm 0z}'|\kr^{-1}$,
only order-unity amplification can be expected for disturbances which
are consistent with the requirements for quasi-steady normal mode
analysis.

Because the normal-mode dispersion relations indicate larger
values of growth rates when $R\kr/m$ is small (which is however not
self-consistent with the WKB treatment), 
it is desirable to extend investigation to
allow for $R\kr/m$ small. The coherent wavelet formalism used for
the MRI in \S 7 suggests that when $m\gg 1$ (or $R\kz \gg 1$ for
$v_{\rm 0z}'\neq 0$ cases), this can be done
by regarding $\kr$ as a 
time-dependent variable according to eq.\ (40) and using
the asymptotic dispersion relations of \S 6
(i.e., eq.\ [43] for PB, eq.\ [45] for ATB, eq.\ [52] for NTB, 
and eq.\ [53] for GPB). 
To verify this argument, we specifically consider the NTB modes 
(which are one of the chief instabilities in $B_\phi$-dominated winds) and 
compare the results with 
the shearing sheet temporal integrations. 
For the latter, we set $B_{\rm z}=0$ and 
integrate eqs.\ (26)$\sim$(31) in time, setting all
of the coefficients to constant values. The resulting instantaneous 
growth rates and time evolutions of variables are plotted in Fig.\ 17 
as functions of the normalized time $\tau = t\Omega$.
We omitted the $\kr$-dependent term in eq.\ (26) in order to
remove rapid oscillations arising from a phase mismatch between
the density and radial velocity; the amplitude evolution
is independent of this term.
We also neglected the vertical velocity shear and
and selected $q=\kr(0)=\kz=0$, $R\Omega = 0.1 \Vap$, and $m=100$. 
As initial conditions, we chose 0.1 for every variable except
$b_{\rm R}=0.01$ and integrated the system of the linearized equations.
Various curves are computed from direct
numerical integrations of shearing sheet equations, while the heavy solid 
lines are
drawn from the normal mode solution, eqs.\ (52) and (86) (see below), 
after taking allowance for the time dependence of $\kr$. 
The rapid fluctuations of the perturbed variables for $\tau<0$ are
due to MHD waves with high $|\kr|$, disappearing after
variables grow substantially. 
Again, most of growth
occurs over a relatively short period of time near $\kr \sim 0$.
Note an excellent agreement between the results from two
different approaches; we have also obtained similar results 
with integration from other initial conditions.
This confirms that our normal mode results can also be applied 
to high-$m$ disturbances if $\kr$ is allowed to vary with time.

Using eq.\ (52) with $\kr(t)$ from eq.\ (40), we integrate $\tom$ over 
time to estimate the net amplification for the NTB modes
\begin{equation}
\frac{\chi_1(t)}{\chi_1(0)} = \exp{\int_0^t {\rm Im}(\tom)\, dt} =
\left[ \frac{3}{2}\Omega t + \sqrt{1+\left(\frac{3}{2}\Omega t\right)^2}
\right]^{(1-q)\Vap/3R\Omega},
\end{equation}
where we put $v_{\rm 0z}'=\kr(0)=0$ and assume a Keplerian rotation.
Thus instead of an exponential growth, at later time of evolution we have
the power-law growth due to the kinematic shear.
This behavior is distinct from the MRI modes, which are strictly
stable for large enough $\kr$. However, the continued local growth
of buoyancy perturbations is offset by the role of kinematic shear in
mixing phases of disturbances. Considering waves on $z$=constant plane 
in a square with sides $L$, the maximum averaged contrast in any variable 
relative to the mean value is $(\lambda/L)\chi_1(t)$,
where $\lambda$ is the local wavenumber of the waves. 
With $\lambda \sim \kr(t)^{-1}$ and using eqs.\ (6) and (86), 
the average contrast for the NTB modes
evolve with time as $\sim (\Omega t)^{\sqrt{2(1-q)}/3-1}$, vanishing as 
$t \rightarrow \infty$ for $0<q<1$. 
From eq.\ (86), amplification factors are essentially scale-free.
Although there may be a significant growth of the NTB modes on large
scales, their dynamical effect on small scales is limited by the
phase mixing due to shear.

In the presence of vertical shear, the evolution of ATB modes is also
affected by the kinematic growth of $\kr$ although they are axisymmetric 
modes. The net amplification of the ATB modes follows a power-law 
growth as that of the NTB mods does. In fact, eq.\ (86) with $\Omega$ replaced 
by $2v_{\rm 0z}'/3$ gives the temporal behavior of the net amplification
for the ATB modes.

\subsection{Discussion of Applications to Protostellar Winds}

In order for a disk wind to overcome the gravitational barrier due to a
central object and to be centrifugally launched from the surface of a
Keplerian disk, the poloidal components of field lines should thread the
disk at an angle of 30$^{\rm o}$ or more from the axis \citep{bla82}.
Once material starts to flow outward along such field lines, it is
accelerated primarily in the radial direction by the centrifugal force or
by the pressure gradient in the toroidal field.
Beyond the \Alf\ surface where the local,
poloidal component of the flow velocity is equal to that of the \Alf\
wave velocity, the magnetic field is not strong enough to play a role
of ``a rigid wire", and the inertia of gas becomes important, winding up
the field lines to be progressively more toroidal. In this process, the
azimuthal flow velocity decreases below the corotation value.
With the increase in the azimuthal component of the magnetic field,
the associated hoop stress provides the collimation of the outflow and
causes the streamlines to bend upward. The radial flow velocity of
the outflow is still positive, although it decreases gradually,
eventually becoming zero at the cylindrical asymptotic limit.
The power-law solutions (with $\Vr =\br =0$) we adopted for
many specific cases represent the asymptotic limit of each
streamline.

\citet{ost97} presented self-similar steady solutions for disk winds
with cylindrical asymptotics and gave the asymptotic fluid and \Alf\
speeds and the location of the asymptotic streamlines, characterized by
$q$ together with $\RA/R_1$ or $R_0/R_1$, where $R_0, \RA$, and $R_1$
denote the radii of the footpoint, the \Alf\ surface, and the asymptote
of each given streamline, respectively. Typical numerical values for
those solutions are 
$\Omega = 0.2\Omega_0$, $\Vap/R = 0.42\Omega_0$, and 
$\Vap/v_{\rm z}=6$ for $q=0.5$, and
$\Omega = 0.1\Omega_0$ and $\Vap/R = 0.45\Omega_0$ for $q=0.9$, where
$\Omega_0$ is the Keplerian rotation rate at the streamline footpoint. 

Using these values we can estimate the growth times of the global
fundamental mode and the fastest growing (with $\kr$ near 0) toroidal 
buoyancy modes. The foregoing analysis suggests that these disturbances
will play the most significant dynamical role, given the ineffectiveness
of MRIs in cold, $B_\phi$-dominated flows.
We define the time to grow by $\Gamma$ orders of magnitude as 
$t_\Gamma$. For FM, $t_{\Gamma,\rm FM} = \Gamma/(|\omega|\log e)$, so from
Fig.\ 2 and eq.\ (25) we have for $q=0.9$
\begin{displaymath}
t_{\Gamma,\rm FM} \sim 
\frac{14\Gamma}{\Omega_{\rm 0,i}} =
25\; \Gamma \;{\rm days} \;
\left(\frac{M}{M_\odot}\right)^{-1/2}
\left(\frac{R_{\rm 0,i}}{0.1\;\rm AU}\right)^{3/2}.
\end{displaymath}
For ATB and NTB, from eqs.\ (45) and (52), we have for 
$i\rightarrow 0$ and $q=0.5$
\begin{displaymath}
t_{\Gamma,\rm ATB} \sim
\frac{10^\Gamma}{0.21\Omega_0} =
24\times 10^\Gamma\; {\rm yrs} \;
\left(\frac{M}{M_\odot}\right)^{-1/2}
\left(\frac{R_0}{10\;\rm AU}\right)^{3/2}\;\;\;{\rm and}
\end{displaymath}
\begin{displaymath}
t_{\Gamma,\rm NTB} \sim 
\frac{10^{1.4\Gamma}}{3\Omega_0} =
1.7\times 10^{1.4\Gamma} \;{\rm yrs} \;
\left(\frac{M}{M_\odot}\right)^{-1/2}
\left(\frac{R_0}{10\;\rm AU}\right)^{3/2},
\end{displaymath}
respectively. Here, $M$ is the mass of the central star and 
$\Omega_{\rm 0,i}$ is the angular speed of a disk
at the footpoint $R_{\rm 0,i}$ of the innermost streamline of winds.
The fact that the growth of the FM by a factor $10^\Gamma$ occurs within 
$\sim \Gamma$ times the rotation period of the disk at the inner radius,
far shorter than the lifetime of winds ($\sim 10^4-10^5$ yrs),
suggests that the FM mode is dynamically important
in the evolution of the disk winds.
When $q$ is small, the radial turbulent mixing of the wind, 
caused by both axisymmetric and non-axisymmetric toroidal buoyancy modes over
a relatively short time, is likely to cascade down into arbitrarily smaller
scales to dissipate when the microscopic processes such as magnetic
reconnection are included. The released energy in the dissipation
processes may heat up the flow,
potentially making a significant contribution to the heating of 
protostellar winds and jets. Because the growth rates of buoyancy modes are
proportional to the equilibrium magnetic forces (cf.\ eqs.\ [45] and [52]),
winds that have approached a force-free magnetic configuration will not be
subject to the ATB and NTB instabilities.

The global fundamental mode affects only the inner region
of the disk winds (e.g., the central tenth for the model shown
in Fig.\ 3). The logarithmic density gradient $\partial\ln\rho/\partial\ln R$
changes relative to the equilibrium value by $\partial(\rho_1/\rho_0)/
\partial\ln R$. Fig.\ 3 shows that as a consequence of the 
fundamental mode, the very central region becomes more steeply
stratified, a surrounding concentric region less steeply
stratified, and the balance (most of the wind) remains nearly
unchanged. Thus, the FM tends to enhance jetlike structure in the
central parts of winds. In addition, because of their tendency
to compress interior gas via the FM mechanism, disk winds may help
to collimate any interior flows into narrow, fast jets, even when
the disk winds themselves have relatively slow motion \cite[cf.][]{ost97}.

What do the present results imply about the likely radial extent 
of protostellar winds? First, we note that observed optical jets are unlikely
to be isolated structures, because if so they would be significantly 
overpressured relative to the ambient medium:
Since magnetocentrifugal jet models typically predict internal \Alf\ speeds
comparable to their flow speeds in the range $150\sim400$ km s$^{-1}$,
they have strong internal magnetic pressure $P_{\rm wind} = B_\phi^2/8\pi \sim
\rho_{\rm w}\Va^2/2 \sim 2.8\times 10^{-7}$ ergs cm$^{-3}$,
which is about 6 orders of magnitude greater than the gas pressure of
ambient medium, $P_{\rm ext} = \cs^2\rho_{\rm ext}
\sim 1.3\times 10^{-13}$ ergs cm$^{-3}$.
Here as reference values we adopted
$\rho_{\rm w} = 350\; m_{\rm H}$ cm$^{-3}$,
$\rho_{\rm ext} = 200\; m_{\rm H}$ cm$^{-3}$,
$\Va = 310$ km s$^{-1}$, and
$\cs = 0.2$ km s$^{-1}$ with $m_{\rm H}$ being the mass of a
hydrogen atom \citep{har99}. Thus the pressure imbalance at the outer
boundary of the jet would cause either the wind as a whole or only its
surface layer to expand until a new balance is attained.
Based on the results of this paper, if the magnetic field at the base 
of the wind is stratified less steeply than $R^{-1}$, then perturbations
of the outer parts of the wind are stable. 
As a consequence, only the surface layers
of such winds would expand in order to achieve a pressure-balanced
condition with the ambient medium. If, on the other hand, the wind's 
magnetic field is stratified more steeply than $R^{-1}$ at its base,
no equilibrium is even possible; the wind would expand as a 
whole to fill the entire $4\pi$ steradians,
with the inner parts having higher density observable as a narrow
optical jet \cite[cf.][]{shu94,sha98}.
Numerical simulations presently underway \citep{lee00}
support previous work indicating that protostellar winds with a wide-angle
component are better able to produce observed molecular outflow
structures than purely jetlike winds \cite[see also][]{li96,
ost97,ost98,mat99}
but further studies are required to determine just how distributed
in angle the wind momentum should be - i.e., to discriminate between
``fully-expanded'' and ``surface-expanded'' models. Recent 
observations \cite[see, e.g.,][]{ric00} showing a correlation
in molecular outflow kinematics with age - with extremely high velocity,
highly-collimated flows seen only in the youngest sources - may indicate
an underlying temporal evolution from more-collimated to more-expanded
protostellar winds.

\acknowledgements

We acknowledge a stimulating report from an anonymous referee, and
helpful comments from N. Turner and S. Balbus.

\clearpage
\begin{figure}
\figurenum{1}
\epsscale{1.0}
\vspace{0.in} \plotone{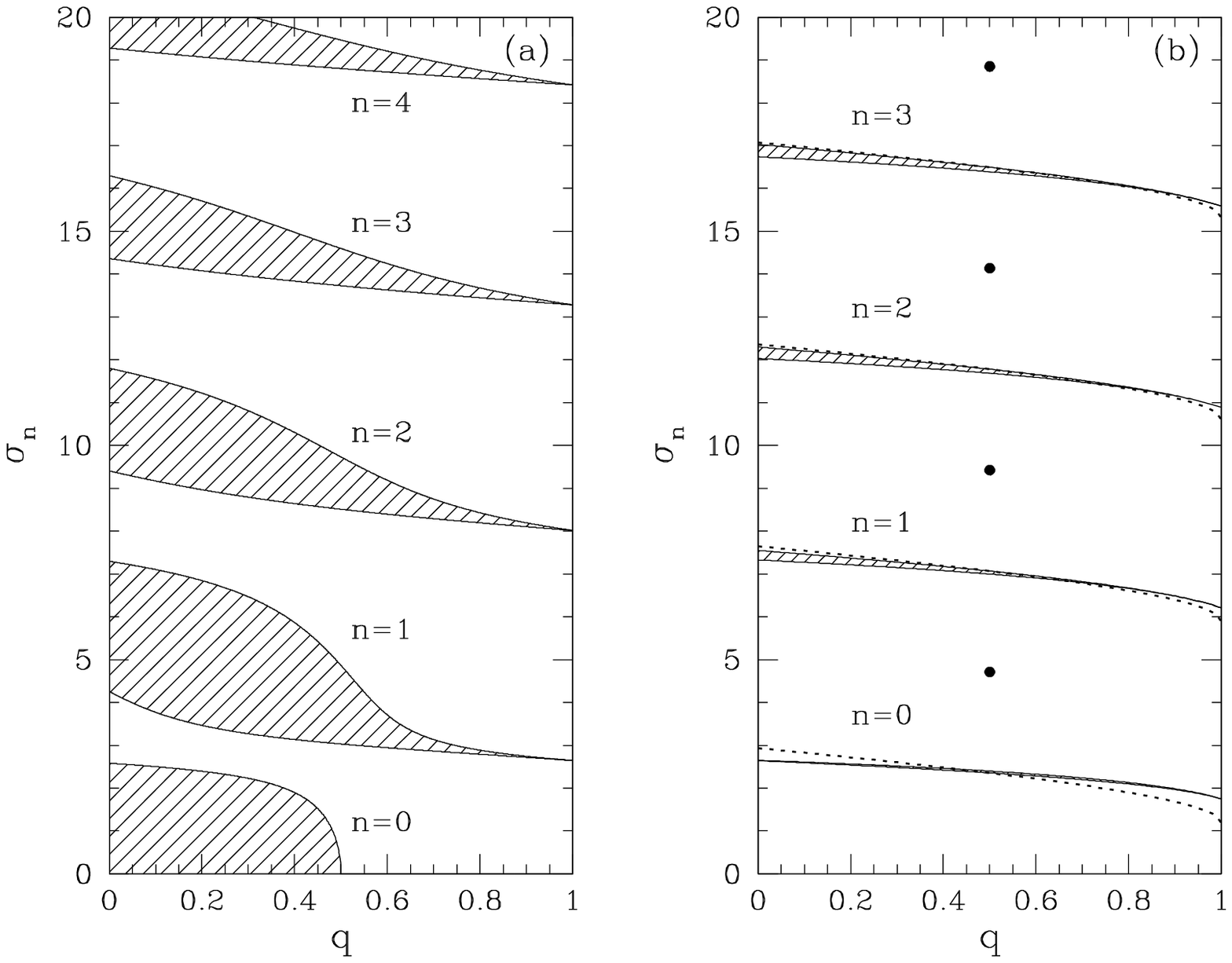} \vspace{-0.5in}
\caption{
Eigenvalues $\sigma = \omega \Ree/\Va(\Ree)$
of stable global fundamental modes for (a) $r_{\rm i}=0.1$
and (b) $r_{\rm i}=10^{-4}$ are plotted as functions of $q$
($\Ree$ is the radius of the outer boundary, $\Ri$ is the
radius of the inner boundary, and $r_{\rm i} \equiv \Ri/\Ree$).
Each shaded envelope is filled with eigenfrequencies having the same $n$ but
different $i<i_{\rm max}$. Since $i_{\rm max}$ is a decreasing function
of $q$, having $i_{\rm max}=0$ when $q=1$, the envelopes
narrow in width as $q$ increases. Eigenvalues with $r_{\rm i}=0.1$ are
more sensitive to $q$ and $i$ than those with $r_{\rm i}=10^{-4}$.
Mode conversion (see text) occurs when a cylindrical wind has a narrow
thickness ($r_{\rm i}$ near 1).
Dotted lines in (b) represent asymptotic eigenvalues with $r_{\rm i} \ll 1$,
which are in good agreement with numerical solutions. Also shown are
the eigenvalues for the case with $q=0.5$ and $i=0$ as filled circles.
The upper boundary of each envelope corresponds to $i=0^{\rm o}$.
}
\end{figure}

\clearpage
\begin{figure}
\figurenum{2}
\epsscale{1.0}
\plotone{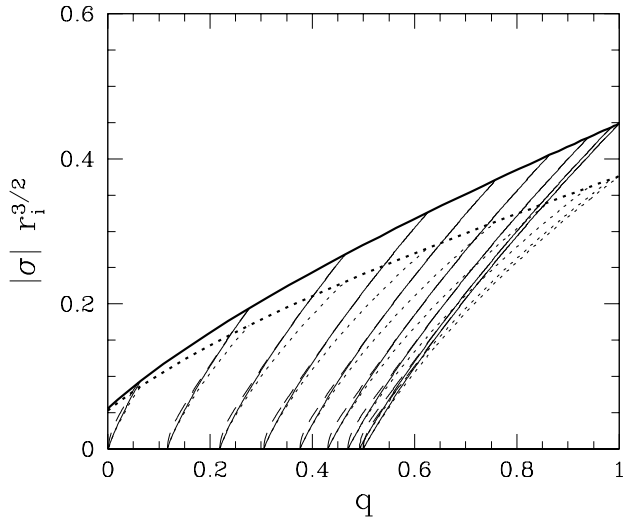} \vspace{-0.5in}  
\caption{
Eigenvalues $|\sigma|r_{\rm i}^{3/2} = |\omega|\Ri/\Va(\Ri)$
of unstable global fundamental modes. Solid lines 
($r_{\rm i}=10^{-4}$) and dashed lines ($r_{\rm i}=10^{-1}$) are
the exact results computed from eqs.\ (20b) and (22). 
Drawn as dotted lines, the approximate, analytic solutions (eq.\ [25]) 
follow the exact values fairly well when $|\sigma|$ is relatively small.
The various curves represent different pitch angles of
the equilibrium magnetic field configuration: 
$i=0^{\rm o}, 5^{\rm o},\cdot\cdot\cdot,35^{\rm o},
40^{\rm o}$ from right to left. 
The uppermost thick lines correspond to the maximum pitch angle, 
$i_{\rm max}$. Note that $|\sigma| r_{\rm i}^{3/2}$ is almost
independent of $r_{\rm i}$.
}
\end{figure}

\clearpage
\begin{figure}
\figurenum{3}
\epsscale{1.0}
\plotone{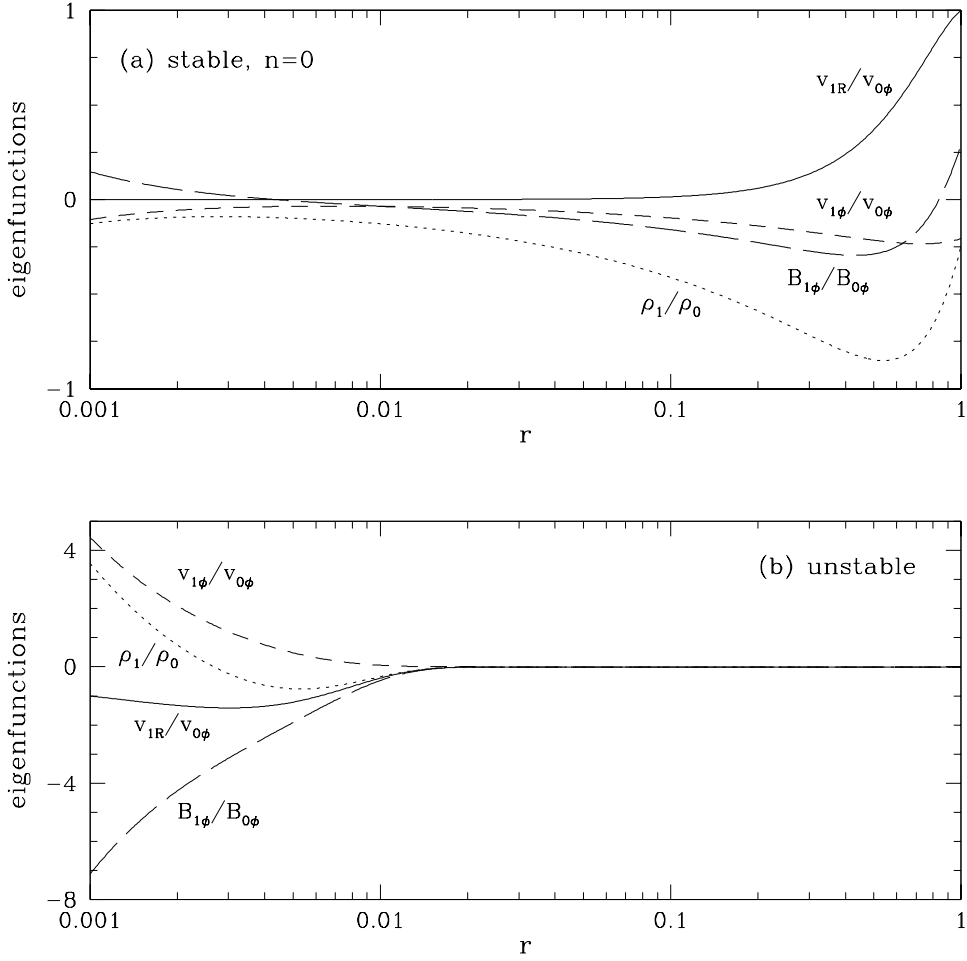} \vspace{-0.5in} 
\caption{
Examples of normalized (a) stable and (b) unstable fundamental modes with
$i=0^{\rm o}$ and $r_{\rm i}=10^{-3}$. We choose
$q=0.4$, $\sigma_0 = 2.42$, and 
$v_{\rm 1R}/v_{0\phi} = 1$ at $r=1$ for the stable modes,
and $q=0.6$, $|\sigma|r_{\rm i}^{3/2}=0.11$, 
$v_{\rm 1R}/v_{0\phi} = -1$ at $r=r_{\rm i}$
for the unstable modes.
Stable eigenfunctions dominate the outer region, 
while unstable ones affect only the inner part of the system.
}
\end{figure}

\clearpage
\begin{figure}
\figurenum{4}
\epsscale{1.0}
\plotone{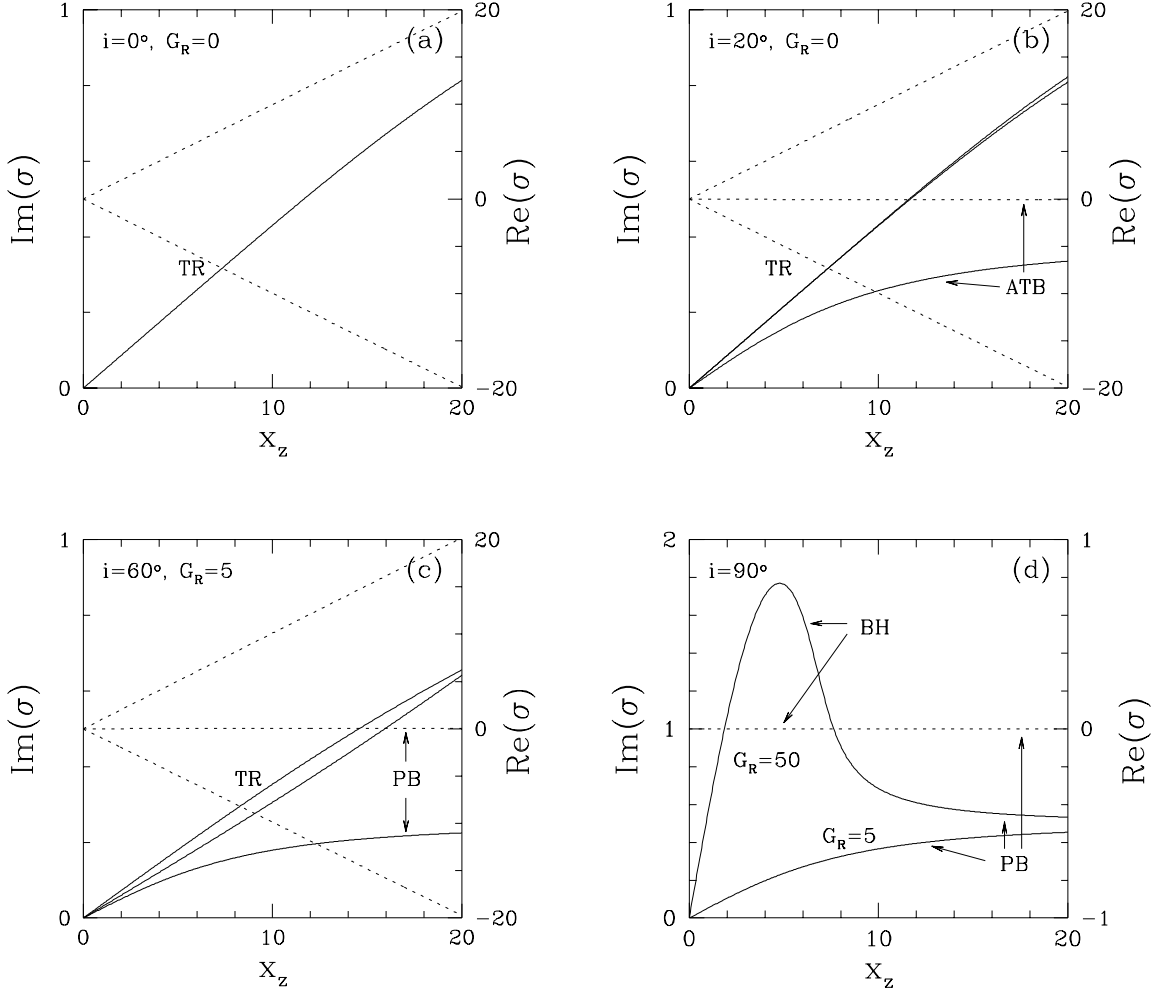} \vspace{-1.0in}
\caption{
Normalized frequencies $\sigma \equiv \tom\Ro/\Va(\Ro)$ 
of unstable and overstable modes of axisymmetric 
perturbations are
plotted against the normalized vertical wavenumber
$\xz=\kz\Ro$. For all 4  cases, $\xr=10$, $q=\zeta=0$ are
adopted. In each frame, solid and dotted lines represent imaginary
and real parts of the solution frequencies, respectively.
(a) When $i=0$, only the toroidal resonance mode (TR) exists, which 
is overstable  with  larger real parts.
(b) and (c) The TR splits into two branches with the inclusion of
poloidal fields.
Axisymmetric buoyancy modes begin to appear; with relatively 
small $i$, the axisymmetric toroidal buoyancy mode (ATB) exists,
whereas the poloidal buoyancy mode (PB) operates with 
predominant poloidal fields.
(d) For a pure poloidal configuration, only BH and PB modes exist.
The BH is dominant at a relatively
larger wavelength region with a higher growth rate when magnetic 
field is weak (larger $\GR$).
}
\end{figure}

\clearpage
\begin{figure}
\figurenum{5}
\epsscale{1.0}
\plotone{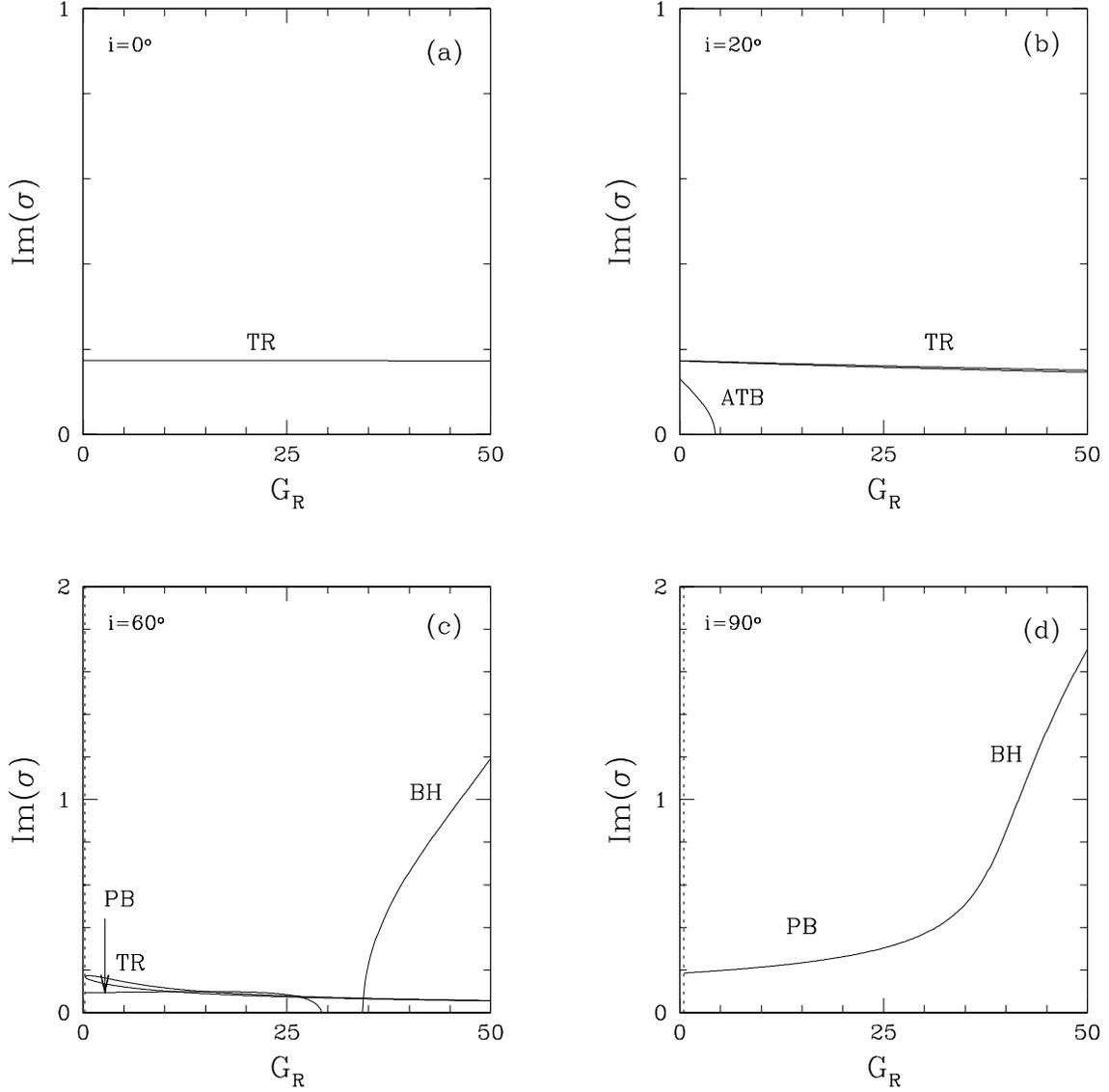}
\caption{
Changes of axisymmetric unstable and overstable modal growth rates 
with $\GR$ and $i$.
Normalized vertical wavenumber is fixed as $\xz\equiv \kz\Ro=4$.
A rotation profile $\Omega \propto R^{-3/2}$ is assumed and
$\xr=10$, $q=\zeta=0$ are taken.
ATB and PB modes need to have smaller $\GR$ to be unstable. Note that
as $\GR$ increases, corresponding to a weak field regime, only
the BH mode prevails. Dotted lines in the frames (c) and (d)
mark the minimum value of $\GR$, available for given $q$ and $i$,
below which no initial equilibrium exists.
For details, see text.
}
\end{figure}

\clearpage
\begin{figure}
\figurenum{6}
\epsscale{1.1} 
\plotone{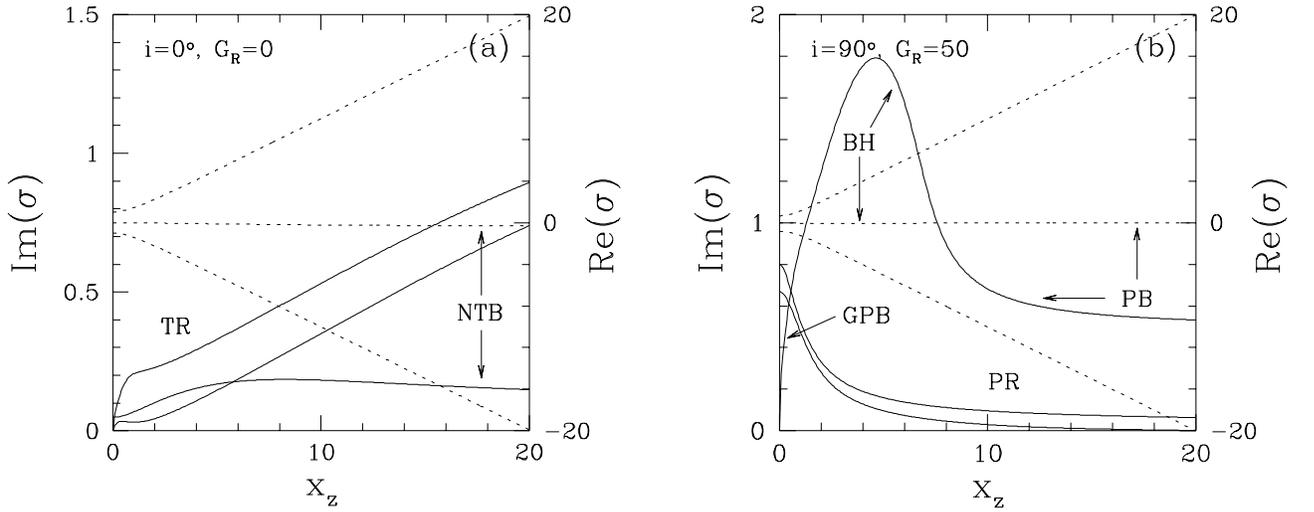} \vspace{-4.in}
\caption{
Frequencies of unstable and overstable modes for non-axisymmetric
perturbations with $m=1$ are plotted against the normalized
vertical wavenumber. For both frames,
$\xr=10$, $q=\zeta=0$ are adopted and an $\Omega \propto R^{-3/2}$ 
rotation profile is assumed.
Solid and dotted lines represent imaginary
and real parts of the normalized wave frequencies, respectively.
(a) In a toroidal
configuration, only TR and NTB exist, with TR being split by
the non-axisymmetric effect. NTB has a nearly constant growth rate
independent of $\xz$.
Note that the real part of TR increases linearly with $\xz$, while
that of NTB is nearly zero.
(b) When $i$ is high, GPB begins to appear.
PB is almost unchanged by the non-axisymmetry.
Like TR, PR is also an overstable mode with a larger real part
proportional to $\xz$. 
}
\end{figure}

\clearpage
\begin{figure}
\figurenum{7}
\epsscale{1.0}
\plotone{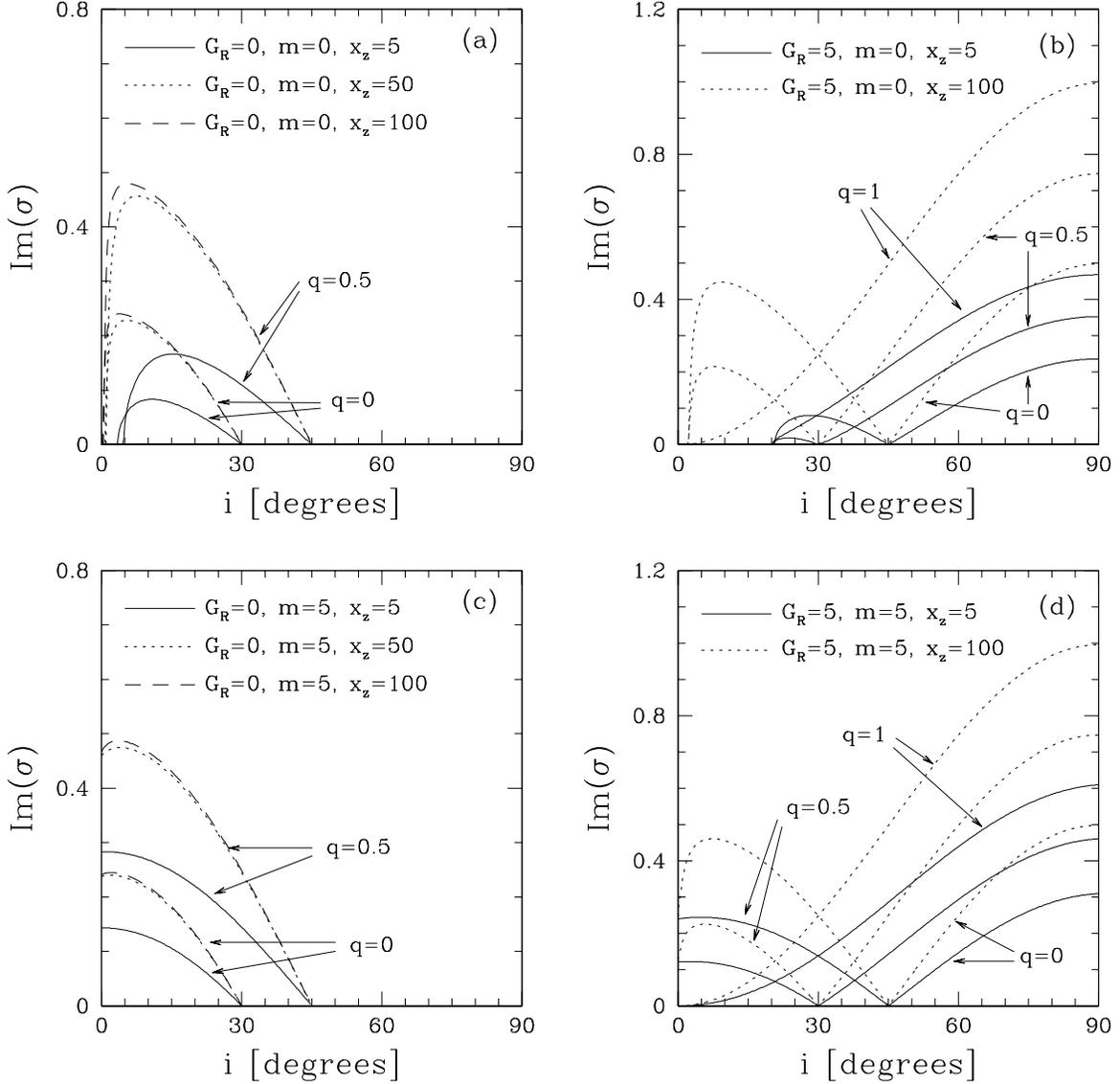}
\caption{
Effects of $i$ and $q$ on the (a and b) axisymmetric  and (c and d) 
non-axisymmetric buoyancy mode frequencies.
A rotation profile $\Omega \propto R^{-3/2}$ is assumed,
and $\xr=10$, $\zeta=0$, and $\GR=5$ are adopted. 
In all four frames, the unstable modes with $i < i_{\rm crit}
\equiv \cos^{-1} \sqrt{(1+q)/2}$ correspond to toroidal buoyancy modes,
whereas poloidal buoyancy modes have $i>i_{\rm crit}$. Note that
at $i=0$, no unstable ATB mode exists while its maximum growth rates
are achieved at $i\rightarrow 0$ as $\xz\rightarrow \infty$.
}
\end{figure}

\clearpage
\begin{figure}
\figurenum{8}
\epsscale{1.0}
\plotone{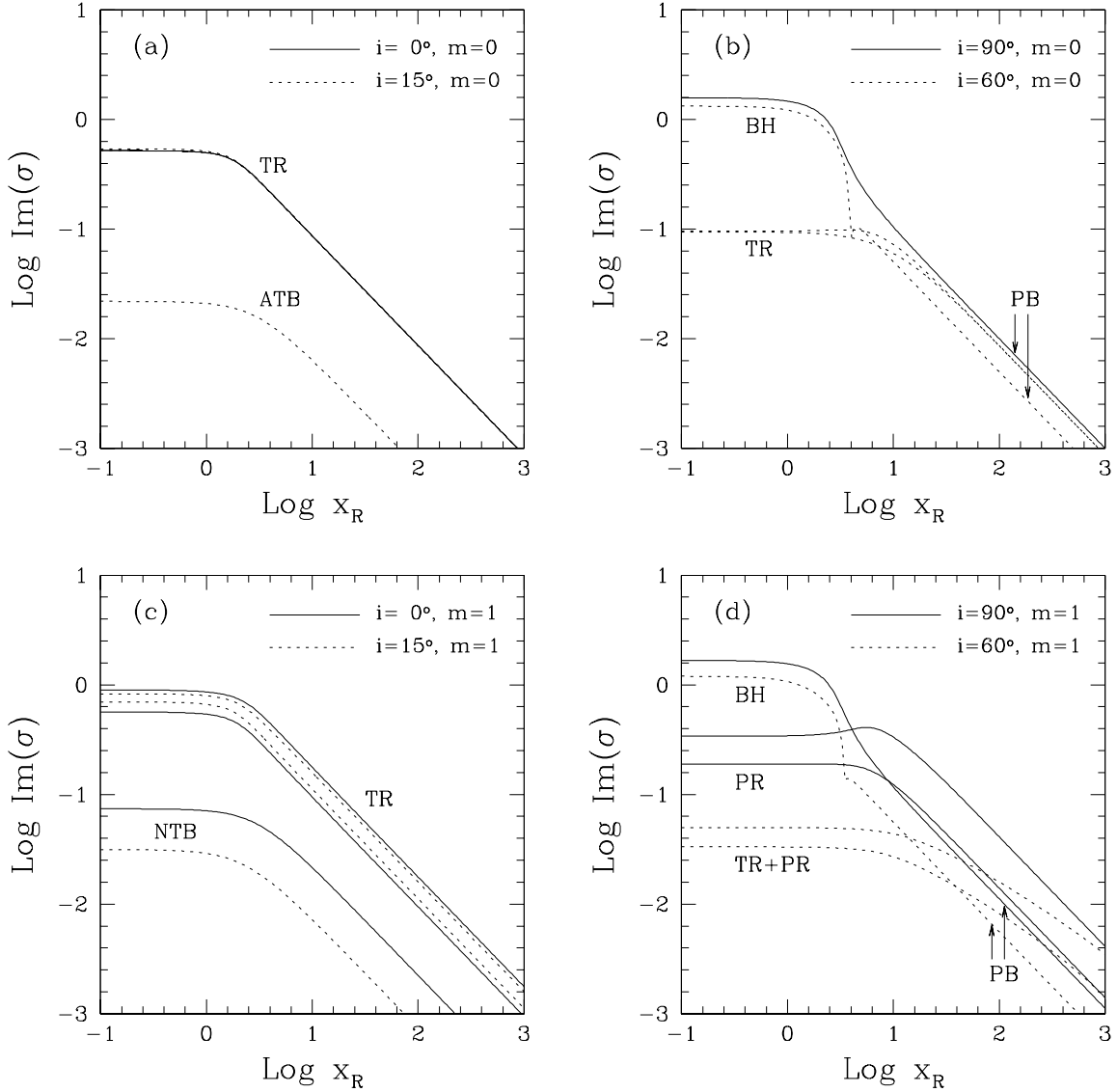}
\caption{
Effects of changing $\kr\equiv \xr/\Ro$ on the 
unstable/overstable axisymmetric (a and b)
and non-axisymmetric (c and d) modes. A rotation profile
$\Omega \propto R^{-3/2}$ is assumed, and
$\xz=2$ and $\zeta=0$ for all cases; $q=0.8$,
$\GR=0$ for left frames  and  $q=0$, $\GR=5$ for right frames are
adopted. BH modes cease to exist when $\xr \gtrsim 3$, but all the
other modes follow Im($\sigma) \sim \xr^{-1}$ at $\xr \gg 1$,
which is in good
agreement with the asymptotic dispersion relations presented in
the text. Since kinematic shear causes $\kr$ to increase linearly
with time, the growth of perturbations would occur in a power-law
fashion rather than an exponential one at later time of evolution.
}
\end{figure}

\clearpage
\begin{figure}
\figurenum{9}
\epsscale{1.0}
\plotone{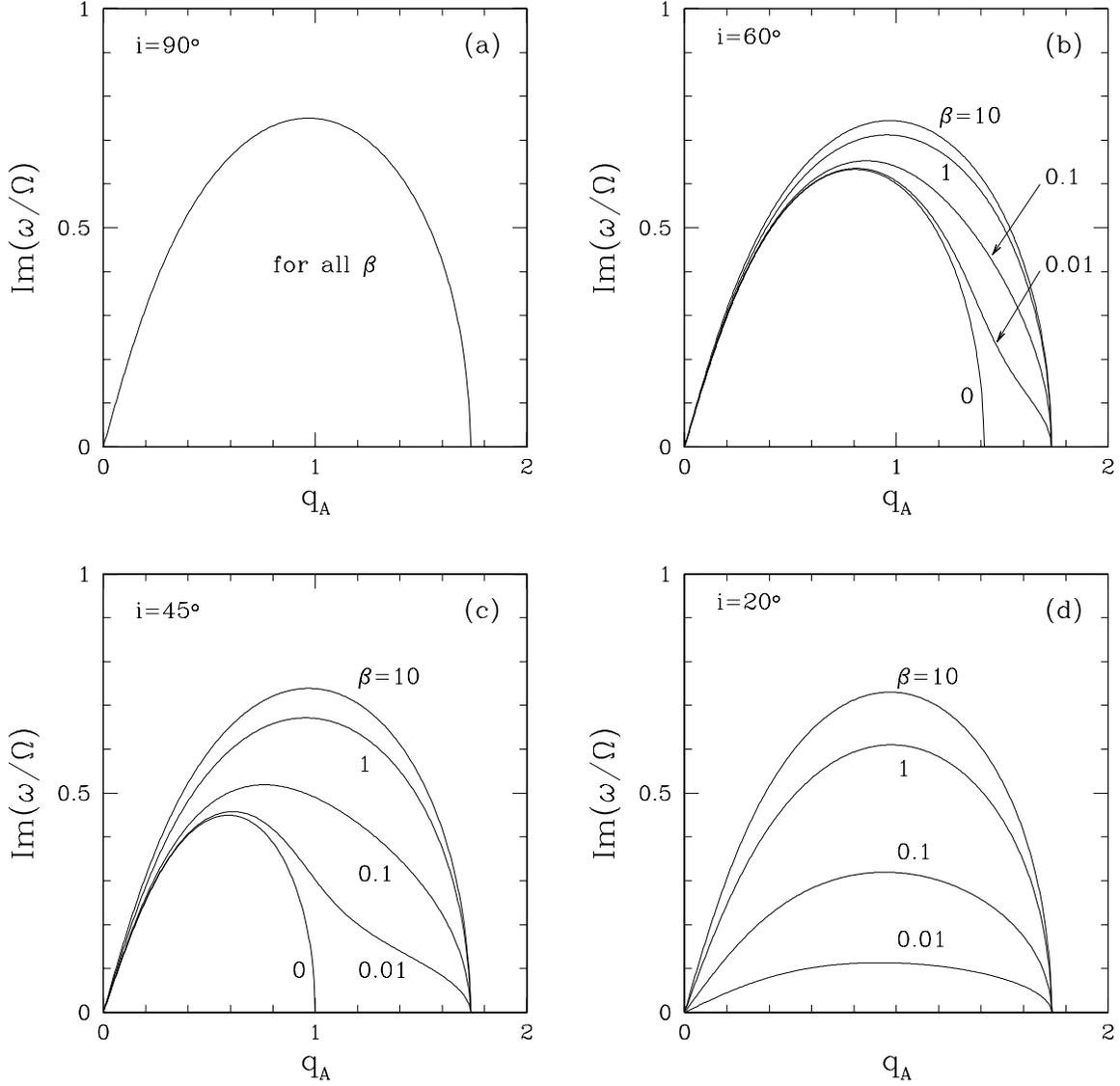}
\caption{
Temperature effect of the compressible BH instability.
The abscissa is the normalized wavenumber $\qA
\equiv (\bold{k}\!\cdot\!\bold{v}_{\rm A})/\Omega$
($=\kz\Vaz/\Omega$ for $m=0$).
A rotation profile $\Omega \propto R^{-3/2}$ is assumed and
no radial perturbation is considered ($\kr=0$).
Curves with different $\beta \equiv \cs^2/\Va^2$ show
how thermal effects modify the BH instability. For $i=90^{\rm o}$,
the BH growth rate is independent of $\beta$,
while $\beta$ has a significant impact
on the growth rates when $i\neq 90^{\rm o}$. In a cold MHD limit ($\beta=0$),
no BH mode is expected if $i<30^{\rm o}$.
}
\end{figure}

\clearpage
\begin{figure}
\figurenum{10}
\epsscale{1.0}
\plotone{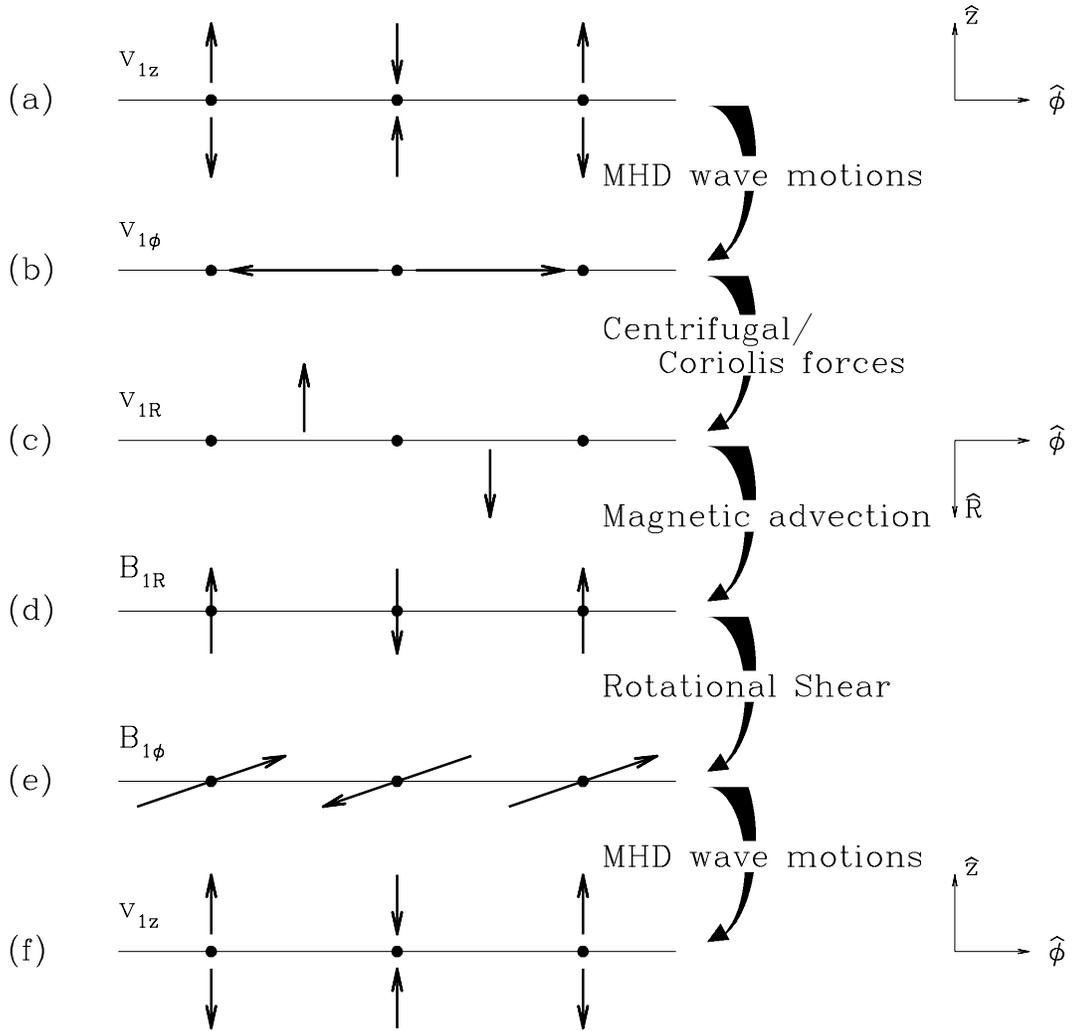}
\caption{
Schematic diagram showing the development of the
NMRI with pure toroidal fields. Magnetic fields, thermal pressure,
and differential rotation with
$d\Omega/dR <0$ all work in cooperation to amplify applied disturbances.
See text for the detailed explanation.
}
\end{figure}

\clearpage
\begin{figure}
\figurenum{11}
\epsscale{1.0}
\plotone{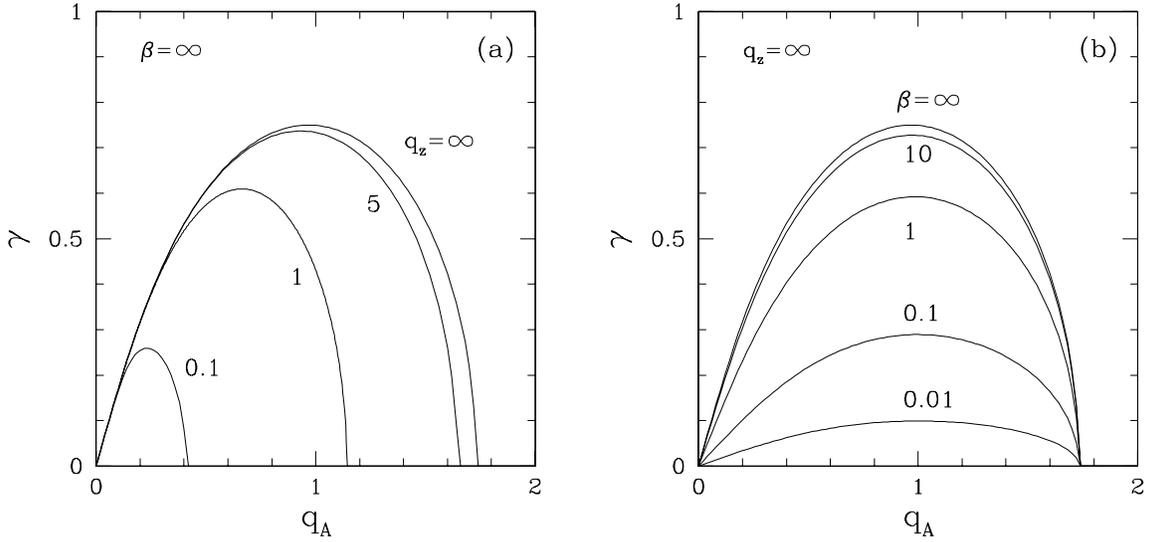}
\caption{
Non-axisymmetric magnetorotational instability of toroidal magnetic
fields. Here, we define
$\qA \equiv (\bold{k}\!\cdot\!\bold{v}_{\rm A})/\Omega$ ($=\Vap m/R\Omega
\equiv \qm$ in the text
for $\Vaz=0$),
$\qz \equiv \Va\kz/\Omega$ ($=\Vap\kz/\Omega$ for $i=0^{\rm o}$),
and $\beta\equiv \cs^2/\Vap^2$.
Rotation with $\Omega\propto R^{-3/2}$ is assumed.
The NMRI instability becomes more unstable with (a) higher $q_{\rm z}$ and
(b) higher sound speed. The critical wavenumber $q_{\rm m}$ is
independent of $\beta$, but the maximum growth rates are sensitive to
temperature for $\beta \lesssim 10$. 
For $\qm \ll 1$, $\gamma = \qm\sqrt{3\beta/(1+\beta)}$.
Eventually at $\beta=0$, no NMRI is expected.
}
\end{figure}

\clearpage
\begin{figure}
\figurenum{12}
\epsscale{1.0}
\plotone{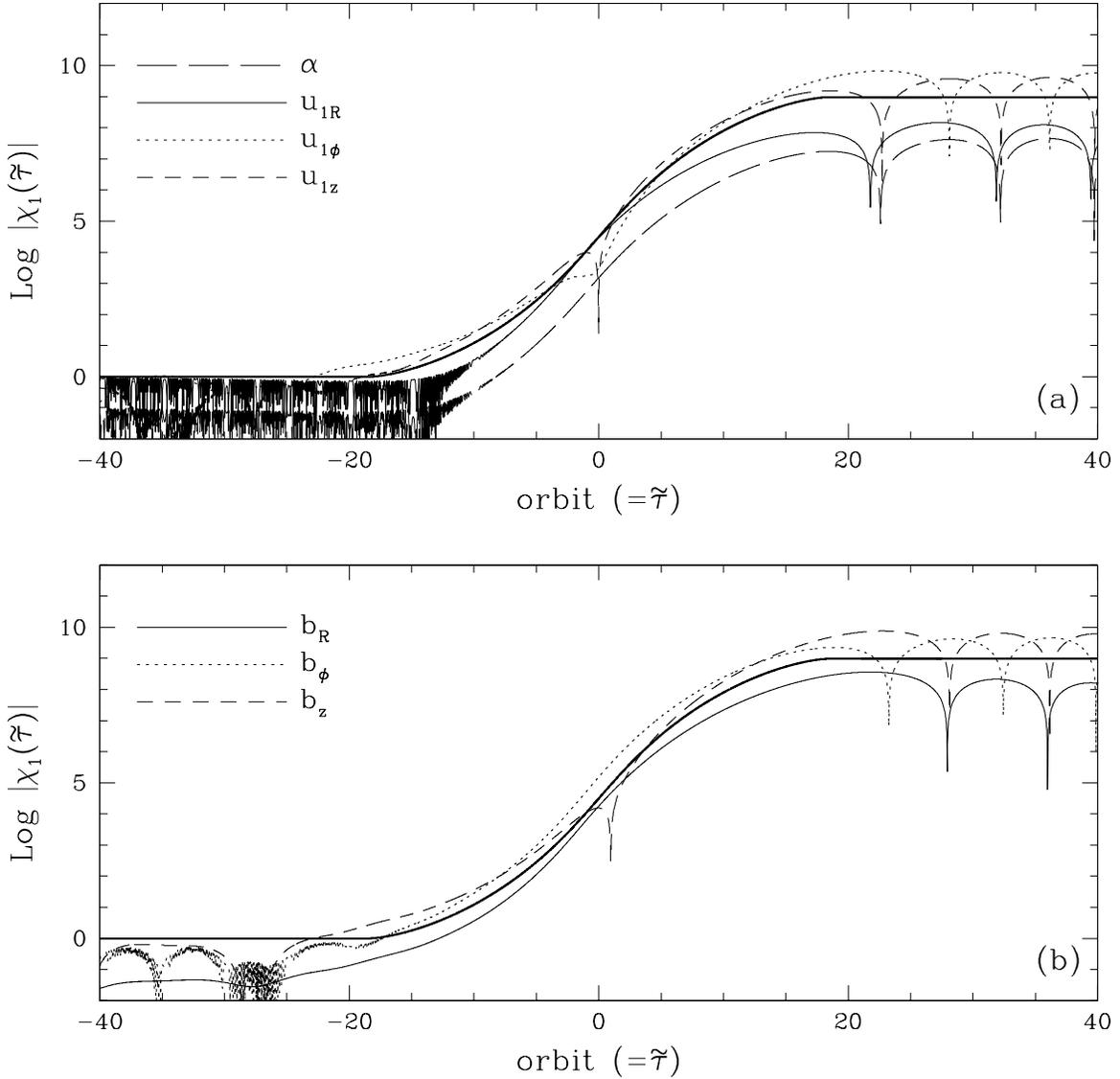}
\caption{
An exemplary run of the NMRI shearing sheet equations for $\qm=0.1$,
$\qz=1$, and $\beta=100$. We choose the initial conditions as
$b_{\rm R}=0.01$, $b_{\rm z}=0.4$, and 0.1 for the other variables,
and integrate the system of equations from $\tlt=-42.44$ with which
the initial perturbed magnetic field is divergence free.
Overall evolution can be divided into three stages: initial 
relaxation phase ($\tlt<-20$), instability phase($|\tlt|<20$), and 
stable oscillation phase ($\tlt>20$). Most of growth occurs when
$|\tlt|$ is relatively small. 
Although kinematic growth
of the radial wavenumber eventually stops the further growth of 
disturbances, the net amplification is tremendous.
The coherent wavelet solution (eq.\ [76]) represented by thick solid lines 
in (a) and (b) is in excellent agreement with the shearing sheet 
results.
}
\end{figure}

\clearpage
\begin{figure}
\figurenum{13}
\epsscale{1.0}
\plotone{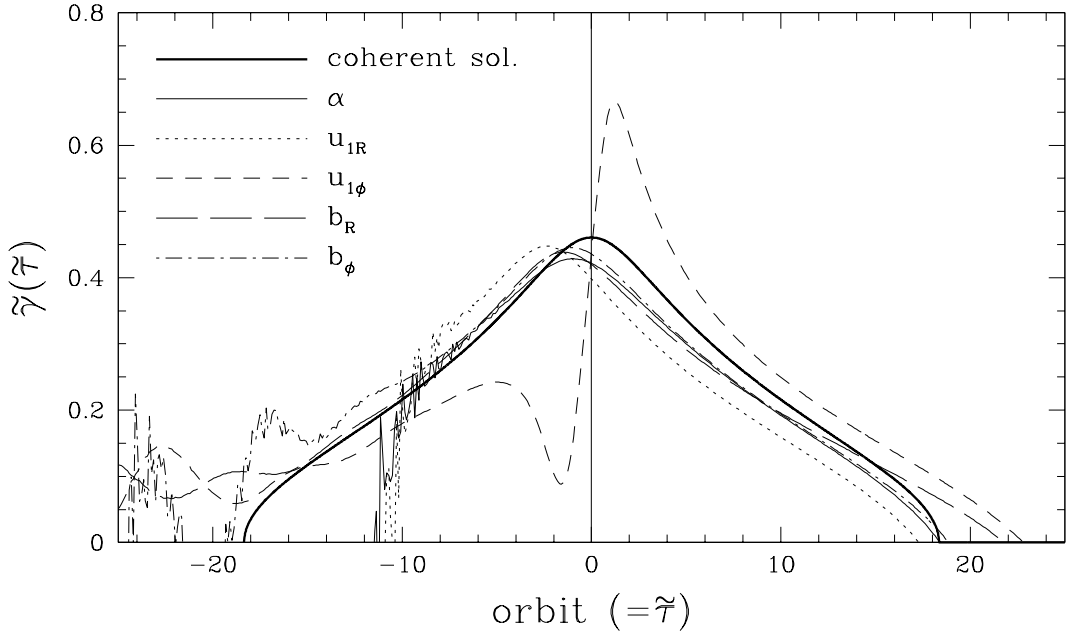}
\caption{
Instantaneous growth rates
$\tilde{\gamma}(\tlt)$  for the NMRI with pure toroidal fields
are plotted against the orbit $\tlt$ $(\equiv t\Omega/2\pi$) for
$\qm (= \Vap m/R\Omega) = 0.1$,
$\qz (= \Vap\kz/\Omega) = 1$, and $\beta=100$.
The thick curve representing the coherent wavelet solution 
agrees fairly well with the behaviors of 
various curves for individual variables 
from the direct numerical
integration of the shearing sheet equations.
The NMRI shows only a temporary growth, but
the net amplification is about 9 orders of magnitude in this example.
Kinematic shear increases $\qR$ with time so that eventually it suppresses
the instability completely at orbit $\tlt\simeq 18.3$.
$\tilde{\gamma}(\tlt)$ is symmetric with respect to $\tlt\sim 0$
and the coherent solution can well be fitted with eq.\ (77).
}
\end{figure}

\clearpage
\begin{figure}
\figurenum{14}
\epsscale{1.0}
\plotone{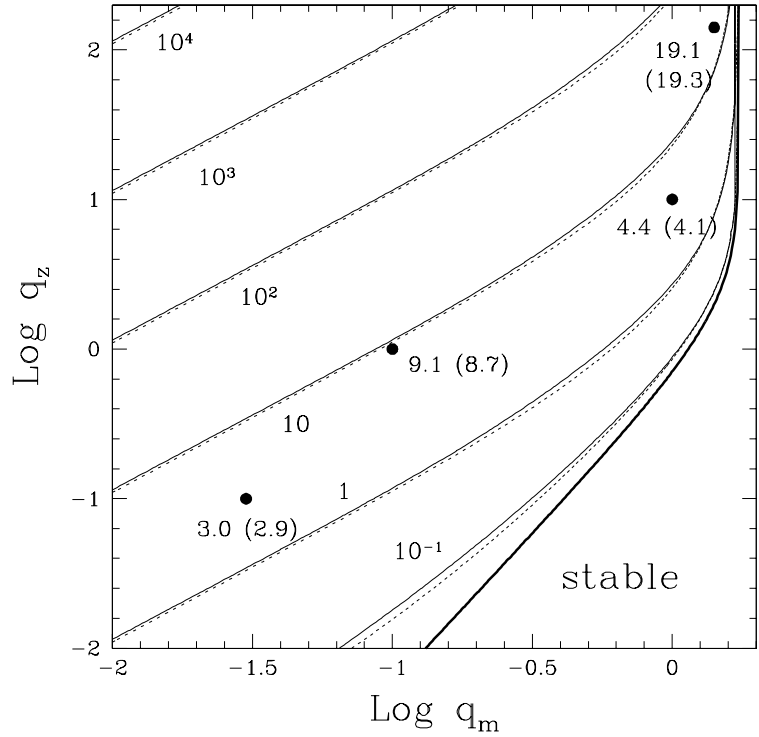} \vspace{-0.7in}
\caption{
The total amplification magnitude $\Gamma(\infty)$ of the 
NMRI with pure toroidal magnetic fields is
drawn on the $\qm\!-\!\qz$ plane.
Solid contours are computed from the approximate analytic estimate,
eq.\ (79), while dotted contours represent the results from direct
numerical evaluation of eqs.\ (75) and (76).
Four dots correspond to
the results of the direct temporal integrations
of the shearing sheet equations: the adapted parameters are
$\beta=100$ and
($\qm$, $\qz$) = (0.03, 0.1), (0.1, 1), (1, 10), and
($\sqrt{2}$, 100$\sqrt{2}$) from lower-left to upper-right.
The numbers labeling a dot is the exact and estimated
(in parentheses) total amplification magnitudes.
Note that the analytic estimate predicts the true amplification
magnitude very closely.
$\Gamma(\infty)$ has a higher value with larger $\qz$ and smaller
$\qm$. The heavy contour corresponds to the locus of the marginal
stability.
}
\end{figure}

\clearpage
\begin{figure}
\figurenum{15}
\epsscale{1.0}
\vspace{-1.3in} \plotone{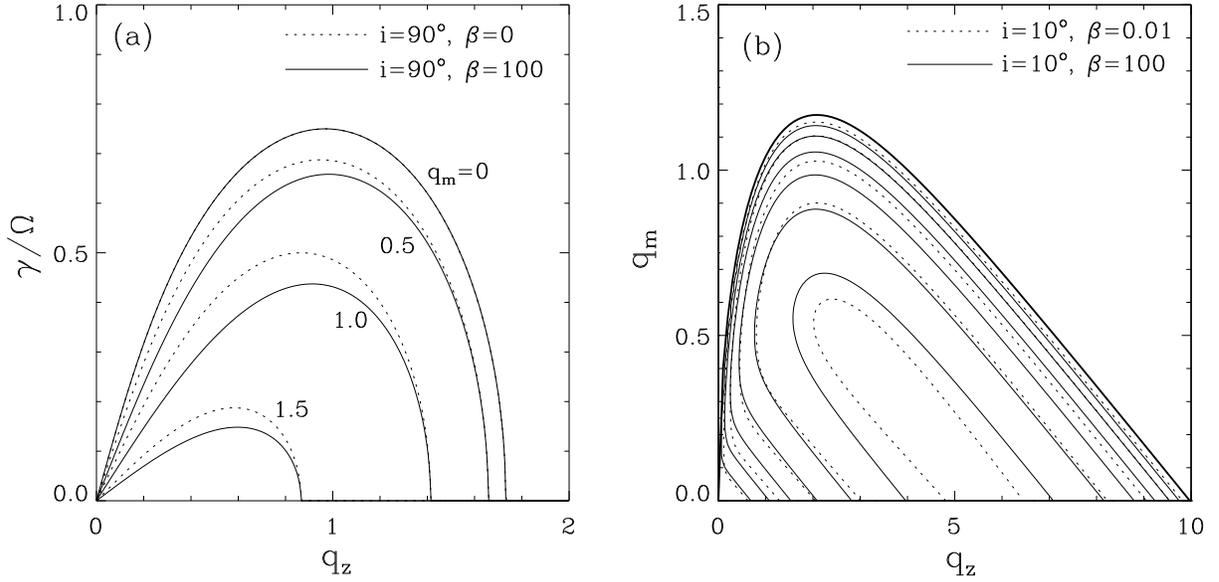} \vspace{-0.2in}
\caption{
Growth rates of the generalized MRI are drawn
as functions of $\qz\equiv\Va\kz/\Omega$ and $\qm\equiv\Va m/R\Omega$,
(a) for $i=90^{\rm o}$ and
(b) for $i=10^{\rm o}$.
We put $\kr=0$ in both frames. 
In frame (b), solid contours corresponding to $\beta=100$ show
$\gamma/\Omega$= 0.7,0.6,...,0.2, from inside to outside, while
dotted contours corresponding to $\beta=0.01$ show
$\gamma/\Omega$= 0.1,0.08,...,0.02, from inside to outside.
As $\qm$ increases, both growth rates
and unstable ranges of $\qz$ decrease. Eventually, if
$\qm >1.73$ for $i=90^{\rm o}$ or $\qm >1.17$ for $i=10^{\rm o}$,
the generalized MRI is completely suppressed by MHD wave motions.
When $i=90^{\rm o}$,
thermal pressure tends to reduce the growth rates
by activating azimuthal wave motions if $\qm \neq 0$;
however, the similarity between the $\beta=0$, 100 curves in (a) shows
its effect is not significant.
The uppermost thick curve in (b) represents
the locus of the marginally critical wavenumbers (cf.\ eq.\ [81b]) above
which no instability can be expected.
}
\end{figure}

\clearpage
\begin{figure}
\figurenum{16}
\epsscale{1.0}
\vspace{-0.7in} \plotone{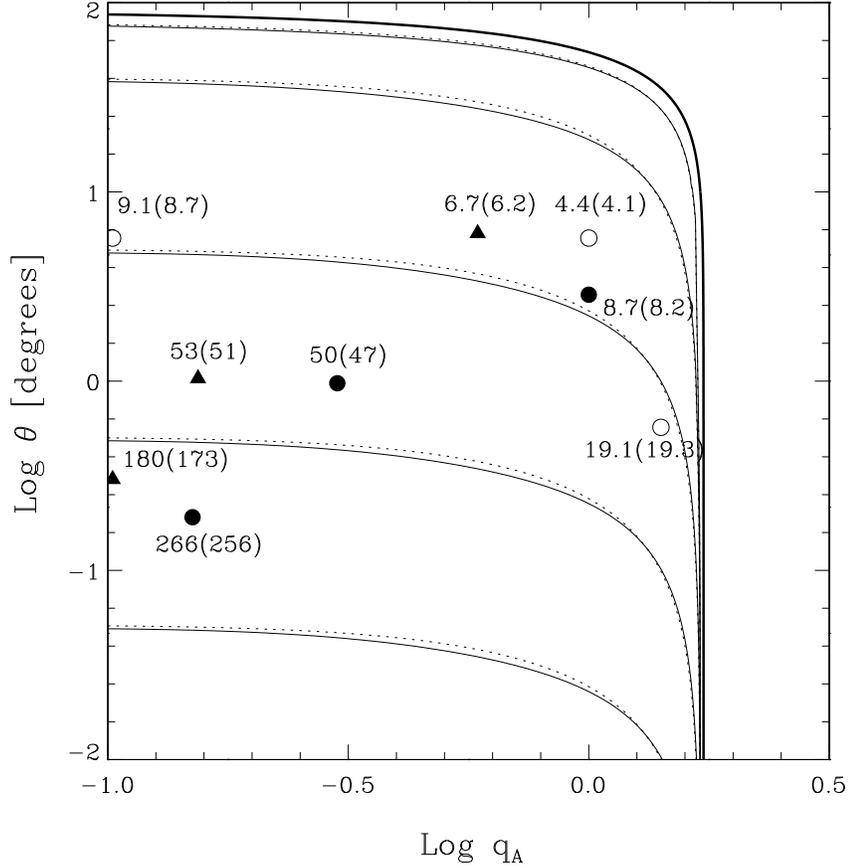} \vspace{-0.5in}
\caption{
The total amplification magnitude $\Gamma(\infty)$ of the generalized
incompressible MRI is drawn as a function of 
$\qA\equiv ({\bf k\cdot\Va})/\Omega$ and $\theta \equiv \tan^{-1} (m/R\kz)$.
Thin solid contours are computed from the approximate analytic estimate,
eq.\ (83), while dotted contours represent the results from
numerical evaluation of eqs.\ (76) and (82). Both types of contours show
$\Gamma(\infty)=10^3, 10^2, 10^1, 1, 0.1$, from bottom to top.
We adopt a Keplerian rotation profile.
The  results of the direct temporal integrations of
the shearing sheet equations with $\beta=100$ are shown with
different symbols.
Open circles corresponding to $i=0^{\rm o}$, adopted from Fig.\ 14,
are for ($\qm$, $\qz) \equiv \Va(m/R,\kz)/\Omega$ 
= (0.1, 1), (1, 10), and ($\sqrt{2}$, 100$\sqrt{2}$),
filled circles corresponding to $i=90^{\rm o}$
for ($\qm$, $\qz$)=($5\times 10^{-4}$, 0.15), (0.005, 0.3), and (0.05, 1),
and filled triangles with $i=30^{\rm o}$
for ($\qm$, $\qz$)=(0.001, 0.2), (0.005, 0.3), and (0.1, 1), from left to right.
The numbers labeling each symbol is the exact and estimated
(in parentheses) total amplification magnitudes.
$\Gamma(\infty)$ has a higher value with smaller
$\theta$ but nearly independent of $\qA \ll 1$. 
The heavy curve represents the locus of the marginal
stability.
}
\end{figure}

\clearpage
\begin{figure}
\figurenum{17}
\epsscale{1.0}
\plotone{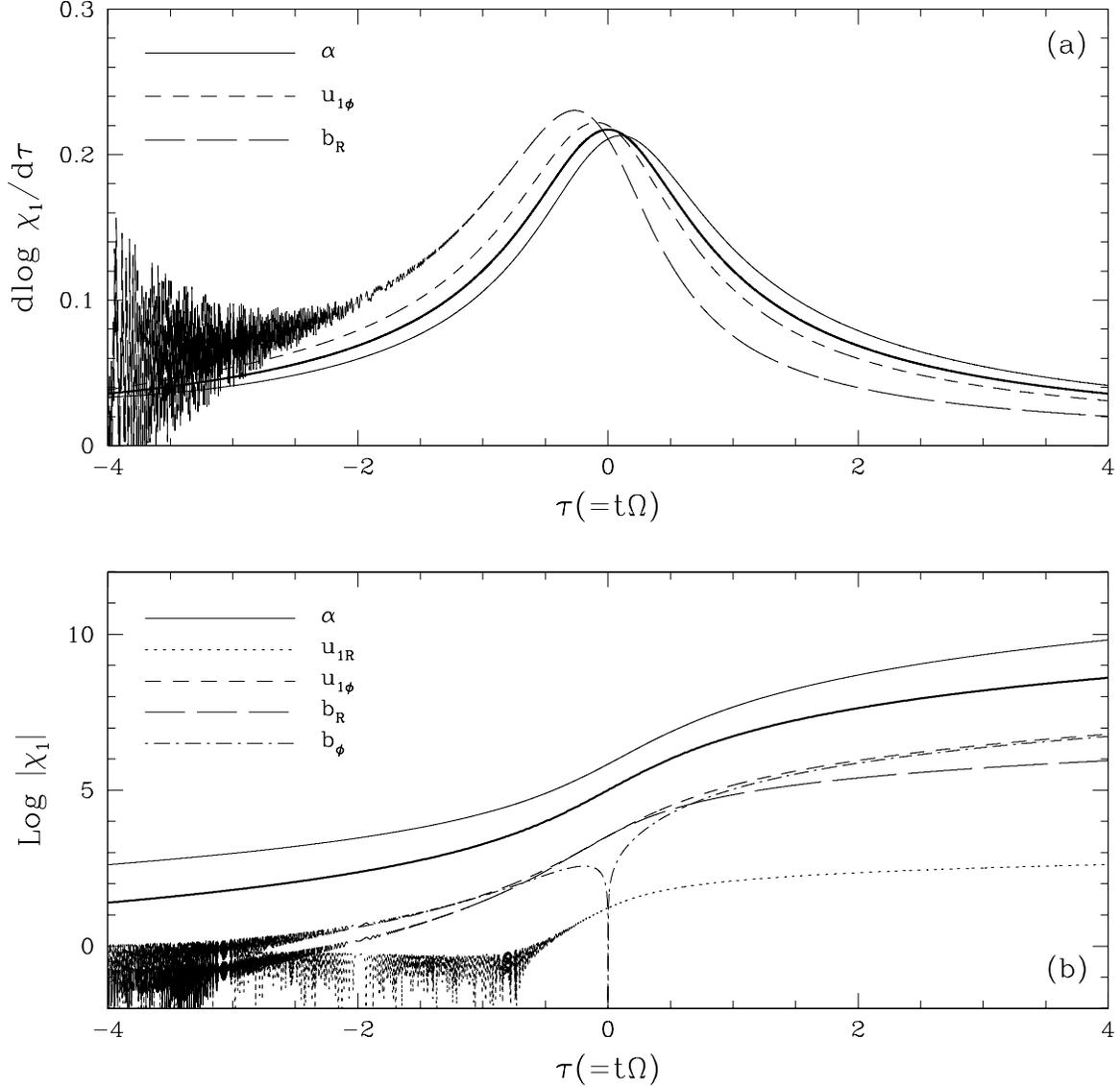}
\caption{
(a) Instantaneous growth rates and (b) evolutionary
behaviors of physical variables are displayed for the 
non-axisymmetric toroidal buoyancy modes
with $q=0$, $R\Omega = 0.1 \Vap$, and $m=100$.
As initial conditions, we choose 0.1 for all variables except $b_{\rm R}=0.01$.
Various curves are computed from the direct temporal integrations
of the shearing sheet equations. Thick solid lines, drawn from the normal
mode solution, eq.\ (52), and
its time integration, eq.\ (86), with  $\kr(t) = -mt\Omega'$ provide
excellent predictions of the numerical results.
A Keplerian rotation is assumed and vertical shear is neglected. 
The most significant growth of the NTB modes occurs when 
$\kr$ is small. With increasing $\kr$, the growth rate is 
gradually reduced. The rapid oscillations for $\tau < 0$ is due
to MHD waves associated with high $|\kr|$, which are smoothed out
as disturbances grow.
}
\end{figure}

\clearpage

\begin{deluxetable}{ccccc}
\footnotesize
\tablecaption{Summary of the unstable/overstable mode properties.
\label{tbl-1}}
\tablewidth{0pt}
\tablehead{
\colhead{Type}                         &
\colhead{
  \begin{tabular}{c}
     Geometry of \\ Perturbation
  \end{tabular}
        }                              &
\colhead{
  \begin{tabular}{c}
     Physical \\  Mechanism
  \end{tabular}
        }                              &
\colhead{
  \begin{tabular}{c}
     Magnetic field \\  Configuration
  \end{tabular}
        }                              &
\colhead{
  \begin{tabular}{c}
     Stability \\  Character
  \end{tabular}
}
}
\startdata
FM  &$m=\kz=0$  & global mode & toroidal & unstable   \\
    &           &             &          &            \\
ATB &           & buoyancy    & toroidal & unstable   \\ 
PB  & $m=0$     & Parker      & poloidal & unstable   \\
BH  &$\kz\neq 0$& MRI         & poloidal & unstable   \\
TR  &           & resonance   & toroidal & overstable \\
    &           &             &          &            \\
NTB &           & Parker      & toroidal & unstable   \\
PR  & $m\neq 0$ & resonance   & poloidal & overstable \\
GPB & $\kz\neq0$&geometric    & poloidal & unstable   \\
NMRI&           & MRI         & toroidal & unstable   \\
\enddata

\end{deluxetable}

\end{document}